\documentclass[fleqn,usenatbib]{mnras}

\newcommand{\dustcode}{\textsc{DustCharge}}

\newcommand{\pe}{\mathrm{pe}}
\newcommand{\Ion}{\mathrm{ion}}

\usepackage{newtxtext,newtxmath}
\usepackage{multicol}
\usepackage{mathtools, cuted}

\usepackage[T1]{fontenc}
\usepackage{ae,aecompl}

\usepackage{booktabs}

\usepackage{graphicx}	
\usepackage{amsmath}	
\usepackage{amssymb}	

\defcitealias{Weingartner2001PhotoelectricHeating}{WD01}	
\defcitealias{Draine1987CollisionalGrains}{DS87}	
\defcitealias{Joshi2018OnConditions}{J18}

\title[Dust charge in the ISM.]{Dust charge distribution in the interstellar medium}

\author[J. C. Ib\'a\~nez-Mej\'{\i}a]{Juan C. Ib\'a\~nez-Mej\'{\i}a,$^{1, 2}$\thanks{E-mail: ibanez@ph1.uni-koeln.de}
Stefanie Walch,$^{1}$\thanks{E-mail: walch@ph1.uni-koeln.de}
Alexei V. Ivlev,$^{2}$
Seamus Clarke,$^{1}$
\newauthor
Paola Caselli$^{2}$
 and 
Prabesh R. Joshi$^{1}$
\\
$^{1}$I. Physikalisches Institut, Universit\"at zu K\"oln, Z\"ulpicher Str. 77, D-50937 K\"oln, Germany\\
$^{2}$Max-Planck-Institute for Extraterrestrial Physics (MPE), Giessenbachstr. 1, D-85748 Garching, Germany
}

\date{Accepted XXX. Received YYY; in original form ZZZ}

\pubyear{2018}

\begin{document}
\label{firstpage}
\pagerange{\pageref{firstpage}--\pageref{lastpage}}
\maketitle

\begin{abstract}
We investigate the equilibrium charge distribution of dust grains in the interstellar medium (ISM). Our treatment accounts for collisional charging by electrons and ions, photoelectric charging due to a background interstellar radiation field, the collection of suprathermal cosmic ray electrons and photoelectric emission due to a cosmic ray induced ultraviolet radiation field within dense molecular clouds. We find that the charge equilibrium assumption is valid throughout the multi-phase ISM conditions investigated here, and should remain valid for simulations with resolutions down to AU scales.
%
%
The charge distribution of dust grains is size, composition, and ISM environment dependent: local radiation field strength, $G$, temperature, $T$, and electron number density, $n_{\mathrm{e}}$.
The charge distribution is tightly correlated with the ``charging parameter'', $G\sqrt{T}/n_{\mathrm{e}}$.
In the molecular medium, both carbonaceous and silicate grains have predominantly negative or neutral charges with narrow distributions. In the cold neutral medium, carbonaceous and silicate grains vary from negative and narrow distributions, to predominantly positive and wide distributions depending on the magnitude of the charging parameter. In the warm neutral medium, grains of all sizes are positively charged with wide distributions.
We derive revised parametric expressions that can be used to recover the charge distribution function of carbonaceous and silicate grains from 3.5~{\AA} to 0.25~$\mu$m as a function of the size, composition and ambient ISM parameters.
Finally, we find that the parametric equations can be used in environments other than Solar neighborhood conditions, recovering the charge distribution function of dust grains in photon dominated regions.

\end{abstract}




\section{Introduction}

Dust grains play an important role in the thermal, and chemical evolution of the interstellar medium (ISM).
%
Grains interact with the gas through collisions, as well as long-range Coulomb forces, and experience gravitational attraction and varying coupling strengths to magnetic field lines depending on their charge \citep{Draine1979OnGas, Lazarian2001GrainGas, Yan2004DustTurbulence}.
The charge distribution of dust grains in the ISM depends on the flux of charged particles onto and from the grain surface, due to collisions with ions and electrons, and photoionization, which  strongly dependent on the local properties of the ISM, such as: temperature, density \citep{Draine1979OnGas, Draine1979DestructionDust, Draine1987CollisionalGrains}, local radiation field \citep{Weingartner2000ForcesFields, Weingartner2001PhotoelectricHeating, Weingartner2006PhotoelectricRadiation} and, within dense molecular clouds and protostellar disks, the cosmic ray (CR) ionization rate \citep{Ivlev2015InterstellarEffects,Padovani2018Cosmic-rayDiscs}. 
Photoelectrons ejected from dust grains constitute the primary heating mechanism in the warm and cold phases of the ISM \citep{Bakes1994TheHydrocarbons, Wolfire1995TheMedium, Weingartner2001PhotoelectricHeating}, where the photoelectron heating efficiency is strongly dependent on the charge of the grain.
Dust grains provide a ``surface'' for simple and complex molecules to form \citep[e.g.][]{Vasyunin2017FormationApproach}, and the depletion of elements from the gas phase onto a population of dust grains is strongly influenced by the grain charge, enhancing or decreasing the collisional cross sections \citep{Weingartner1999InterstellarGrains, Ferrara2016TheGalaxies, Zhukovska2018IronDepletions}.
Dust charge also influences the growth and destruction of dust grains, affecting dust-dust collisions, as well as modifying the sputtering yields \citep{Barlow1978TheSputtering,Draine1979DestructionDust,Tielens1994TheShocks}.

Sticking collisions with ions and electrons charge the grain positively or negatively, respectively. 
The collisional rate of change of the dust charge depends on the relative velocity distribution of the collisional partners and their local abundances.
Given that the electrons are lighter than ions, collisions with them are more frequent and negatively charge the grain \citep{Draine1987CollisionalGrains, Draine2011PhysicsMedium}. 
On the other hand, energetic photons which impact the grain can eject an electron, and positively charge the grain \citep[hereafter WD01]{Weingartner2001PhotoelectricHeating}, also known as photoemission. 
The photoemission rate of electrons exponentially decays with visual extinction \citep{Bakes1994TheHydrocarbons} and may be completely absent within dense molecular clouds.
In these dense environments, cosmic rays produce suprathermal electrons and ions that can collisionally ionize the grains, as well as induce an ultraviolet radiation field leading to photoelectric charging, strongly influencing the charging of dust grains within molecular clouds  \citep{Ivlev2015InterstellarEffects}.

Charged dust grains may have a strong impact in the properties and propagation of shocks in the interstellar medium, affecting the gas-grain coupling, and influencing the dynamics and chemistry of shock propagation \citep{Pilipp1994GrainsShocks,Caselli1997AstronomyAstrophysics., Ciolek2004MultifluidShocks, Chapman2005DustClouds, Guillet2007ShocksClouds, Guillet2009ShocksClouds, Guillet2011ShocksClouds}.
The momentum transfer between dust grains and gas is tightly coupled to the charge state of the grains in the pre-shock and shocked regions \citep{Flower2003TheClouds}.
In particular, charge fluctuations during the passage of the shock can strongly influence the coupling and processing of dust grains in shocks \citep{Guillet2007ShocksClouds}.

In this paper, we analyze the distribution of grain charges in the warm neutral (WNM), cold neutral (CNM) and cold molecular (CMM) phases of the ISM.
%
The structure of this paper is as follows. 
In Section \ref{sec:charge}, we describe the process of dust charging for a grain population in the ISM, and describe the properties of the colliding flow simulations used to generate the turbulent multi-phase ISM.
In Section \ref{sec:results}, we compute the charge distribution function for carbonaceous and silicate grains and explore their dependence with grain size, composition and ISM properties.
Section \ref{sec:discussion} proposes revised parametric equations to estimate the charge distribution of dust grains as a function of the environment, and tests these equations for grain sizes and ISM environments outside the ones used to calibrate them.
Final remarks and conclusions are in Section \ref{sec:conclusions}.

\section{Charging of dust grains}
\label{sec:charge}

Charging of dust grains is a discrete process that can be divided into two main channels: collisional charging, accounting for both negative and positive charging, when electrons and ions collide and stick to the surface of the grain \citep[hereafter DS87]{Draine1987CollisionalGrains}, and radiative charging, positively charging the dust as photoelectrons are kicked out of the grain from the absorption of an energetic photon \citep{Weingartner2001PhotoelectricHeating}. 
Cosmic ray-induced (CR) charging processes are encompassed within these two main channels.
CRs ionize the gas producing protons and electrons that can collisionally charge the grains, as well as induce an ultraviolet radiation field, due to the fluorescence of H$_{2}$, which can radiatively charge the grains \citep{Ivlev2015InterstellarEffects}.

\subsection{Collisional charging}

We adopt the collisional charging rate, $J_{i}$, introduced by \citetalias{Draine1987CollisionalGrains}, assuming the colliding partners have a Maxwellian distribution, given by:
\begin{eqnarray}
	J_{i}(Z) = n_{i}s_{i}\left(\frac{8 k T}{\pi m_{i}}\right)^{1/2} \pi a^{2} \tilde{J}\left( \tau, \nu \right),
    \label{eq:Ji}
\end{eqnarray}
where the subscript $i$ represents the collisional partner (electron or ion, e.g. H$^{+}$, C$^{+}$), $Z$ is the charge in units of proton charge, $\tau = akT/q_{i}^2$ is the ``reduced temperature'', where $a$ corresponds to the grain size, and $q_{i}$ to the charge of the collisional partner, $\nu = Ze/q_{i}$ is the ratio of the charge on the grain to the charge on the colliding particle, where $e$ is the proton charge, $n_{i}$, $q_{i}$, and $m_{i}$ are the density, charge, and mass of the colliding partners, and $\tilde{J}(\tau, \nu)$ is the reduced rate coefficient --accounting for the attractive/repulsive Coulomb interaction between the charged grain and the colliding partners-- given by equations (3.2), (3.3), (3.4), and (3.5) in \citetalias{Draine1987CollisionalGrains}. 
For the sticking coefficient, $s_{i}$, we adopt $s_{\Ion}=1.0$ for ions and $s_{e}$ given by a combination of parameters accounting for the probability of inelastic scattering and radiative stabilization of colliding electrons, given by equations (27) through (30) in \citetalias{Weingartner2001PhotoelectricHeating}.
Collisional charging is weakly dependent on the grain composition affecting the sticking coefficients. 
We describe the implementation and benchmarking of the collisional charging rates in Appendix \ref{appendix_collisional_charging}.

\subsection{Radiative charging}

When an energetic photon impacts the surface of a dust grain, it may release a photoelectron from the dust grain \citep{Bakes1994TheHydrocarbons, Weingartner2001PhotoelectricHeating, Ivlev2015InterstellarEffects}.
The photoemission rate of electrons is given by
\begin{equation}
\begin{split}
\qquad	J_{\pe}(Z)= \pi a^{2} \int_{\tiny{\nu_{\mathrm{pet}}}}^{\tiny{\nu_{\mathrm{max}}}} d\nu Y(h\nu,& Z, a) Q_{\mathrm{abs}}(\nu) \frac{c u_{\nu}}{h\nu} + \\
    &\int_{\nu_{\mathrm{pdt}}}^{\nu_{\mathrm{max}}}d\nu \sigma_{\mathrm{pdt}}(\nu) \frac{cu_{\nu}}{h\nu}, 
   	\label{eq:Jpe}
\end{split}
\end{equation}
where $\nu_{\mathrm{pet}}(Z, a)$ is the photoelectric threshold (pet), or minimum photon frequency required to ionize the grain, and $\nu_{\mathrm{max}}$ is the maximum photon frequency of the incident radiation field,
$Y(h\nu, Z, a)$ is the photoelectric yield, the probability that an electron will be ejected when a photon with energy $h\nu$ is absorbed,  $Q_{\mathrm{abs}}(\nu)$ is the absorption efficiency \footnote{Absorption efficiencies are taken from calculations by \citet{Laor1993SpectroscopicNuclei}, and can be found at http://www.astro.princeton.edu/$\sim$draine/dust/dust.diel.html}, $u_{\nu}$ is the radiation energy density, and $c$ the speed of light.
The second term of equation \ref{eq:Jpe} is only necessary for negatively charged grains, $Z<0$, where $\nu_{\mathrm{pdt}}$ is the photodetachment threshold (pdt) frequency of excess attached electrons \footnote{When the valence band of a dust grain is full, energy levels above it are occupied by excess attached electrons.} and $\sigma_{\mathrm{pdt}}$ is the photodetachment cross section. 

For the photoelectric yield, we use the model provided by \citetalias{Weingartner2001PhotoelectricHeating}, 
based on \citet{Bakes1994TheHydrocarbons} and the refractive indices for carbonaceous (and silicate) grains from \citet{Draine1984OpticalGrains}. 

The radiation energy density, $u_{\nu}$, contains the information of the spectrum of the incident radiation field on the dust grain.
In this work, we account for two sources of radiation, the interstellar radiation field  \citep{Mezger1982TheEmission,Mathis1983InterstellarClouds}, and CR induced H$_{2}$ fluorescence \citep{Prasad1983UVImplications, Ivlev2015InterstellarEffects}, $u_{\nu} = u_{\nu, 0} + u_{\nu, \text{\tiny{CR}}}$, such that the total photoelectric emission rate is given by the combination of the radiative emission by the interstellar radiation field and by the CR-induced radiation field:
\begin{eqnarray}
J_{\pe, \mathrm{tot}} = J_{\pe, 0} + J_{\pe, \text{\tiny{CR}}}.
\label{eq:Jpe_tot}
\end{eqnarray}

The fluorescence of H$_{2}$ was proposed by \citet{Prasad1983UVImplications} for the generation of ionizing radiation within dense molecular clouds. 
Collisions of CRs and secondary electrons with molecular hydrogen produce excitations followed by spontaneous emission of ultraviolet photons, resulting in a wealth of narrow lines in the energy range between 11.2 and 13.6 eV.
Following the calculation by \citet{Cecchi-Pestellini1992CosmicClouds} and its implementation by \citet{Ivlev2015InterstellarEffects}, we compute the CR induced UV flux (photon cm$^{-2}$ s$^{-1}$) as:
\begin{eqnarray}
F_{\text{\tiny{UV}}} \simeq 960 \left( \frac{1}{1-\omega} \right) \left( \frac{\zeta}{10^{-17} \mathrm{s}^{-1}} \right) \left( \frac{N_{\text{H}_{2}}/A_{v}}{10^{21} \mathrm{cm}^{-2} \mathrm{mag}^{-1}} \right) \left( \frac{R_{V}}{3.2} \right)^{1.5},
\label{eq:CR_FUV}
\end{eqnarray}
where $\omega$ is the dust albedo, $\zeta$ is the cosmic ray ionization rate, $N_{\mathrm{H}_{2}}/A_{v}$ is the gas-to-extinction ratio, and $R_{V}$ is a measure of the slope of the extinction curve at visible wavelengths.
We assume the following values in this paper: $\omega = 0.5$ and $R_{V}=3.1$ \citep{Cecchi-Pestellini1992CosmicClouds}, $N_{\mathrm{H}_{2}}/A_{v}=1.87 \times 10^{21}$~cm$^{-2}$~mag$^{-1}$, and CR ionization rate of H$_{2}$, $\zeta$, given by Equation \ref{eq:CR_ionization} in appendix \ref{appendix_CR_charging}.
For the remainder of this paper, we characterize the strength of the CR-induced ultraviolet radiation field in units of the Habing field, $G_{CR} = F_{\text{\tiny{UV}}}*12.4~{\mathrm{eV}} / u^{\tiny{UV}}_{\mathrm{\tiny{Hab}}} c_{\mathrm{light}}$, assuming an average energy of 12.4~eV per CR-induced photon, and where $u^{\tiny{UV}}_{\mathrm{\tiny{Hab}}}=5.33\times10^{-14}$~ergs~cm$^{-3}$ is the energy density in the starlight radiation field between 6 and 13.6~eV estimated by \citet{Habing1968TheA}, and $c_{\mathrm{light}}$ is the speed of light.

The contribution of the CR-induced photoelectric emission can be approximately calculated by \citep{Ivlev2015InterstellarEffects}:
\begin{eqnarray}
	J_{\pe, \text{\tiny{CR}}} (Z) \simeq \pi a^{2} F_{\text{\tiny{UV}}} \langle Y(\nu, Z) Q_{abs}(\nu) \rangle_{\text{\tiny{UV}}},
\label{eq:Jpe_CR}
\end{eqnarray}
where $\langle Y(\nu, Z) Q_{abs}(\nu) \rangle_{{\text{\tiny{UV}}}}$ is averaged over the Lyman and Werner bands.
In contrast to the calculations by \citet{Ivlev2015InterstellarEffects}, we include the dependence of the grain charge through the dependence of the photoemission yield on $Z$. \\

Appendix \ref{appendix_photoelectric_charging}, contains a more detailed explanation of the interstellar radiation field spectra, calculation of photoelectric yields, and benchmarks of charge distribution calculations.

\subsection{Cosmic Ray charging}

We include the charging of dust by CRs following the implementation by \citet{Ivlev2015InterstellarEffects}.
CRs have a dual role in the charging of dust grains. 
On the one hand they influence the ionization state of the gas, generating a collection of suprathermal electrons and protons capable of collisionally charging the grains.
On the other hand, the suprathermal electrons are capable of exciting H$_{2}$ without dissociating it, which then produces a wealth of emission in the UV, capable of photoionizing dust grains, introduced in equation \ref{eq:Jpe_CR}.

As shown by \citet{Ivlev2015InterstellarEffects}, for energies up to $10^{6}$~eV, the
flux of CR-induced electrons is orders of magnitude higher than the flux of CR-induced protons. 
In addition, protons have a smaller contribution to the collisional charging of dust grains due to the large particle mass difference between ions and electrons.
For these two reasons, we ignore the contribution of suprathermal protons on the dust charge distribution calculation.
Following \citet{Draine1979OnGas}, the suprathermal electron collisional charging rate is given by:
\begin{eqnarray}
	J_{e, \text{\tiny{CR}}} = \pi a^{2} \int_{E_{int}}^{\infty} dE 4\pi j_{e}(E)[s_{e}(E) - \delta_{e}(E)],
    \label{eq:J_e_CR}
\end{eqnarray}
where $j_{e}(E)$ is the CR generated electron spectrum, $s_{e}(E)$ is the sticking coefficient, $\delta_{e}(E)$ is the yield of the secondary electrons and $E_{int} = 1.5 \times 10^{-2}$~eV is the intersection energy where the flux of CR-induced electrons drops below the flux of thermal electrons in the plasma. \\

For details of the implementation and benchmarking of the charging effects due to CRs see Appendix \ref{appendix_CR_charging}.

\subsection{Charge distribution function}
\label{sec:charge_distribution}

The competition of the positive and negative charging rates results in charge fluctuations, but given sufficient time to relax, dust grains will converge to a statistical distribution of charges around a mean.
Let $f(Z)$ be the probability of finding a grain with net charge $Ze$.
Assuming only single charge changes at a time, the distribution function can be derived using the master charging equation \citepalias{Draine1987CollisionalGrains},
%
%
\begin{equation}
\begin{split}
\qquad\;	f(Z)[J_{\pe,\mathrm{tot}}(Z) + J_{\Ion}(Z)&] =  \\
    f(Z+1)&[J_{e}(Z+1)+J_{e, \text{\tiny{CR}}}] ,
    \end{split}
    \label{eq:chargedist}
\end{equation}
where $J_{\pe,\mathrm{tot}}$ is the total radiative charging rate, accounting for the interstellar radiation field, and CR-induced radiative charging, $J_{\Ion}$ and $J_{e}$ are the collisional charging by a Maxwellian distribution of ions and electrons in the plasma, and $J_{e, \text{\tiny{CR}}}$ is the collisional charging rate of CR electrons. \\

The charge distribution, $f(Z)$, needs to be computed between a minimum and maximum charge.
The limits on the grain charge depend on the grain size, composition and maximum photon energy of the incident radiation field. 
If a grain is too negatively charged, internal Coulomb forces would eject a surface electron, thus the minimum allowed charge is given by:
\begin{eqnarray}
	Z_{\mathrm{min}} \approx \left( \frac{U_{\mathrm{ait}}}{14.4~\mathrm{V}} \frac{a}{{\textrm{\AA}}} \right),
\end{eqnarray}
where $U_{\mathrm{ait}}$ is the autoionization threshold potential\footnote{We take $-U_{\mathrm{ait}}/\mathrm{V} = 3.9 + 0.12(a/{\textrm{\AA}}) + 2({\textrm{\AA}}/a)$ for carbonaceous and $-U_{\mathrm{ait}}/\mathrm{V} = 2.5 + 0.07(a/{\textrm{\AA}}) + 8({\textrm{\AA}}/a)$ for silicate grains (\citetalias{Weingartner2001PhotoelectricHeating}).}. 
%
The maximum possible charge is given by:
\begin{eqnarray}
	Z_{\mathrm{max}} \approx \left[ \left( \frac{h\nu_{\mathrm{\small{max}}} - W}{14.4 ~\mathrm{eV}} \frac{a}{{\textrm{\AA}}} + \frac{1}{2}  \right)  \right],
\end{eqnarray}
where $h\nu_{\mathrm{max}}$ is the maximum energy of the photons composing the incident radiation field and $W$ is the work function\footnote{We take $W = 4.4$~eV for carbonaceous \citep{Smith1961CapacitiveHydrocarbons, Bakes1994TheHydrocarbons} and $W=8$~eV for silicate grains \citepalias{Weingartner2001PhotoelectricHeating}.}. \\

To obtain the charge distribution, $f(Z)$, we successively apply equation \ref{eq:chargedist} between the minimum and maximum allowed charges; thus
\begin{align}
\qquad	f(Z>0) &= f(0) \prod_{Z^{'} = 1}^{Z} \left[ \frac{J_{\pe, \mathrm{tot}}(Z^{'}-1) + J_{\Ion}(Z^{'}-1)}{J_{e}(Z^{'}) + J_{e,\text{\tiny{CR}}}(Z^{'})} \right] \\
    	f(Z<0) &= f(0) \prod_{Z^{'} = Z}^{-1} \left[ \frac{J_{e}(Z^{'}+1) + J_{e, \text{\tiny{CR}}}(Z^{'}+1)}{J_{\pe, \mathrm{tot}}(Z^{'}) + J_{\Ion}(Z^{'})} \right]
\end{align}
where the multiplicative constant $f(0)$ is determined by the normalization condition
\begin{eqnarray}
	\sum_{Z=Z_{\mathrm{min}}}^{Z_{\mathrm{max}}} f(Z) = 1.
\end{eqnarray} 

We have put together all the charging processes and calculation of the charge distribution function previously discussed as a {\sc{Python}} code {\dustcode}\footnote{The code is available for download at https://github.com/jcibanezm/DustCharge}.

\subsection{Charging timescale}

In this work, we assume that the charge distribution for all grains has sufficient time to relax to the equilibrium charge distribution.
This assumption may not always be valid, and could play a crucial role in the dynamical evolution of dust grains.
\citet{Ciolek2002TimedependentTests, Ciolek2004MultifluidShocks} have studied the dynamics of dust grains in C-type shocks and have seen that dust charges vary much slower than the time it takes for the grains to cross the shock.
Therefore, dust grains of equal size and composition follow different trajectories due to  variations in their net charge. 
If this is also the case for the dynamics of dust grains in the ISM, it would be necessary to explicitly compute the time evolution of the dust charge in order to accurately model the dynamics of dust grains in the Galaxy.


In order to test the validity of the charge equilibrium assumption for ISM simulations of molecular cloud formation, evolution and collapse, we calculate their charging timescale and compare it to the simulation timestep, constrained by the Courant-Friedrichs-Levy (CFL) condition \citep{Courant1928UberPhysik}.

The time-averaged charge centroid, $\langle Z \rangle$ is given by
\begin{eqnarray}
	\langle Z \rangle = \sum_{-\infty}^{\infty} Z f(Z),
\end{eqnarray}
and the timescale for charge fluctuations around this mean, 
$\tau_{f_{Z}}$,  
\citep{Draine1998ElectricGrains} is given by\footnote{This equation is valid for Gaussian charge distributions, which is a valid approximation for large grains in the cold and warm neutral medium phases. However, small grains have only discrete charge distributions.},
\begin{eqnarray}
	\tau_{f_{Z}} = \frac{\langle(Z - \langle Z \rangle)^{2}\rangle}{\sum_{\sc{z}}f(Z)J_{tot}(Z)}
    \label{eq:charging_timescale}
\end{eqnarray}
where $J_{tot}(Z) = J_{e}(Z)+J_{e, \text{\tiny{CR}}}+J_{\Ion}(Z)+J_{\pe, \mathrm{tot}}(Z)$ is the total charging rate.

\subsection{Colliding flow simulations}
\label{sec:colliding_flows}

In order to investigate the charge distribution of dust grains and its dependence on the local ISM properties, we post-process a three-dimensional (3D) hydrodynamic simulation by \citet[][hereafter J18]{Joshi2018OnConditions}.
The simulation setup is known as the ``colliding flows'' setup \citep{Vazquez-Semadeni2006MolecularFormation,Heitsch2006TheFlows,Vazquez-Semadeni2007MolecularConditions,Heitsch2008RapidMovies,Banerjee2009ClumpFormation,Heitsch2011Flow-drivenSimulations, Micic2013CloudFunction}, where two streams of warm neutral medium (WNM) converge, forming cold, dense, turbulent structures at the collision interface. 
The simulations were performed using the 3D magnetohydrodynamics code {\sc{flash4}} \citep{Fryxell2000FLASH:Flashes}, including gas self gravity \citep{Wunsch2018Tree-basedDepths}, radiative heating and cooling coupled to a chemical network to follow the non-equilibrium formation of H$_{2}$ and CO \citep{Seifried2018SILCC-Zoom:Clouds, Walch2015TheISM,Glover2011ApproximationsApproaches, Glover2009ModellingMedium}, diffuse heating and its attenuation by dust shielding \citep{Wunsch2018Tree-basedDepths,Walch2015TheISM,Clark2011TreeCol:Simulations}.
The interstellar radiation field is included by means of a uniform background radiation field with an intensity of $G_{0}=1.7$, \citep[appropriate for Solar neighbourhood conditions;][]{Walch2015TheISM,Girichidis2016LAUNCHINGMEDIUM,Seifried2018SILCC-Zoom:Clouds}.
For every cell in the simulation we know the local visual extinction $A_{v}$, which we use to compute the local effective strength of the radiation field, $G_{\mathrm{eff}} = G_{0}\mathrm{exp}(-2.5 A_{v})$.

The colliding flows simulations presented here correspond to \citetalias{Joshi2018OnConditions}  ``CF-R8'' run, which has a domain of $ 128~\mathrm{pc}\times 32~\mathrm{pc} \times 32~\mathrm{pc}$ with a base-grid resolution of $\Delta x = 0.25$~pc, and a maximum adaptive-mesh-refinement (AMR) resolution of $\Delta x_{\mathrm{max}} = 0.032$~pc.
The AMR criterion uses a threshold on the second spatial derivative of the density up to a resolution of $\Delta x=0.125$~pc, and an additional Jeans refinement for smaller $\Delta x$ \citep{Truelove1998SelfgravitationalFragmentation}.

Initially the domain is filled with warm neutral medium (WNM) with a uniform density of atomic hydrogen atoms of $n_{\mathrm{H}, 0}=0.75$~cm$^{-3}$, and temperature equal to the equilibrium temperature, $T=4082$~K.
The chemical composition of the ISM is set to Solar metallicity, with abundances of: $x_{\mathrm{He}} = 0.1$, $x_{\mathrm{C}} = 1.4\times 10^{-4}$, $x_{\mathrm{O}} = 3.2 \times 10^{-4}$ and $x_{\mathrm{M}} = 1.0\times 10^{-7}$ for Helium, Carbon, Oxygen and all heavier Metals respectively, relative to total hydrogen atoms. 
The chemical network is initialized with the equilibrium abundances for the WNM, e.g. at this density and temperature hydrogen is fully atomic and there is no CO. 
The WNM streams meet at a perturbed interface in the middle of the box with equal and opposite velocity of $v_{x} = 13.6$~km s$^{-1}$, such that the collision occurs immediately at the start of the simulation.
Shortly after the simulation starts, the shocked gas starts to cool and form H$_{2}$.
After 14~Myr of evolution, CO begins to form within the dense filaments and cores, of which some are gravitationally bound and undergo gravitational collapse.

\begin{figure}
\centering 
\vspace*{-1mm}
\includegraphics[width=0.49\textwidth]{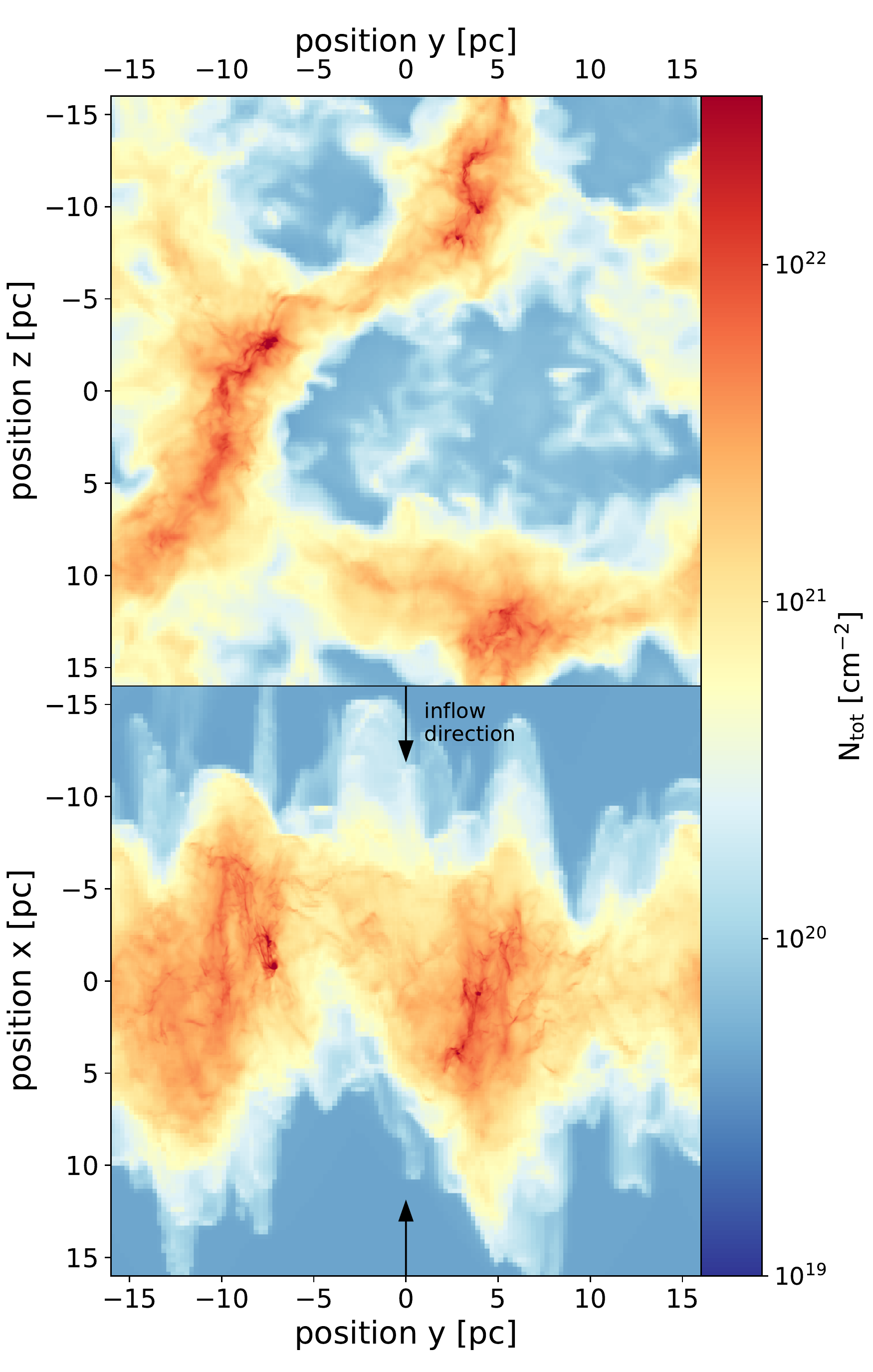} 
\vspace*{-6mm}
\caption{Column density projection of total gas in the colliding flow simulation by \citetalias{Joshi2018OnConditions} at an evolutionary time of $t=18$~Myr. {\emph{Top}}: projection perpendicular to the collision front, showing the full extent of the simulation in the $y-$ and $z-$directions.
{\emph{Bottom}}: projection parallel to the collision front showing only $\pm16$~pc of the simulated box along the $x$ axis. 
\label{fig:CF_projection}} 
\end{figure}

Figure \ref{fig:CF_projection} shows the morphology of the total gas column density projected parallel and perpendicular to the collision plane, at $x=0$, of the simulation by \citetalias{Joshi2018OnConditions}.
The snapshot used for the analysis in this work is taken at time $t=18$~Myr, where 36\% of the mass in the simulation is in molecular hydrogen and 2\% in carbon monoxide.
This snapshot is close to the end of the simulations and it was selected because of the high dynamic range in temperature, density and extinction, key quantities for the calculation of the charge distribution of dust grains.

\begin{figure*}
\centering 
\includegraphics[width=1.0\textwidth]{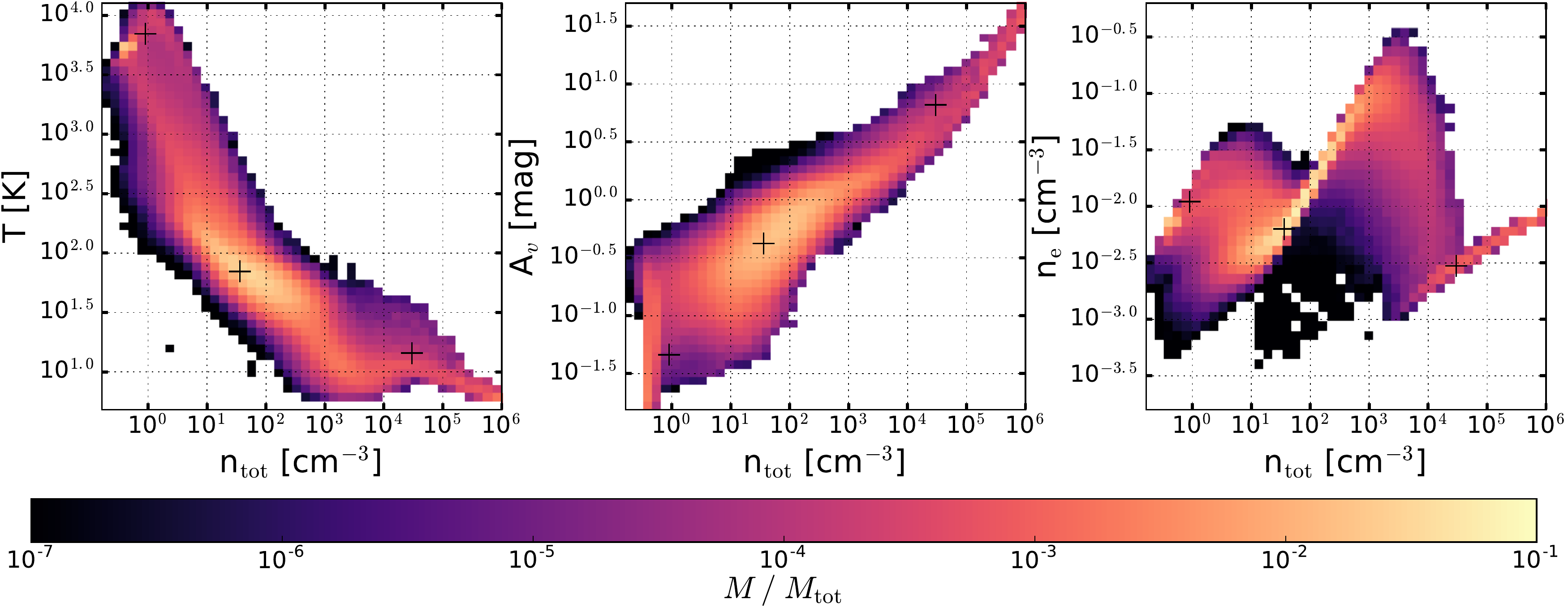} 
\vspace*{-6mm}
\caption{Mass weighted 2D-probability density function (PDF) of the ({\emph{left}}) temperature, ({\emph{middle}}) visual extinction and ({\emph{right}}) revised electron number density as a function of the total gas number density in the ``CF-R8'' colliding flow simulation by \citetalias{Joshi2018OnConditions} at $t=18$~Myr. 
The PDF is normalized to the total mass in a box with dimensions 32~pc$^{3}$ centered around the collision interface at $x=0$.
Three black plus signs show the location of the WNM, CNM and CMM conditions where the charge distribution is calculated in Figure \ref{fig:fz_sample}.
The revised electron number density includes the updated ionization fraction calculated in a post-processing step, the original electron density from the simulation is shown in figure \ref{fig:ionizationfraction_comp} in the appendix \ref{appendix_ionization_fraction}.
\label{fig:CF_phasePlots}} 
\end{figure*}

Figure \ref{fig:CF_phasePlots} shows the distribution of gas number density, temperature, extinction and electron number density in a 32~pc$^3$ box around $x=0$ (shown in Figure \ref{fig:CF_projection}) at $t=18$~Myr.
In these Figure we show an updated version of the electron number density compared to the one provided by the chemical network in the simulation as discussed in the next subsection.

\subsubsection{Ionization state of the gas}
\label{sec:ionization_state}

In spite of the complexity of the chemical network implemented in the simulations, it overestimates the recombination rates of electrons in the cold-molecular phase of the ISM.
In the network, the electron recombination rate is a grain-assisted ion recombination, proportional to $G_{\mathrm{eff}}\sqrt{T}/n_{e}$, suggested by \citet{Weingartner2001ElectronIonHydrocarbons}.
Due to the fast recombination rates in the network, the ionization state of the gas in the simulation quickly drops at $n_{\mathrm{H}} > 10^{3}$~cm$^{-3}$, corresponding to visual extinctions of $1-3$~mag, once the interstellar radiation field is attenuated and ionized carbon recombines.
However, the ionization state within dense molecular regions is determined by a balance between CR ionization and various recombination processes. 

In order to correct for the low abundance of free electrons within dense regions, we update the ionization state of the gas in a post-processing step.
We employ the empirical relation derived by \citet{Caselli2002MolecularDegree} for the ionization fraction $\chi_{e} = n_{e}/n_{\mathrm{H_2}}$ in molecular regions
\begin{eqnarray}
\chi_{e} = 6.7 \times 10^{-6} \left( \frac{n_{\mathrm{\tiny{H_2}}}}{\text{cm}^{-3}} \right)^{-0.56} \sqrt{\frac{\zeta (N(\text{H}_{2}))}{10^{-17} \text{s}^{-1}}},
\label{eq:caselli}
\end{eqnarray}
where $\zeta (N(\text{H}_{2}))$ is the cosmic ray ionization rate, which depends on the molecular hydrogen column density.
We get both $n_{\mathrm{H_{2}}}$ and $N(\mathrm{H_{2}})$ from the simulations for every cell in the domain such that these computation is quickly performed.
For a discussion on the updates to the ionization fraction and the electron abundance within dense regions, see appendix \ref{appendix_ionization_fraction}.

\section{Results}
\label{sec:results}

\subsection{Charge distribution function}

As introduced in section \ref{sec:charge_distribution}, the discrete charge distribution, $f(Z)$, for dust grains depends on the ratios of the ``forward'' and ``backward'' charging rates, equation \ref{eq:chargedist}.
These rates depend on the grain composition (silicate or carbonaceous), grain size, and the environment.
In this work we investigate the charge distribution for a range of grains sizes: 3.5~{\AA}, 5~{\AA}, 10~{\AA}, 50~{\AA}, 100~{\AA}, 500~{\AA} and 1000~{\AA}.
These sizes are within the sizes encountered in the sizes distribution of dust grains in the ISM \citep{Weingartner2001DustCloud}.
However, instead of showing $f(Z)$ for each grain size at each figure, we present the results and subsequent analysis for a small (5~{\AA}), intermediate (100~{\AA}), and large (1000~{\AA}) grain population.
Nonetheless, at the end of section \ref{sec:discussion}, we provide the parameters of the parametric equations for reconstructing the charge distribution of all the grain sizes and compositions analyzed, and show the distribution of charge centroids and widths for all sizes in appendices \ref{appendix_silicate_grains}, and \ref{appendix_carbonaceous_grains}. 

\begin{figure*}
\centering 
\includegraphics[width=1.0\textwidth]{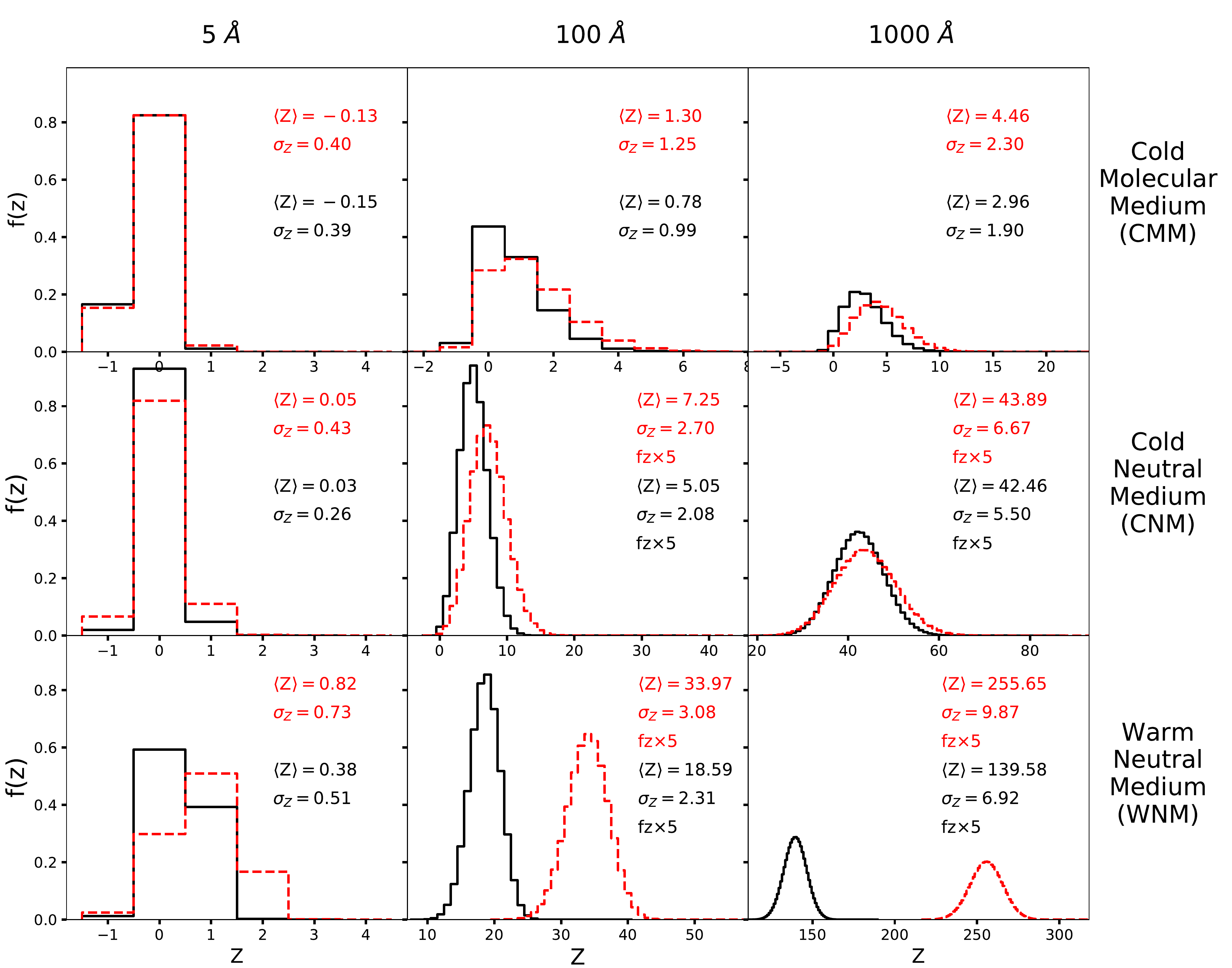} 
\vspace*{-6mm}
\caption{Charge distribution of $5$~{\AA}, $100$~{\AA} and $0.1~\mu$m size, carbonaceous ({\emph{dashed red}}) and silicate ({\emph{solid black}}) grains, in warm neutral medium, cold neutral medium and cold molecular medium conditions. Each panel includes the charge centroid, $\langle Z \rangle$, and the charge width, $\sigma_{Z}$, of the distributions. The local environment parameters for the calculations are: 
{\textbf{WNM:}} $n_{\mathrm{H}} = 0.9$~cm$^{-3}$, $T = 7000$~K, $\chi_{e}= 0.012$, $G_{\mathrm{tot}}=1.52$, $x_{\mathrm{H_{2}}} = 4.6\times10^{-5}$; 
{\textbf{CNM:}} $n_{\mathrm{H}} = 36$~cm$^{-3}$, $T = 70$~K, $\chi_{e}=1.8\times 10^{-4}$, $G_{\mathrm{tot}}=0.60$, $x_{\mathrm{H_{2}}} = 0.15$; 
{\textbf{CMM:}} $n_{\mathrm{H}} = 3 \times 10^{4}$~cm$^{-3}$, $T = 14.4$~K, $\chi_{e}=9.0\times10^{-8}$, $G_{\mathrm{tot}}=7.9\times10^{-4}$, $x_{\mathrm{H_{2}}} = 0.99$. 
Where $G_{\mathrm{tot}}=G_{\mathrm{eff}}+G_{\text{\tiny{CR}}}$ is given in units of Habing field, $\chi_{e}=n_{\mathrm{e}}/n_{\mathrm{H}}$ is the ionization fraction, and $x_{\mathrm{H}_{2}}=n_{\mathrm{H}_{2}}/n_{\mathrm{H}}$ is the molecular hydrogen fraction relative to total hydrogen atoms.
\label{fig:fz_sample}} 
\end{figure*}

Figure \ref{fig:fz_sample} shows the equilibrium charge distribution of the small, intermediate, and large, carbonaceous (\emph{dashed red}) and silicate (\emph{black}) grains in three different environments: WNM ($300$~K$\leq$\textit{T}<$10^{4}$~K), cold neutral medium (CNM, \textit{T}<$300$~K), and cold molecular medium (CMM, \textit{T}$\leq$$30$~K, with an additional restriction on the H$_{2}$ fraction $x_{\mathrm{H}_{2}}>0.5$).
In each panel, the charge centroid, $\langle$Z$\rangle = \sum Z f(Z)$, and the width, $\sigma_{Z}^{2} = \sum f(Z) (Z - \langle Z \rangle)^{2}$, of each of the distributions is included.
In the CMM ({\emph{top row}}) we see that: ({\emph{left}}) small carbonaceous and silicate grains tend to have mostly neutral charge with narrow widths; ({\emph{middle}}) intermediate size grains display some positive charges with a centroid around $\langle Z \rangle \approx 0-1$  and a width of order unity; ({\emph{right}}) large grains are mostly positively charged, $\langle Z \rangle \approx 3-5$, with wider distributions, $\sigma_{Z}\approx 2-3$.
As discussed by \citet{Ivlev2015InterstellarEffects}, CR induced molecular hydrogen excitation and UV emission is the main charging mechanism within dense molecular clouds, resulting in the predominantly positive charge observed in intermediate and large dust grains.
In the CNM ({\emph{middle row}}), small grains remain mostly neutral, while intermediate and large grains have predominantly positive charges and wide distributions.
In the WNM ({\emph{bottom row}}), small silicate grains tend to be neutral, while carbonaceous grains tend to have positive charges of order unity; intermediate and large grains are always positively charged and have wide charge distributions. 
Carbonaceous grains are always more positively charged than silicate grains and have wider charge distributions.
This is because carbonaceous grain are ionized more easily due to their lower work function, resulting in higher photoelectric charging rates.
 \\

Figure \ref{fig:fz_sample} gives us an idea of the behavior of the charge distribution as a function of grain size, composition, and environment. 
However, in order to understand the transition of the charge distribution between environments, we perform our calculations for a population of dust grains across the ISM conditions captured by the colliding flow simulation by \citetalias{Joshi2018OnConditions} (see Figure \ref{fig:CF_phasePlots}).
The snapshot that we use in our calculation has a total of $\approx 9.7 \times 10^{6}$ cells, and calculating the full charge distribution in each and every single cell is too computationally expensive. 
So instead, we stochastically sample $1\%$ of the cells in the simulation and use them to obtain an estimate of the behavior of the charge distribution.
However, even calculating 1\% of the cells in the simulation is too computationally expensive for the 500~{\AA} and 1000~{\AA} grains, such that we reduce even further the number of stochastically sampled cells to 0.1\% of the total number of cells in the simulation ($\sim 10^{4}$ cells).
For a completeness analysis on how accurate our results are with respect to the percentage of cells sampled, see appendix \ref{appendix_completeness_test}.

For the remainder of this paper we refer to the distribution of charge centroids, $\langle$Z$\rangle = \sum Z f(Z)$, and charge widths, $\sigma_{Z}^{2} = \sum f(Z) (Z - \langle Z \rangle)^{2}$, as shown in the top right corner of each panel in Figure \ref{fig:fz_sample}, instead of discussing the full probability distribution function of charge.
This simplification assumes that the full charge distribution may be recovered by assuming a Gaussian distribution with mean and standard deviation equal to the charge centroid and charge width.
This simplification is not always valid, especially for the small grains, and for grains in the cold molecular phase.
Dust grains have discrete and sometimes skewed charge distributions, that may not always be accurately described by a Gaussian distribution.
For example, the skewness of the charge distribution in the CMM is mostly due to dust charging being dominated by CR-induced processes.
Collisional cross sections with other dust grains and ions depend on the charge state of the grain, and a skewed charge distribution would influence grain growth and depletion rates.
However, in this work we are interested in understanding the global behaviour of the charge distribution of dust grains in the ISM, for which approximating the charge distribution with a normal distribution is a valid approximation. \\

\begin{figure*}
\centering 
\includegraphics[width=0.95\textwidth]{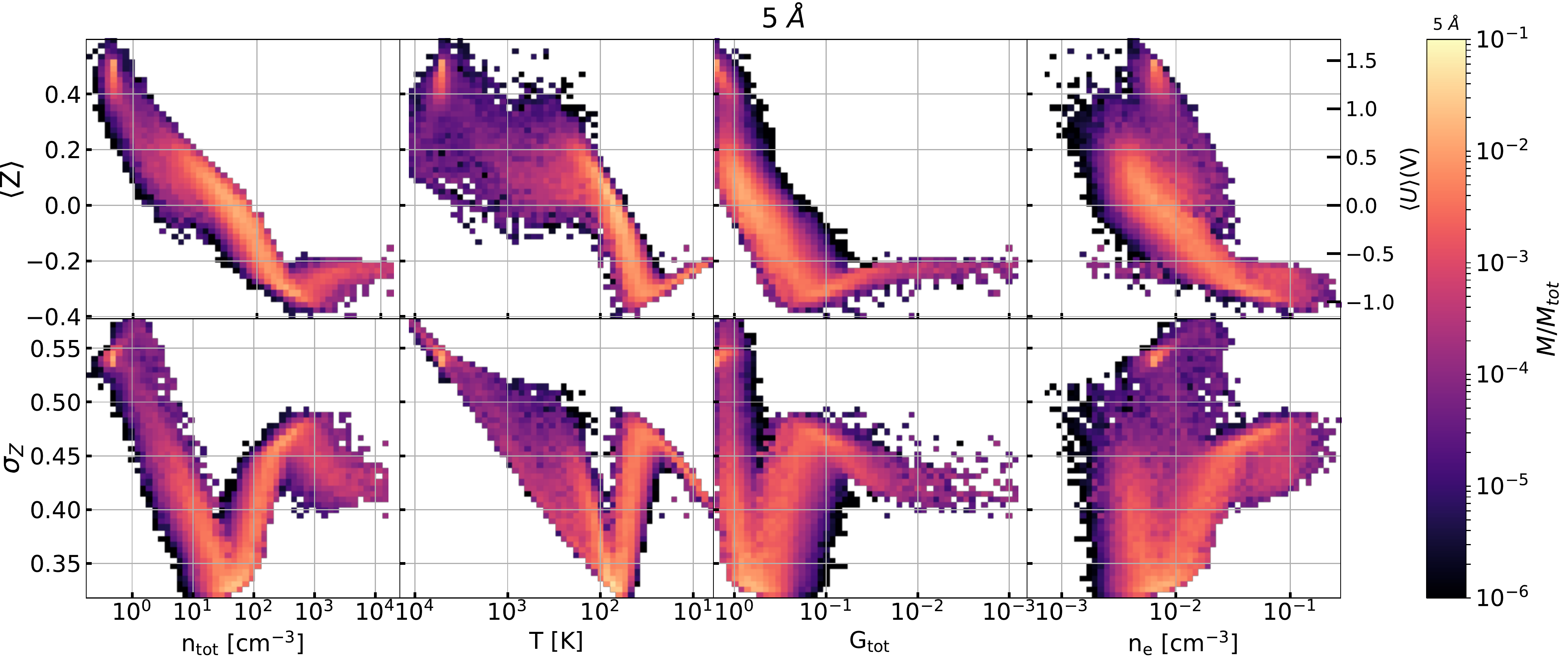} 
\includegraphics[width=0.95\textwidth]{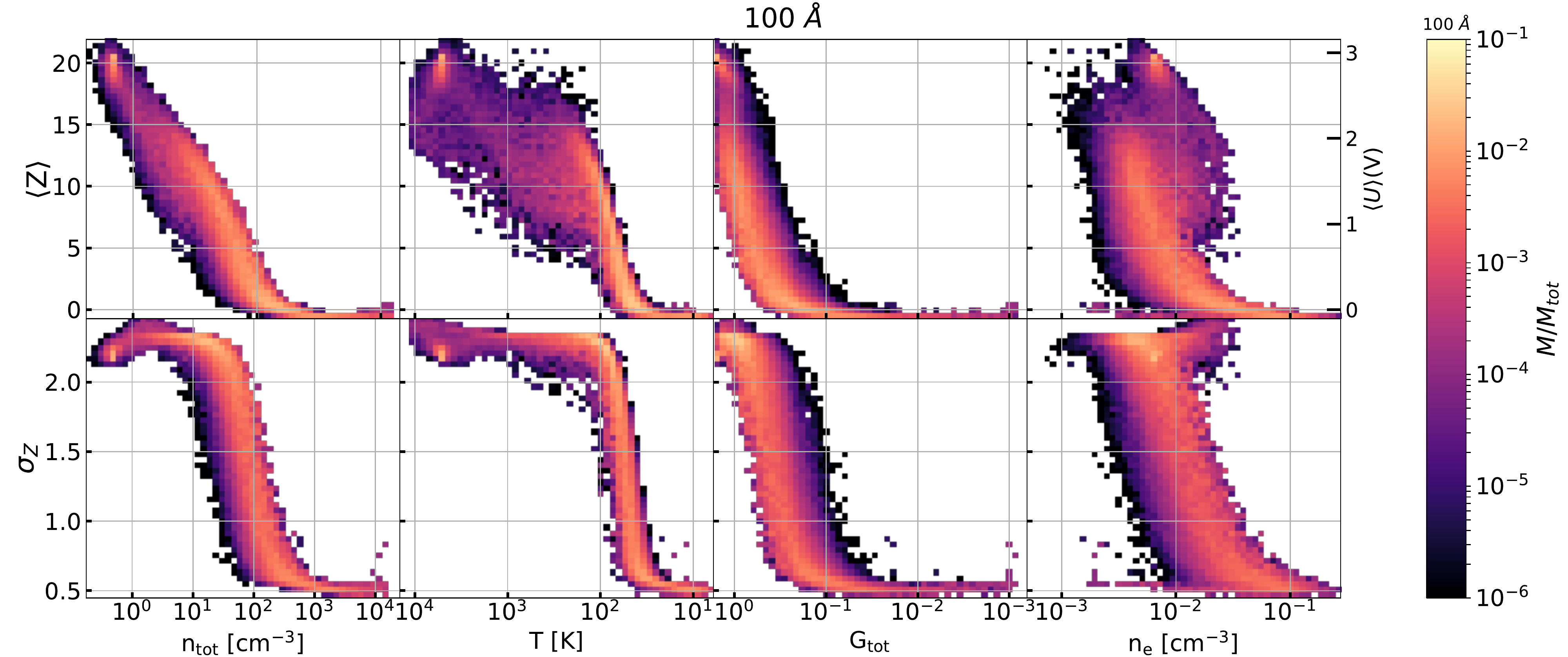} 
\includegraphics[width=0.95\textwidth]{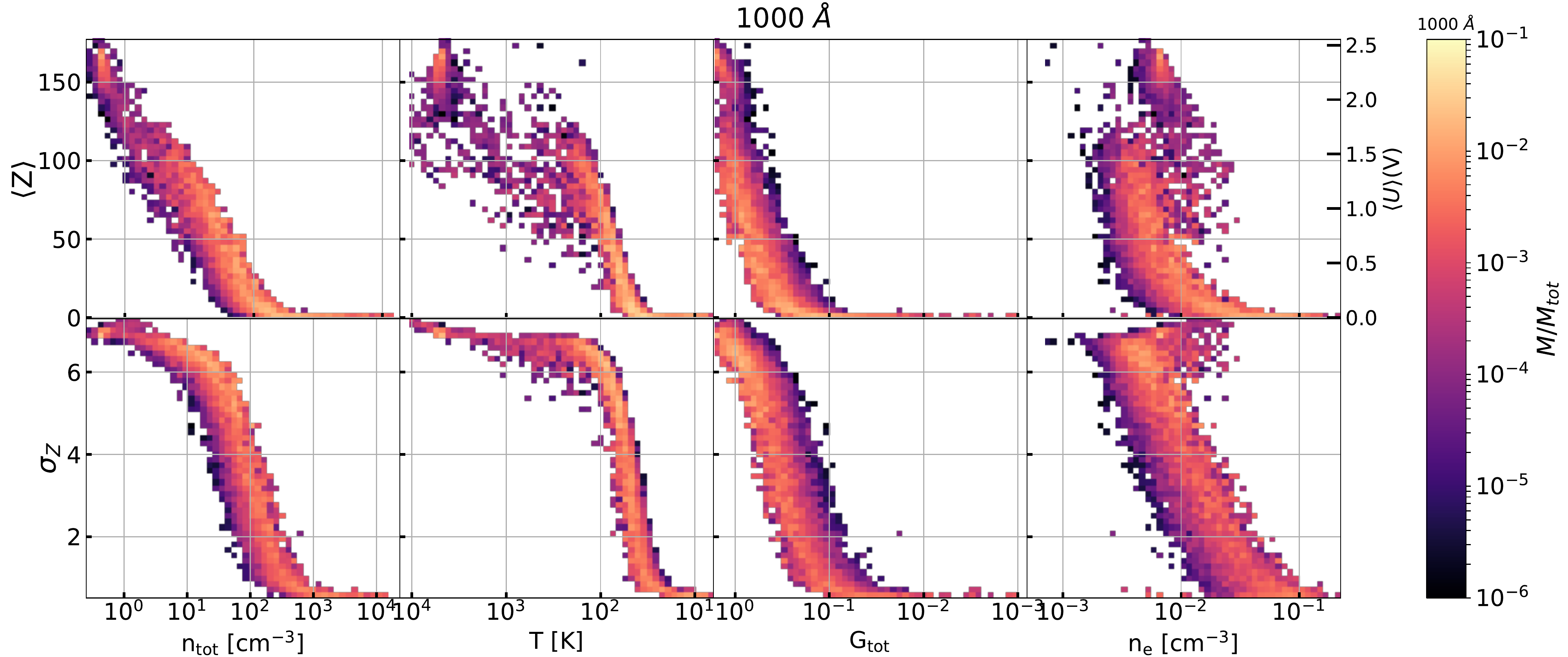} 
\caption{Distribution of charge centroids, $\langle Z \rangle$ ({\emph{upper rows}}), and widths, $\sigma_{Z}$ ({\emph{lower rows}}), as a function of total number density, temperature, local strength of the radiation field, $G_{\mathrm{tot}}$, and electron number density  ({\emph{from left to right}}), for the stochastically sampled cells within the colliding flow simulation, for silicate grains of of sizes 5~\textrm{\AA} (\emph{top}), 100~\textrm{\AA} (\emph{middle}), and $1000~$\textrm{\AA} (\emph{bottom}).
At the right hand side of the charge centroid distribution rows, the average electrostatic potential, $\langle U \rangle$(V), is also shown.  
\label{fig:fz_sil}} 
\end{figure*}

Figure \ref{fig:fz_sil} shows the centroid and width of the distribution for a small, intermediate, and large silicate grain as a function of various ISM parameters.
For grains of all sizes, the charge centroid calculated is predominantly positive, although there are some cases towards high densities, $n > 50$~cm$^{-3}$, where small and intermediate grains tend towards negative charge.
For the three sizes displayed in Figure \ref{fig:fz_sil}, the centroid and width have a similar behavior with respect to the local ambient properties, but span different ranges.
The distribution of centroids and widths are inversely proportional to the density of the environment and the electron density, and proportional to the temperature and strength of the radiation field.
This behavior is to be expected given that as one moves towards higher densities, the ambient temperature decreases and the extinction increases, reducing the photoelectric charging rate.

The more positively charged grains with wider distributions are found in the WNM, where the photoelectric charging rate is strong, but in these regions electron densities and ambient temperatures are high resulting in efficient negative charging as well.
The competition between positively and negatively charging the grain results in wide charge distributions, although the photoelectric charging process generally dominates resulting in overall positive charges.
The least charged grains with narrow widths are found within dense clouds, where even the CR-induced radiative charging is low.
These are the regions where some grains may attain negative charge where collisional charging with electrons dominates.

Of particular interest is the distribution of charge centroids of the small grains, where at densities of $n_{\mathrm{H}}\approx 50$~cm$^{-3}$, the distribution becomes negative, reaching a minimum at a density of $n_{\mathrm{H}}\approx 1000$~cm$^{-3}$, where it starts moving towards charge neutrality again.
This transition occurs once the ISRF is slightly attenuated and there is not yet a strong CR-induced UV field, collisional charging dominates. 
However, at higher densities radiative charging by the CR-induced UV field balances with the collisional charging by electrons, decreasing the magnitude of the charge centroid.
It is also important to point out that charge is a discrete quantity. 
Thus having a centroid of $0.2$ or $-0.2$, and a width of $\sigma_{Z} \approx 0.3$, means that the charge distribution is peaked at $\langle Z \rangle = 0$ and only has small wings towards positive and negative charges.

The distribution of charge centroids and widths of the intermediate and large grains have almost the exact same shape.
However, they differ in the range of the charge centroid and width, as 100~{\AA} grains have a maximum charge centroid of $\langle Z \rangle \approx 20$, whereas 1000~{\AA} grains have a maximum charge centroid of $\langle Z \rangle \approx 180$.
Similarly for the widths, 100~{\AA} grains have a maximum width of $\sigma_{Z} \approx 2.2$, whereas 1000~{\AA} have maximum widths of $\sigma_{Z} \approx 7$.

The distribution of charge centroids and widths for the carbonaceous grains are very similar in shape to the distributions for silicate grains.
Figures for the centroid and width distribution for carbonaceous grains are shown in appendix \ref{appendix_carbonaceous_grains}.

%

\subsection{Charge equilibrium assumption}

The calculations of the charge distribution performed here correspond to the equilibrium charge distribution.
%
We now investigate the validity of this assumption for the results presented in this paper, by comparing the charging timescale of our dust population, equation \ref{eq:charging_timescale}, to the simulation timestep, set by the Courant-Friedrichs-Lewy (CFL) condition \citep{Courant1928UberPhysik}.


\begin{figure}
\centering 
\includegraphics[width=0.48\textwidth]{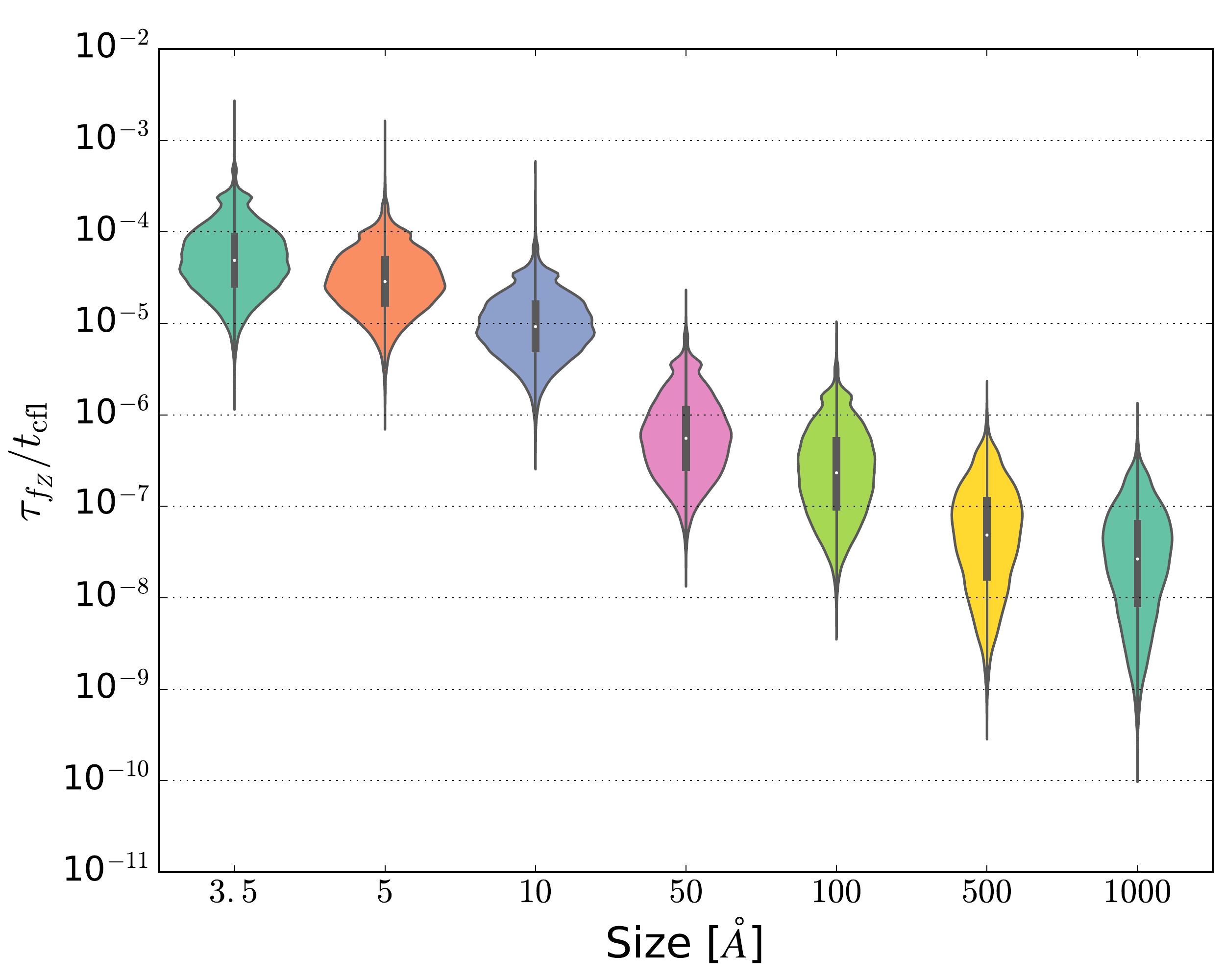} 
\vspace*{-5mm}
\caption{Ratio of charging timescale and local CFL timescale at each point where the dust charge distribution was evaluated for silicate grains with sizes between 3.5~{\AA} and 1000~{\AA}. The shape of each object in the violin plot represents the density distribution of relative timescales. The central white point corresponds to the median, the thick black line to the inter-quantile range and the thin black line to the 95\% confidence interval.}
\label{fig:charging_equilibrium} 
\end{figure}

Figure \ref{fig:charging_equilibrium} shows the ratio of the local charging timescale, $\tau_{f_{Z}}$, and the CFL timescale, $t_{\textrm{\tiny{CFL}}}$, for silicate grains of various sizes.
Given that $t_{\textrm{\tiny{CFL}}}$ is the same for the different grain sizes and composition, variations in the relative timescales are solely due to variations in the charging timescale of the grains.
It is observed that the charging timescale decreases with increasing grain size due to the increased width of the charge distribution.
From Figure \ref{fig:charging_equilibrium}, it is clear that the charging timescales are significantly smaller than the timescale restricted by the CFL condition in the simulation, confirming that the assumption of charge equilibrium is valid in our calculations.
The smallest grains, both carbonaceous and silicate, are the ones that have the longest charging timescales, with a median of 0.5~yr.
However, this charging timescale is still $4\times10^{4}$ times smaller than the timestep in the simulation with a resolution of $\Delta x = 0.032$~pc.
Given that  $t_{\textrm{\tiny{CFL}}}$ depends on the resolution of the simulation and the turbulent velocity field, going to higher resolutions or environments with faster turbulent velocities, e.g. shocks, would decrease this timescale up to a point where the charge equilibrium assumption is no longer valid.
If the properties of the simulation were fixed, such that the timestep decreased linearly with the resolution only, at a resolution of $\Delta x \sim 1$~AU, 5\% of the 3.5~\textrm{\AA} grains would have a charging timescale approximately equal to the simulation timestep.
At this point, one could argue that it would be necessary to follow the non-equilibrium evolution of the charge, at least for the very small grains, in order to accurately estimate the charge state of all dust grains in the simulation.

\section{Discussion}
\label{sec:discussion}

Knowing the complete charge distribution function of dust grains is important as it influences the photoelectric heating efficiency, the collisional cross sections, the depletion rates of ions and the dynamical evolution of dust grains. 
However, calculating it live within a simulation is almost impossible given that this computation is very expensive.

We simplify the description of the charge distribution by assuming it to be well represented by a Gaussian distribution, which is fully described by two parameters: the charge centroid $\langle Z \rangle$, and the width of the distribution, $\sigma_{Z}$.
\citetalias{Weingartner2001PhotoelectricHeating} suggest that the electrostatic potential, $U = Ze/a$, of a dust grain can be approximated via a charging parameter.
We now investigate how accurately the charging parameter approximates the charge centroid for a range of grain sizes and ISM ambient conditions, and then suggest new parametric equations for the charge centroid and width that can be used to recover the charge distribution of dust grains across the ISM.

\subsection{Ambient conditions and the charging parameter}

\begin{figure*}
\centering 
\includegraphics[width=0.31\textwidth]{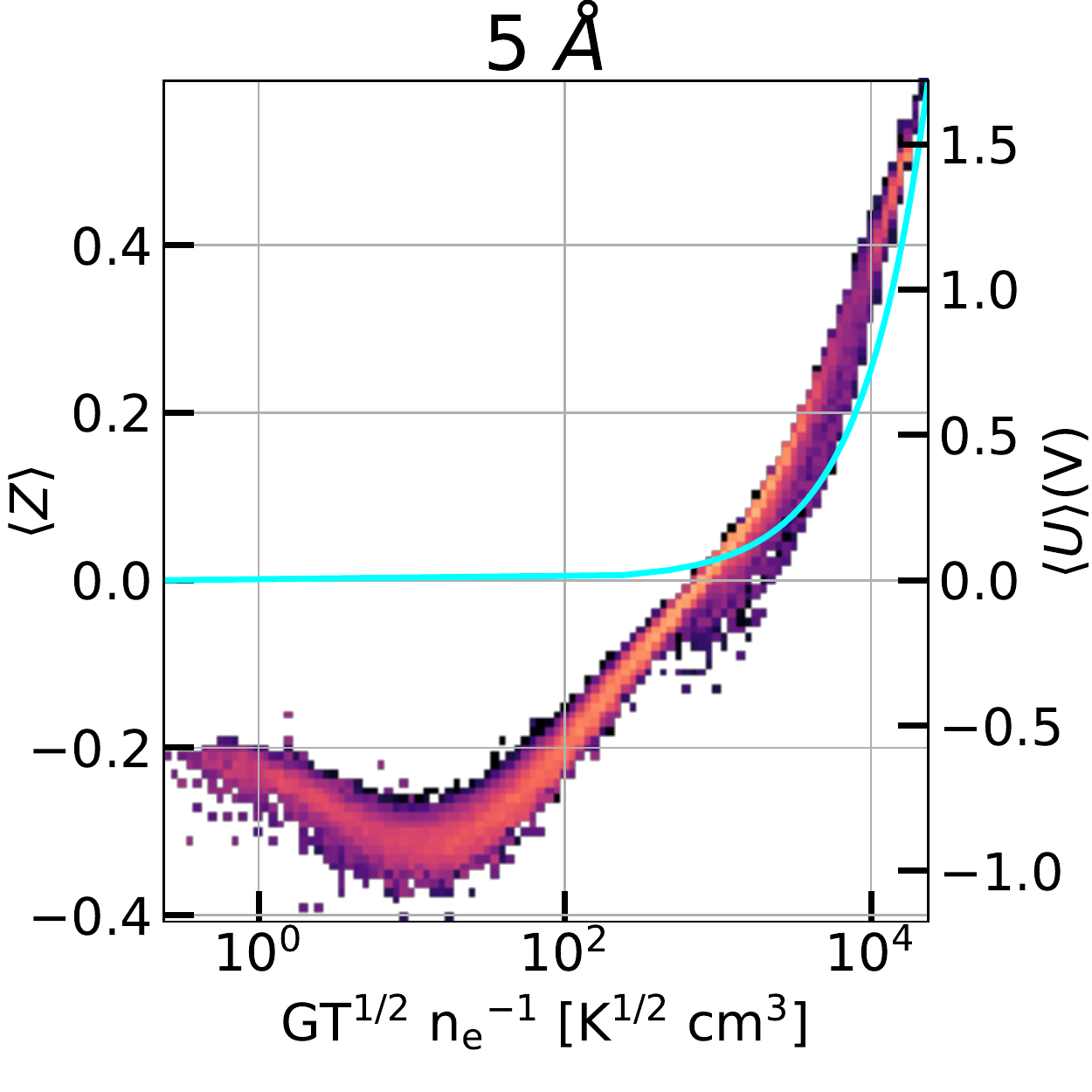}
\includegraphics[width=0.31\textwidth]{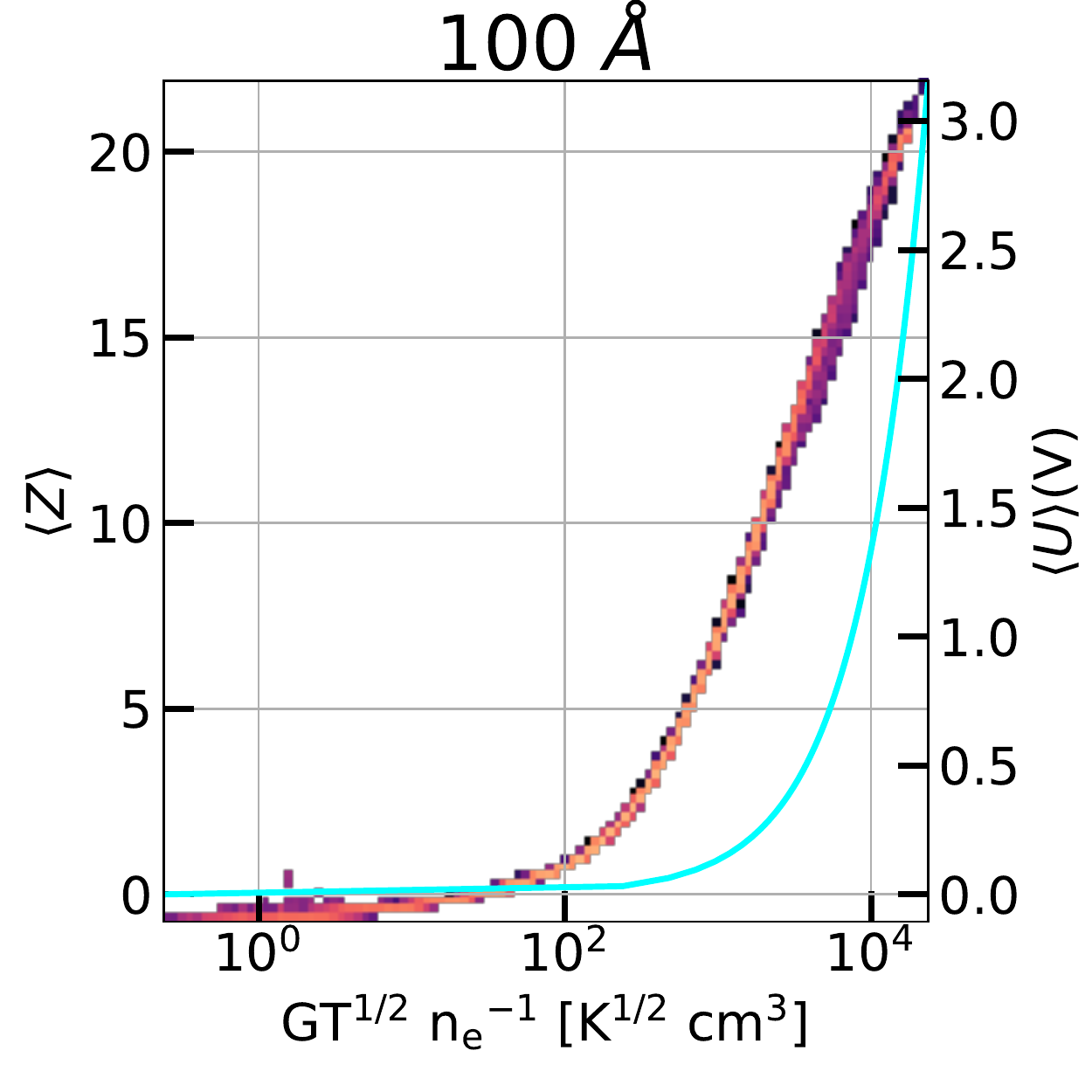}
\includegraphics[width=0.31\textwidth]{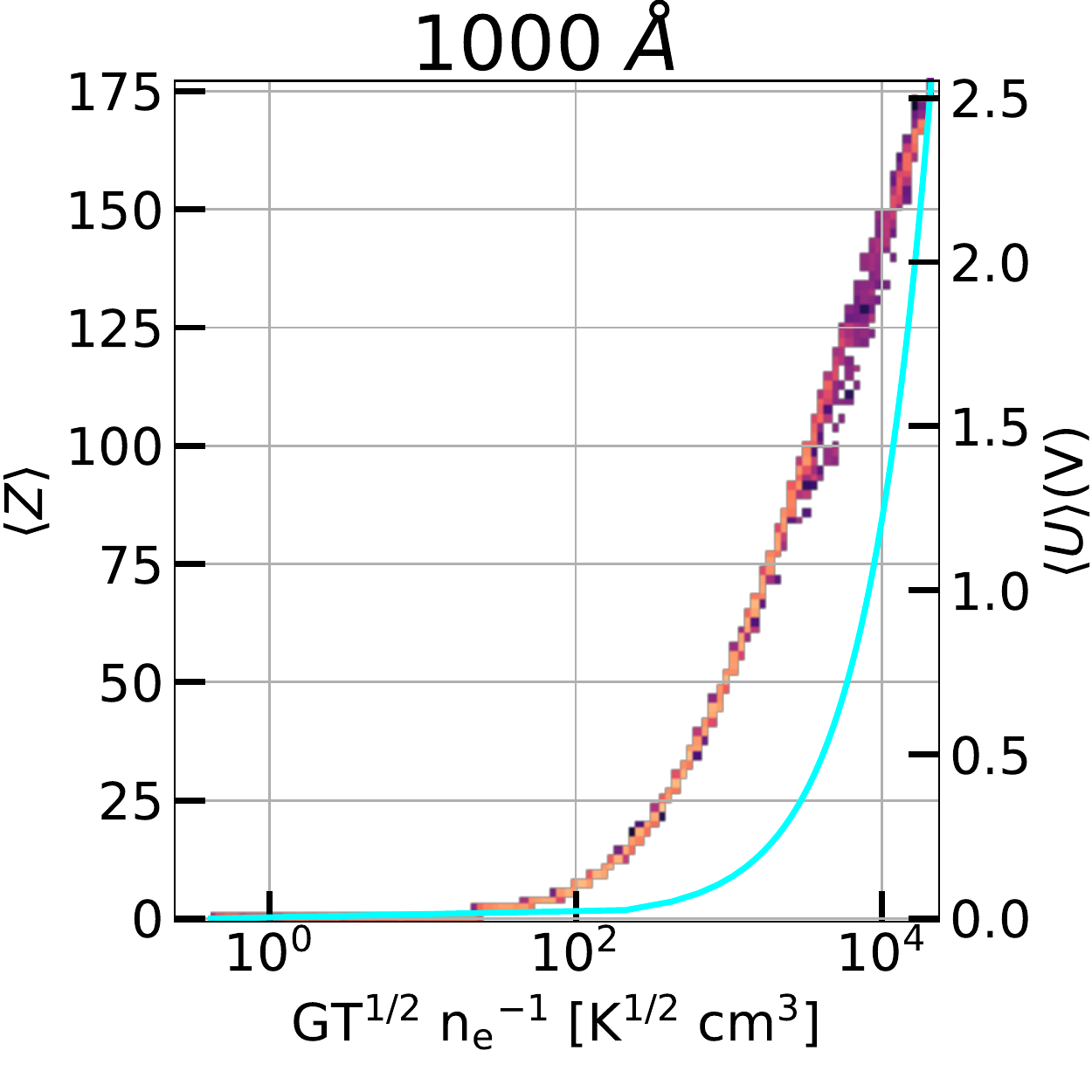} 
\caption{Distribution of charge centroids, and average electrostatic potential, as a function of the charging parameter suggested by \citetalias{Weingartner2001PhotoelectricHeating}, $GT^{0.5} / n_{e}$, for 5~\textrm{\AA} (\emph{left}), 100~\textrm{\AA} (\emph{middle}), and 0.1~$\mu$m  (\emph{right}) silicate grains.
The cyan line shows the charge centroid as a linear function of the charging parameter scaled to the maximum value of $\langle Z \rangle$ for each grain size.
The color bar is the same as in Figure \ref{fig:fz_sil}}
\label{fig:DrainechargingPar} 
\end{figure*}

Given that electrons are less massive than ions, they would collide more often with dust grains, dominating the collisional charging of the grain, which is proportional to $n_{e}/\sqrt{T}$.
Photoelectric charging of dust grains is proportional to the intensity of the local radiation field, $G_{\mathrm{tot}}$.
Therefore, the average electrostatic potential of a grain should behave proportional to the relative intensity of electron collisional charging and photoelectric charging, given by:
\begin{eqnarray}
U_{\mathrm{\tiny{WD01}}} \propto (1 + Z^{-1/2})^{-1} g(U,a) \langle Q_{abs} \rangle \frac{G_{\mathrm{tot}}\sqrt{T}}{n_{e}},
\label{eq:Draineapprox}
\end{eqnarray}
where $g(U, a)$, is described as a decreasing function of U, associated with the yield $Y$, introducing a dependence on the size. 
In their discussion, \citetalias{Weingartner2001PhotoelectricHeating} note that this relation applies only when $G_{\mathrm{tot}}/n_{\mathrm{e}}$ is large. 
However, as it can be observed in Figure \ref{fig:DrainechargingPar}, we are exploring environments with very low values of $G_{\mathrm{tot}}/n_{\mathrm{e}}$, where equation \ref{eq:Draineapprox} should not hold.
Additionally, we are including CR-induced charging processes in this work, which dominate the charging of dust grains at low values of the charging parameter.

Figure \ref{fig:DrainechargingPar} shows the centroid of the charge distribution as a function of the \citetalias{Weingartner2001PhotoelectricHeating} charging parameter.
As a reference, Figure \ref{fig:DrainechargingPar} also shows the centroid as a linear function of the charging parameter (\emph{cyan line)}, scaled to the maximum value of $\langle Z \rangle$ for each grain.
It seems that this combination of parameters results in a tight distribution of the charge centroid with little scatter.
However, the distribution of charge centroids does not vary linearly with the charging parameter.
We now derive a new functional functional form of the charging parameter that will enable us to recover the value of the charge centroid as a function of the local properties of the ISM.\\

We perform a non-linear least squares weighted polynomial regression of the charge centroid distribution as a function of the electron density, temperature and radiation field intensity.
The new combination of parameters describing the behavior of the charge centroid as a function of ISM ambient conditions is
\begin{eqnarray}
\langle Z \rangle &= k \left[ 1 - exp\left(- \frac{G_{\mathrm{tot}}T^{0.5} n^{-1}_{\mathrm{e}}}{h_{Z}}\right) \right] \left( \frac{G_{\mathrm{tot}} T^{0.5}}{n_{e}} \right)^{\alpha} + b,
\label{eq:charging_par}
\end{eqnarray}
where $k$, $b$, $\alpha$ and $h_{Z}$ are the fitting parameters of the regression, and $G_{\mathrm{tot}} = G_{0} + G_{\text{\tiny{CR}}}$ is the total radiation field intensity given by a combination of the intensity of the interstellar radiation field and the cosmic ray induced FUV field, eq. \ref{eq:CR_FUV}, in units of the Habing field.

Additionally, we perform a non-linear least squares regression of the charge distribution width as a function of the charge centroids.
We divide the polynomial fit in two parts depending on the sign of the charge centroid:
\begin{align}
\qquad \qquad \qquad \sigma_{\text{\tiny{Z}}}^{+} &= c^{+} \left(1 - e^{\frac{-\langle Z \rangle}{\eta^{+}}} \right) + d,  \qquad \qquad \langle Z \rangle > 0 \label{eq:width_par_pos} \\
\sigma_{\text{\tiny{Z}}}^{-} &= c^{-} \left(1 - e^{\frac{-|\langle Z \rangle|}{\eta^{-}}} \right) + d, \qquad \quad \;\;\, \langle Z \rangle < 0
\label{eq:width_par_neg}
\end{align}
%
where $c^{+,-}$, $d$ and $\eta^{+,-}$ are the fitting parameters. 
The only constraint to the positive and negative fits of the charge width is that they meet at $\sigma^{+}(\langle Z \rangle = 0) = \sigma^{-}(\langle Z \rangle = 0)$.
Tables \ref{table:silicate} and \ref{table:carbonaceous} give the best fit parameters to the charge centroid and charge width distributions for silicate and carbonaceous grains, respectively.

\begin{table*}
  \centering
  \begin{tabular}{c*{9}{l}}
   	\multicolumn{2}{c}{} & \multicolumn{5}{c}{Silicate Grains} \\
    \toprule
    & \multicolumn{4}{c}{$\langle Z \rangle$} & \multicolumn{2}{c}{$\sigma_{z}^{+}$} &\multicolumn{1}{c}{} & \multicolumn{2}{c}{$\sigma_{z}^{-}$} \\
    \cmidrule(lr){2-5} \cmidrule(lr){6-7} \cmidrule(lr){9-10}
    grain size ({\rm{\AA}}) & $\alpha$ & $k$ & $b$ & $h_{Z}$ & $c^{+}$ & $\eta^{+}$ & $d$ & $c^{-}$ & $\eta^{-}$ \\
    \midrule

    3.5  & 0.3263 & 0.0149 & -0.1212 & 57  & 0.4123 & 0.2513 & 0.1891 & 0.4845 & 0.3532 \\
	5    & 0.3141 & 0.0372 & -0.3043 & 86 & 0.2734 & 0.2925 & 0.3233 & 0.3615 & 0.6532\\
	10   & 0.3535 & 0.0494 & -0.4865 & 73 & 0.4353 & 0.7459 &  0.4451 & 0.1053 & 0.5803\\
	50   & 0.5115 & 0.0717 & -0.4106 & 107 & 1.0758 & 1.7832 & 0.5860 & -1.0379e3 & 7.7069e3  \\
	100  & 0.3525 & 0.6591 & -0.1649 & 384  & 1.6245 & 2.8390 & 0.6346& -4.2075e2 & 1.9840e3  \\
	500  & 0.3643 & 2.6283 & 0.5217 & 345  & 4.0732 & 11.0200 &  0.6797 & -0.2418 & 0.5910 \\
	1000 & 0.3927 & 3.6493 & 0.8389 & 372  & 5.9813 & 20.6410 & 0.6961 & -0.1885 & 0.4237 \\
   
    \bottomrule
  \end{tabular}
  \caption{Fitted parameters of the centroid and width of the distribution for silicate grains, with sizes between 3.5~{\AA} to 1000~{\AA}, for equations \ref{eq:charging_par}, \ref{eq:width_par_pos} and \ref{eq:width_par_neg}.}
  \label{table:silicate}
\end{table*}
%
%

\begin{table*}
  \centering
  \begin{tabular}{c*{9}{l}}
     \multicolumn{2}{c}{} & \multicolumn{5}{c}{Carbonaceous Grains} \\
    \toprule
    & \multicolumn{4}{c}{$\langle Z \rangle$} & \multicolumn{2}{c}{$\sigma_{z}^{+}$} &\multicolumn{1}{c}{} & \multicolumn{2}{c}{$\sigma_{z}^{-}$} \\
    \cmidrule(lr){2-5} \cmidrule(lr){6-7} \cmidrule(lr){9-10}
    grain size ({\rm{\AA}}) & $\alpha$ & $k$ & $b$ & $h_{Z}$ & $c^{+}$ & $\eta^{+}$ & $d$ & $c^{-}$ & $\eta^{-}$ \\
    \midrule
    
    3.5 &	 0.4699 & 0.0085 & -0.1162 & 48 & 0.3103 & 0.2744 & 0.2551 & 0.3766 & 0.5241\\
    5   &    0.4386 & 0.0195 & -0.3084 & 95 & 0.3699 & 0.5654 & 0.4158 & 0.2890 & 1.6241 \\
	10  &	 0.4994 & 0.0199 & -0.4959 & 78  & 0.6511 & 0.9839 & 0.5275 & -0.0213 & 0.0977 \\
	50  &    0.6009 & 0.0523 & -0.4092 & 218&  1.6536 & 2.6688 & 0.6671 & -9.5138 & 35.3519  \\
	100 &	 0.2900 & 2.2310 & -0.2061 & 1063 & 2.5445 & 4.3352 & 0.7010  & -2.5341e3 & 8.1962e3 \\
	500 &	 0.3400 & 5.8944 & 0.1727 & 1034 & 5.9455 & 18.3186 &  0.8377 & -2.4189e3 & 4.9424e3\\
  	1000 &	 0.3500 & 9.6536 & 0.4183 & 1273 & 8.7003 & 36.1014 & 0.9094 & -2.6009e3 & 4.7029e3\\

    \bottomrule
  \end{tabular}
  \caption{Fitted parameters of the centroid and width of the distribution for carbonaceous grains for equations \ref{eq:charging_par}, \ref{eq:width_par_pos} and \ref{eq:width_par_neg}.}
  \label{table:carbonaceous}
\end{table*}

\begin{figure*}
\centering 
\includegraphics[width=1.0\textwidth]{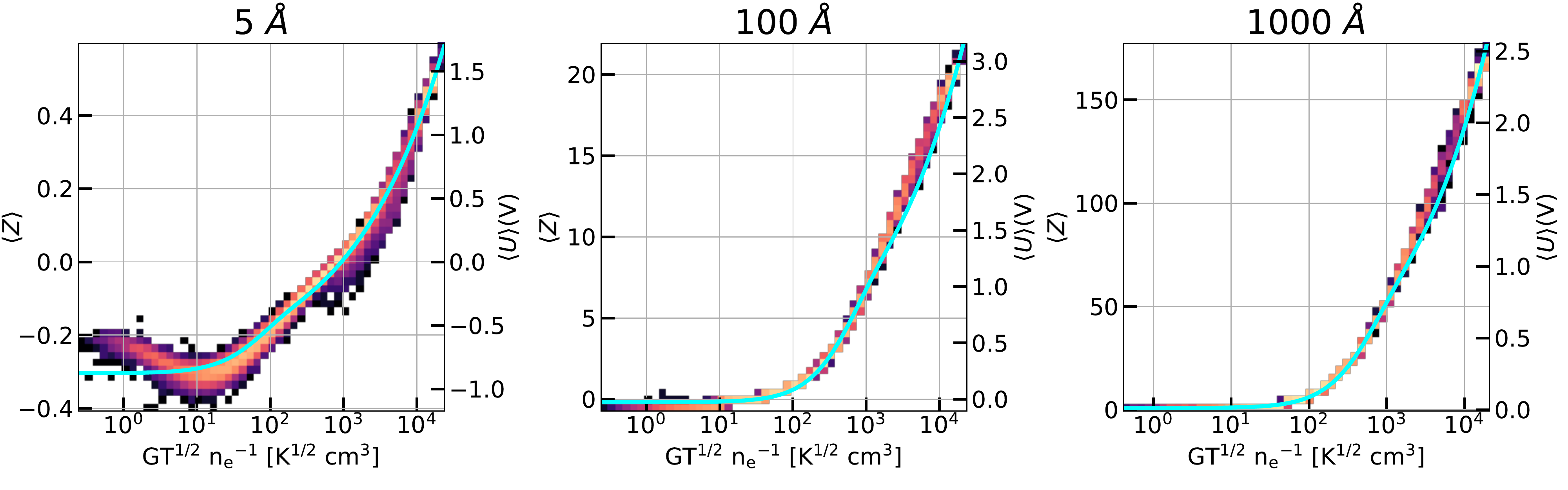}
\includegraphics[width=1.0\textwidth]{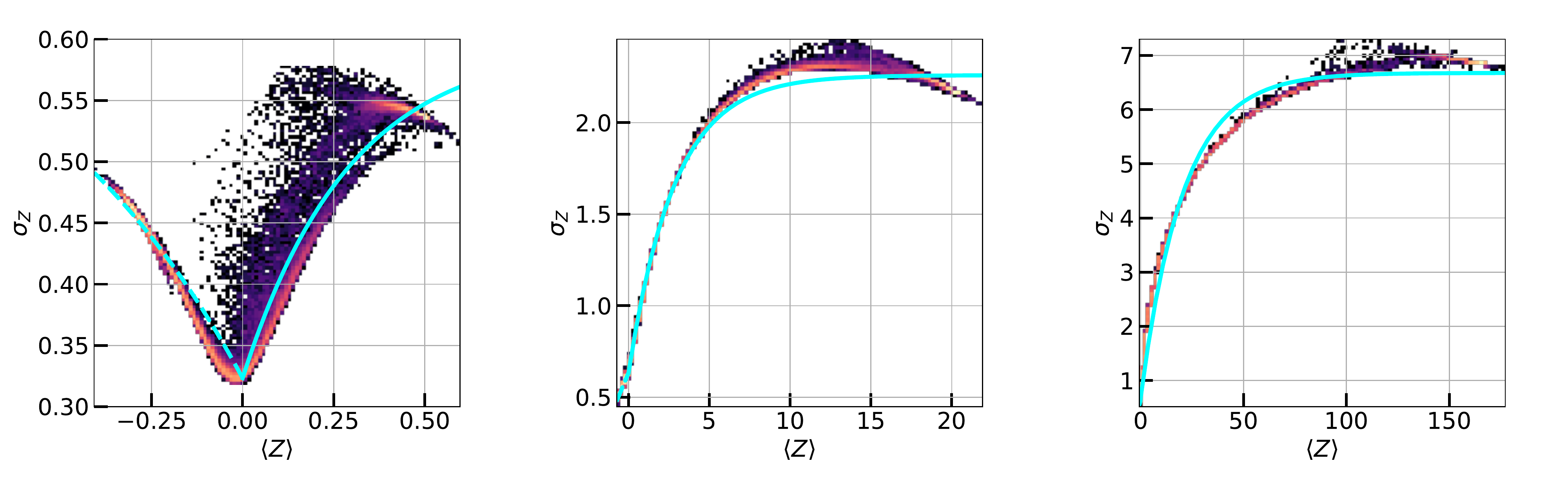}
\caption{(\emph{Top row}) Distribution of charge centroids, and average electrostatic potential, as a function of the charging parameter, and (\emph{bottom row}) distribution of widths as a function of the centroid, for silicate dust grains with sizes 5~\textrm{\AA}, 100~\textrm{\AA} and 1000~\textrm{\AA}. 
Top panels include the polynomial fit to the centroid following equation \ref{eq:charging_par}, while the bottom panels include the polynomial fit to the width following equations \ref{eq:width_par_pos} and \ref{eq:width_par_neg}.
\label{fig:newChargingPar_sil}} 
\end{figure*}
Figure \ref{fig:newChargingPar_sil} shows the distribution of charge centroids and widths for three different sizes of silicate grains, including the new parametric equations for the charge centroid, eq. \ref{eq:charging_par}, and the parametric equations for the width, eqs. \ref{eq:width_par_pos} and \ref{eq:width_par_neg}.
Figures of the centroid distribution and width distribution including the fitted parameters for all the silicates and carbonaceous grains analyzed are shown in appendices \ref{appendix_silicate_grains} and \ref{appendix_carbonaceous_grains}.

\begin{table}
  \centering
  \begin{tabular}{c*{4}{l}}
  	 \multicolumn{4}{c}{Silicate Grains} \\
    \toprule
    \multicolumn{1}{c}{} & \multicolumn{3}{c}{percentile} \\
    \cmidrule(lr){2-4}
	grain size (\AA) &	 25 &	 50 &	 75 \\
	\midrule
	3.5 &   5.4\% &	 13.1\% &	 20.1\% \\
	5 &	    2.4\% &	 4.4\% &	 9.1\% \\
	10 &    1.3\% &	 3.8\% &	 7.7\% \\
	50 &	7.8\% &	 14.6\% &    21.6\% \\
	100 &	2.6\% &	 7.5\% &	 32.8\% \\
	500 &	1.5\% &	 3.6\% &	 9.4\% \\
	1000 &  1.5\% &	 4.0\% &	 9.7\% \\
   \bottomrule
  \end{tabular}
  \caption{The 25, 50, and 75 percentiles of the error comparing the distribution of charge centroids from the full calculation of the charge distribution, with the charge centroid calculated using the parametric equation, eq \ref{eq:charging_par}, for silicate grains.}
  \label{table:silicateError}
\end{table}

\begin{table}
  \centering
  \begin{tabular}{c*{4}{l}}
  	\multicolumn{4}{c}{Carbonaceous Grains} \\
    \toprule
    \multicolumn{1}{c}{} & \multicolumn{3}{c}{percentile} \\
    \cmidrule(lr){2-4}
    	grain size (\AA) &	 25 &	 50 &	 75 \\
	\midrule
	3.5 &	     2.0\% &	 5.1\% &	 9.8\% \\
	5 &	     2.1\% &	 5.3\% &	 9.4\% \\
	10 &	    1.6\% &	         3.9\% &	 8.0\% \\
	50 &	     5.9\% &	 14.8\% &	 30.0\% \\
	100 &    1.2\% &	 3.7\% &	 13.7\% \\
	500 &    1.0\% &	 2.4\% &	 6.6\% \\
	1000 &   2.7\% &	 4.9\% &	 12.8\% \\
   \bottomrule
  \end{tabular}
  \caption{Error distribution for the calculations of the charge centroids for carbonaceous grains.}
  \label{table:carbonaceousError}
\end{table}

In order to test the reliability of the parametric calculation of the charge centroid, we compare the relative error between the centroid recovered using equation \ref{eq:charging_par} and charge centroids calculated from the full calculation of $f(Z)$. Tables \ref{table:silicateError} and \ref{table:carbonaceousError}, show the 25, 50 and 75 percentile of the error distribution for silicate and carbonaceous grains for a range of sizes.
The errors are small, with a median error around 3\%--6\% for both grain compositions and most of the grain sizes.
Only for the 3.5~{\AA} silicate grains, and the 50~{\AA} silicate and carbonaceous grains, the median error goes up to 14\%.


\subsubsection{Testing interpolation of the parametric equations}

We have provided a parametric equation to instantaneously calculate the charge centroid and width for selected grain sizes, which can be used to instantaneously recover the charge distribution function of those grains.
However, what happens if we need the distribution function of a grain with a size not present in our tables?

In this case we perform a linear interpolation or extrapolation of the parameters, and use them to calculate the centroid and width of the distribution.
We present a comparison of the charge centroid for 6 different sizes (7.5~{\AA}, 25~{\AA}, 75~{\AA}, 250~{\AA}, 750~{\AA} and 2500~{\AA}) of silicate grains, in three different ISM environments, WNM, CNM and CMM.

\begin{figure}
\centering 
\includegraphics[width=0.45\textwidth]{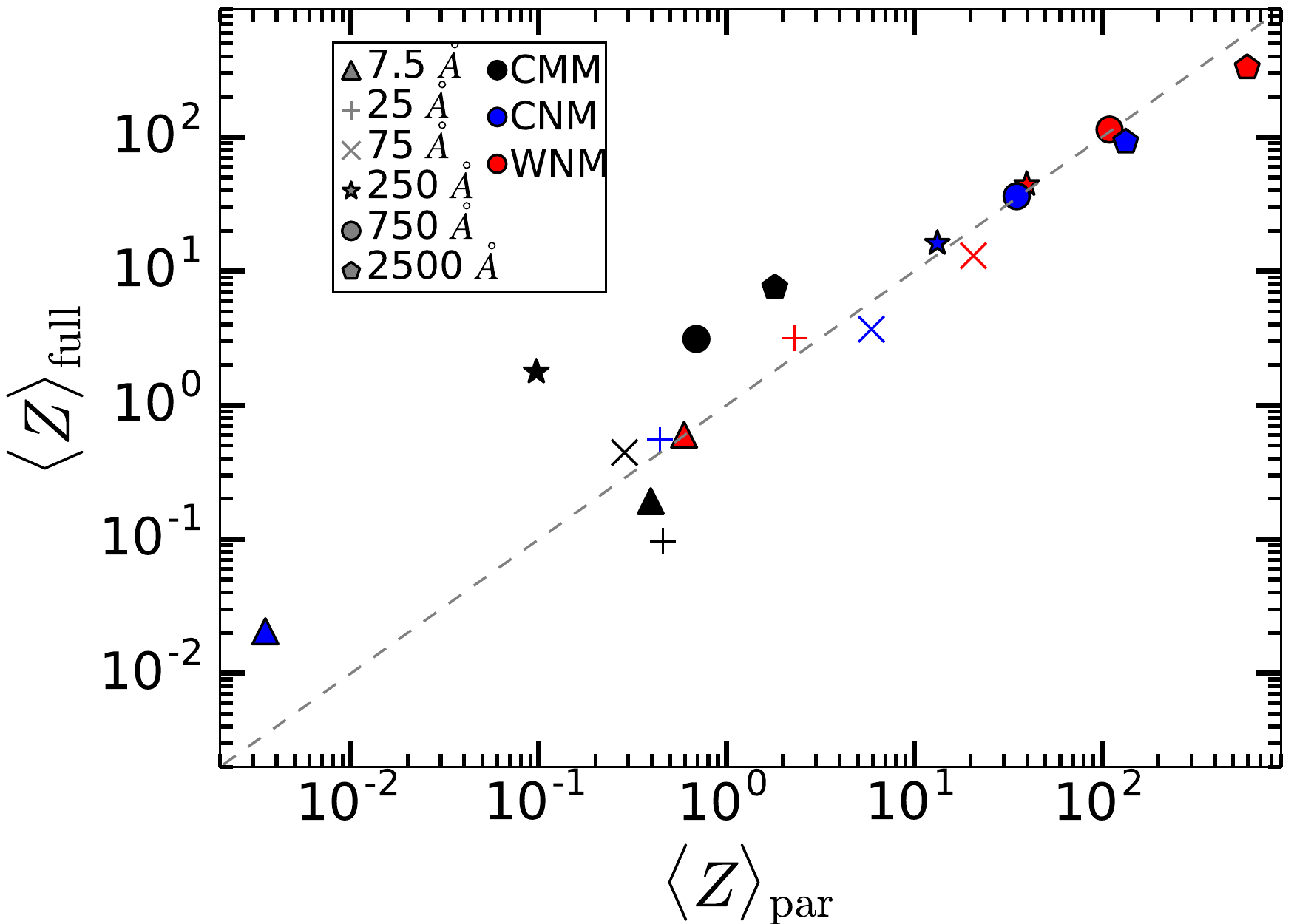}
\caption{Charge centroid calculated using the full calculation of the charge distribution, $\langle Z \rangle _{\mathrm{full}}$, versus the one calculated using the parametric equation, eq. \ref{eq:charging_par}, for silicate grains of five different sizes in three different environments. Grain sizes are differentiated by markers, and ISM ambient parameters by colors. WNM, CNM, and CMM, conditions used for the calculation are the same as the ones used in figure \ref{fig:fz_sample}. 
\label{fig:test}} 
\end{figure}

Figure \ref{fig:test} shows the charge centroid computed using the full calculation of the charge distribution, $\langle Z \rangle _{\mathrm{full}}$, versus the centroid calculated using the parametric equation, eq. \ref{eq:charging_par}, $\langle Z \rangle _{\mathrm{par}}$.
The parameters for the grains between 3.5~{\AA} and 1000~{\AA} are calculated using linear interpolation of the values in table \ref{table:silicate}, and the parameters for the grain with size 2500~{\AA} extrapolates these parameters.

We find good agreement of the charge centroids between the full calculation of the charge distribution and the parametric equation.
The centroids in the cold and warm neutral medium recover the expected charge centroids almost perfectly for all sizes.
Only some deviations between the full calculation and the parametric equation values of the centroid are observed in the cold molecular medium, where the is the larger scatter of the distribution of centroids as a function of the charging parameter.
We observe that for grains above 250~{\AA}, the parametric equation systematically underestimates the charge centroid. 
We also note that, besides the underestimation of the centroid in the CMM, the extrapolation of parameters for the 2500~{\AA} silicate grain does a good job recovering the charge centroid, suggesting that this procedure could be implemented up to these grain size when necessary.


\subsection{Parametric equations applied to PDR conditions.}

The parametric equations provided here have been optimized for recovering the charge distribution of dust grains in solar neighborhood ISM conditions. 
However, how do this parametric equations behave if we apply them 
in an environment outside the parameter space used to calibrate our models?
Will the model predict reasonable charge distributions, or will it predict dust charges far outside from what is expected from the equilibrium charge distribution?

We now test the behavior of the parametric equation in photon dominated regions (PDRs), where the radiation field is about three orders of magnitude stronger than the one used to calibrate our models. 

\begin{figure*}
\centering 
{\Large{Full Calculation}}
\includegraphics[width=0.95\textwidth, trim={0.1cm 0.8cm 0 2cm },clip]{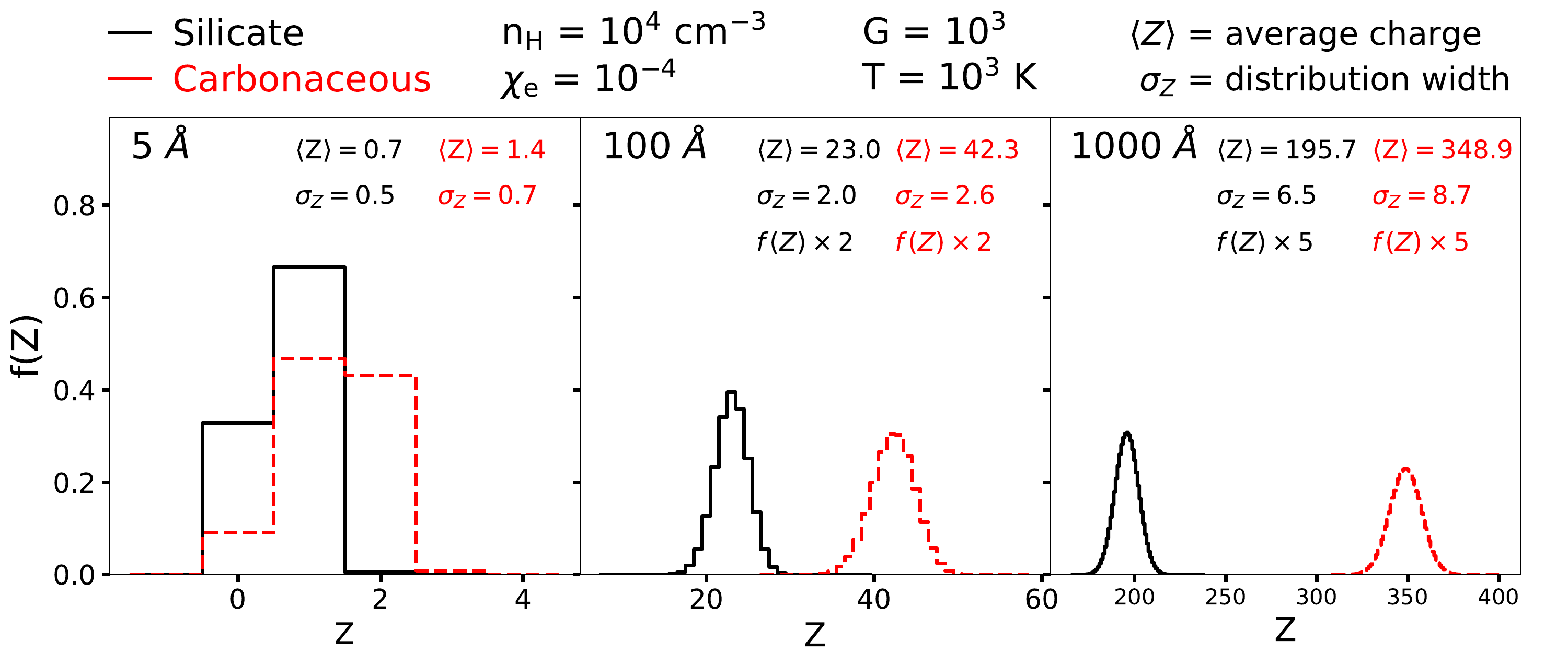}\\

\vspace{0.25cm}
{\Large{Parametric Calculation}}
\includegraphics[width=0.95\textwidth, trim={0 0 0 2cm },clip]{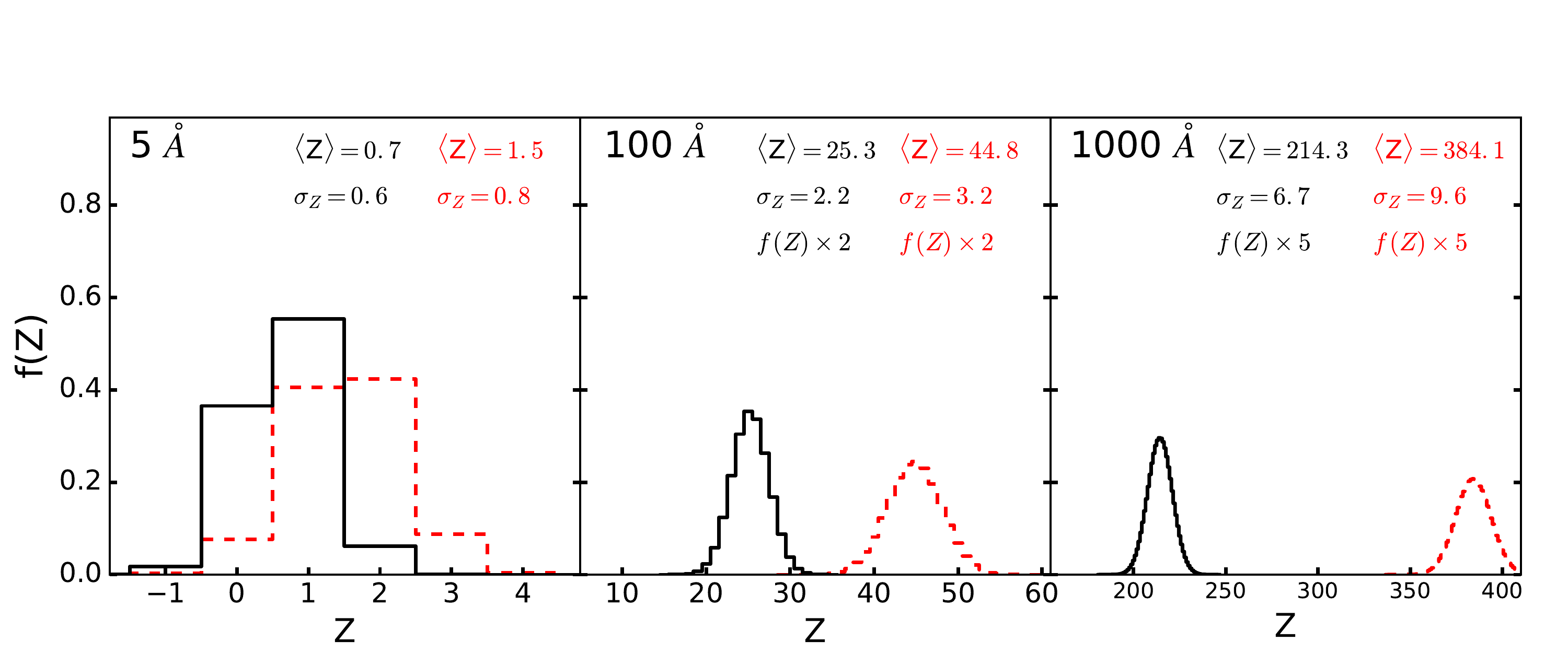}
\vspace*{-0.2cm}
\caption{Charge distribution of dust grains for the ({\emph{Top}}) full calculation using {\dustcode} and (\emph{Bottom}) reconstructed distribution using the parametric equation. 
Distributions for a $5$~{\AA}, $100$~{\AA} and $0.1~\mu$m size, carbonaceous ({\emph{dashed red}}) and silicate ({\emph{solid black}}) grains, in PDR conditions. 
The local environment parameters for the calculations are: $n = 10^{4}$~cm$^{-3}$, $T = 10^{3}$~K, $\chi_{e}= 10^{-4}$, and $G_{\mathrm{tot}}=10^{3}$.
Each panel includes the centroid, $\langle Z \rangle$, and width, $\sigma_{Z}$, of the distributions. 
\label{fig:fz_PDR}} 
\end{figure*}

Figure \ref{fig:fz_PDR} shows the comparison between the full calculation of the charge distribution using {\dustcode}, and the reconstructed distribution 
using the parametric equation, of small, intermediate and large silicate and carbonaceous grains in PDR conditions.
The reconstructed charge distributions for all grain sizes and compositions in the figure were built using the centroid and width from the parametric equations assuming Gaussian distributions. 
If the centroid would be an integer value, then the distribution would be perfectly symmetric. 
However, since the charge centroid, which is the peak of the distribution, is a real number and the distribution is sampled only in discrete bins of unit charge, the distribution might appear non-symmetric.
%
%
Tables \ref{table:PDRsilicate} and \ref{table:PDRcarbonaceous} show the comparison of the charge centroid and width between the full calculation of the charge distribution using {\dustcode} and the predicted values using the parametric equation.
It is remarkable how well the parametric equation predicts the charge distributions, with only small deviations of the order of 6--10\% in the charge centroid, but slightly larger errors in the calculation of the width of the distribution, with errors of the order of 10--20\%.

\begin{table}
  \centering
  \begin{tabular}{c*{7}{l}}
  	\multicolumn{7}{c}{Silicate Grains} \\
    \toprule
    \multicolumn{1}{c}{} & \multicolumn{2}{c}{full calculation} & \multicolumn{2}{c}{parametric eq.} & \multicolumn{2}{c}{error} \\
    \cmidrule(lr){2-3}  \cmidrule(lr){4-5} \cmidrule(lr){6-7} 	
    grain size (\AA) &	 $\langle Z \rangle$ &	 $\sigma_{Z}$ &	  $\langle Z \rangle$ &	 $\sigma_{Z}$ & $\langle Z \rangle$ &	 $\sigma_{Z}$ \\
	\midrule
	5    & 0.7   & 0.5  &	0.7   & 0.6 & 7\% & 12\% \\
	100  & 23.0  & 2.0  &	25.3  & 2.2 & 6\% & 24\% \\
	1000 & 195.7 & 6.5  &	214.3 & 6.7 & 13\% & 10\% \\
   \bottomrule
  \end{tabular}
  \caption{Error distribution for the calculations of the charge centroids for carbonaceous grains in PDR conditions.}
  \label{table:PDRsilicate}
\end{table}

\begin{table}
  \centering
 \begin{tabular}{c*{7}{l}}
  	\multicolumn{7}{c}{Carbonaceous Grains} \\
    \toprule
    \multicolumn{1}{c}{} & \multicolumn{2}{c}{full calculation} & \multicolumn{2}{c}{parametric eq.} & \multicolumn{2}{c}{error} \\
    \cmidrule(lr){2-3}  \cmidrule(lr){4-5} \cmidrule(lr){6-7} 	
    grain size (\AA) &	 $\langle Z \rangle$ &	 $\sigma_{Z}$ &	  $\langle Z \rangle$ &	 $\sigma_{Z}$ & $\langle Z \rangle$ &	 $\sigma_{Z}$ \\
	\midrule
	5    & 1.4   & 0.7  &	1.5  & 0.8	& 9\%	& 12\% \\
	100  & 42.3  & 2.6  &	44.8  & 3.2 & 6\%	& 24\% \\
	1000 & 348.9 & 8.7  &	384.1 & 9.6 & 10\%	& 10\% \\
   \bottomrule
  \end{tabular}
  \caption{Error distribution for the calculations of the charge centroids for carbonaceous grains in PDR conditions.}
  \label{table:PDRcarbonaceous}
\end{table}

\subsection{Implications}

In addition to an accurate estimate of the charge centroid as a function of the local ambient parameters (radiation field intensity, temperature and electron density), knowing the width of the charge distribution as a function of charge centroid allows us to reconstruct the full charge distribution of any dust grain instantaneously without having to calculate it on the fly.

The dust charge determines the coupling strength of grains with the interstellar magnetic field \citep{Lazarian2001GrainGas,Yan2004DustTurbulence}, where an overly simplified calculation of the dust charge, as in \citet{Lee2016TheClouds}, may lead to an overestimation of the coupling efficiency of dust and gas, and thus to overestimation of dust--gas segregation.
Equations \ref{eq:charging_par}, \ref{eq:width_par_pos} and \ref{eq:width_par_neg}, together with tables \ref{table:silicate} and \ref{table:carbonaceous}, can be easily implemented in 3D MHD codes, to accurately estimate Coulomb and Lorentz forces on dust grains. We provide {\sc{Python}}, C and {\sc{Fortran}} functions of the charge distribution parametric calculation ready to be implemented in numerical simulations and chemical models. \\

Knowledge of the charge distribution of dust grains not only influences the evolution of dust grains themselves, but also impacts the thermal, chemical, dynamical, and spectroscopical properties of the ISM.
For example: dust charges modify the non-ideal magneto-hydrodynamic resistivities of the plasma \citep{Elmegreen1979MagneticGrains, Marchand2016ACalculations}.
Variations in the non-ideal magneto-hydrodynamics resistivities modify the coupling between the neutral gas and the interstellar magnetic field, which eventually influences  the dynamics of gravitational collapse of molecular clouds and the formation of stars and planets \citep{Nakano2002MechanismClouds, Dominik2006GrowthFormation, Shu2006GravitationalDissipation,Wurster2017TheFormation,Wurster2018TheMagnetohydrodynamics}. 
The charge of a dust grain modifies the cross sections of the ion accretion on dust, this critically influences the surface chemistry, which in turn influences the formation and growth of grain mantles \citep{Ferrara2016TheGalaxies, Hensley2016ThermodynamicsNanoparticles, Zhukovska2018IronDepletions}.
Grain charge influences the scattering of light on the grain surface, affecting the extinction efficiency and frequency of extinction features in the infrared \citep{Heinisch2012MieParticle,Heinisch2013OpticalPlasma}.


\section{Summary and Conclusions}
\label{sec:conclusions}

We present a new parametric expression for the charge distribution of silicate and carbonaceous dust grains to be implemented in numerical simulations and chemical models\footnote{we provide {\sc{Python}}, C++ and {\sc{Fortran}} functions of the charge distribution parametric calculation ready to be implemented in numerical simulations and chemical models, and can be downloaded from https://github.com/jcibanezm/DustCharge}.
We calculate the equilibrium charge distribution of dust grains using the master charging equation accounting for collisional charging by electrons and ions, and photoelectric charging from an incident radiation field and cosmic ray charging processes.
The ambient conditions, e.g. temperature, density, ionization fraction, molecular hydrogen column density, dust column density, and local visual extinction, were taken from a 3D HD simulation of molecular cloud formation in a colliding flow, including gas self-gravity, non-equilibrium chemical evolution of the gas, and shielding of the ISRF by gas and dust.
The simulations capture a space of parameters of temperature $T\approx 6$~K -- $10^{4}$~K, number density $n_{\mathrm{H}} \approx 0.1$~cm$^{-3}$ -- $5\times 10^{4}$~cm$^{-3}$, radiation field intensity $G \approx 10^{-4}$ -- $1.7$, and a visual extinction of $\mathrm{Av} \approx 10^{-2}$~mag -- $10$~mag.

We find that the charge distribution of dust grains strongly depends on the grain composition and size, and some ISM ambient parameters, such as temperature, electron density, and local strength of the radiation field.
From each of the calculated charge distributions, we extract the charge centroid and width, assuming the distribution is well represented by a Gaussian.
We parameterize the distribution of charge centroids as a non-linear function of the ``charging parameter'', $G_{\mathrm{tot}} \sqrt{T} n_{\mathrm{e}}^{-1}$, where $G_{\mathrm{tot}}$ accounts for the attenuated ISRF and the CR-induced H$_{2}$ fluorescence within molecular clouds, motivated by the dominant mechanisms involved in  ``forward'' and ``backward'' charging processes.
We find that the shape of the charge centroid distribution is well represented by an exponential growth at low values of the charging parameter, followed by a polynomial growth, 
$\langle Z \rangle = k [ 1 - \mathrm{exp}(G_{\mathrm{tot}}T^{0.5} n^{-1}_{\mathrm{e}}h_{Z}^{-1})] (G_{\mathrm{tot}} T^{0.5} n_{e}^{-1})^{\alpha} + b $, where $k$, $b$, $\alpha$, and $h_{Z}$ are the fitting parameters determining the scale factor of the exponential growth, exponent of the power law growth, and minimum charge centroid. 
The best fit values to the charge centroid parameterized expression for various silicate and carbonaceous grains with sizes ranging from $3.5~$\textrm{\AA} to $0.1$~$\mu$m, are given in tables \ref{table:silicate} and \ref{table:carbonaceous}.
We find that interpolating the parameters in the tables to calculate the centroid for grains with sizes not included in our analysis yields results consistent to the charge centroids obtained from the full calculation of the charge distribution.
We find that the distribution of charge widths is tightly correlated to the charge centroid with a plateau at high charges, following an exponential growth, $\sigma_{\text{\tiny{Z}}} = c [1 - \mathrm{exp}(-\langle Z \rangle \eta^{-1})] + d$, where $c$, $\eta$, and $d$ are the fitting parameters describing the scale factor of the exponential growth and minimum width of the charge distribution. 
We divide the parametrized width distribution in two, depending on the sign of the charge centroid, eq. \ref{eq:width_par_pos} for positive and eq. \ref{eq:width_par_neg} for negative charge centroids.
%
We also test the parametric equation in PDR conditions, which lie outside the parameter space  used to calibrate our models, and obtain charge distribution functions consistent to the full calculation of the charge distribution.

Finally we test the validity of the assumption of the charge equilibrium used to calculate the distribution function by comparing the charging timescale of dust grains with the hydrodynamical timescale constrained by the simulation timestep.
We find that the equilibrium charge distribution assumption is valid for ISM simulations with sub-parsec resolution, and would remain valid for simulations with AU scale resolution. 

\section*{Acknowledgements}

We thank Laurent Verstraete and Bruce Draine, for their useful comments on the charge distribution calculation and benchmarking the code. 
We also thank Bo Zhao, Franta Dinbier, Seyit Hocuk, Daniel Seifried and Eric Pellegrini for useful
discussions and comments on the manuscript.
JCIM, SW, and PJ acknowledge funding by the supported by the Deutsche Forschungsgemeinschaft (DFG) via the Priority  Program 1573 ``The physics of the interstellar medium''  and the DFG Collaborative  Research Center SFB 956 ``Conditions and  Impact  of  Star  Formation'' (subproject  C5).
SW and SC  acknowledges support via the ERC starting grant No. 679852 ``RADFEEDBACK''. 
AVI acknowledges support by the Russian Science Foundation (project 18-12-00351).
The  authors  acknowledge  the  Leibniz Rechenzentrum Garching, as  well  as  the  Gauss Center for Supercomputing e.V. (www.gauss-centre.eu) for providing computing time on SuperMUC  via  the  project ``pr62su''. 
The software used in this work was developed in part by the DOE NNSA ASC- and DOE Office of Science ASCR-supported Flash Center for Computational Science at the University of Chicago. 
Parts of the analysis are carried out using the {\sc{YT}} analysis package \citep[][yt-project.org]{Turk2010AData}.







\bibliographystyle{mnras}
\bibliography{Mendeley}

\appendix

\section{Collisional charging rates}
\label{appendix_collisional_charging}

We adopt the collisional charging rates by \citet{Draine1987CollisionalGrains}, accounting for collisions with ions and electrons.
These revisited charging rates account for the effects due to the electrostatic polarization of the grain by the Coulomb field of the approaching charged particle. 
and energy-dependent capture cross sections.

In their work, \citet{Draine1987CollisionalGrains} reduced the complexity of the problem to two dimensionless parameters: the "reduced temperature", $\tau = akT/q$, and the ``effective atomic weight'' of the colliding ions, $\mu = (n_{\mathrm{e}} / n_{\mathrm{I}})^{2}(m_{\mathrm{e}}/m_{p})$, where $a$ is the grain size, $k$ is Boltzmann constant, $T$ is the temperature of the plasma, and $q$ is the charge of the collisional partner, $n_{\mathrm{e}}$ is the volume density of electrons, $s_{\mathrm{e}}$ their sticking coefficient, $n_{\mathrm{I}}$ is the ion volume density, $m_{\mathrm{I}}$ the average ion mass and $m_{\mathrm{p}}$ is the proton mass. 
With these definition, $\mu$ not only accounts for the relative mass between electrons and ions, but also for variations in the composition of the plasma.

Given that electrons have much lower masses compared to any colliding ion, collisional charging will ultimately result in grains with negative charges. 
For low values of the reduced temperature $\tau < 0.2$, grains will be in a dominant charge state of $Z=-1$, while in the limit of $\tau \rightarrow \infty$ grains tend to large negative charges.

\begin{figure*}
\centering 
\includegraphics[width=0.32\textwidth]{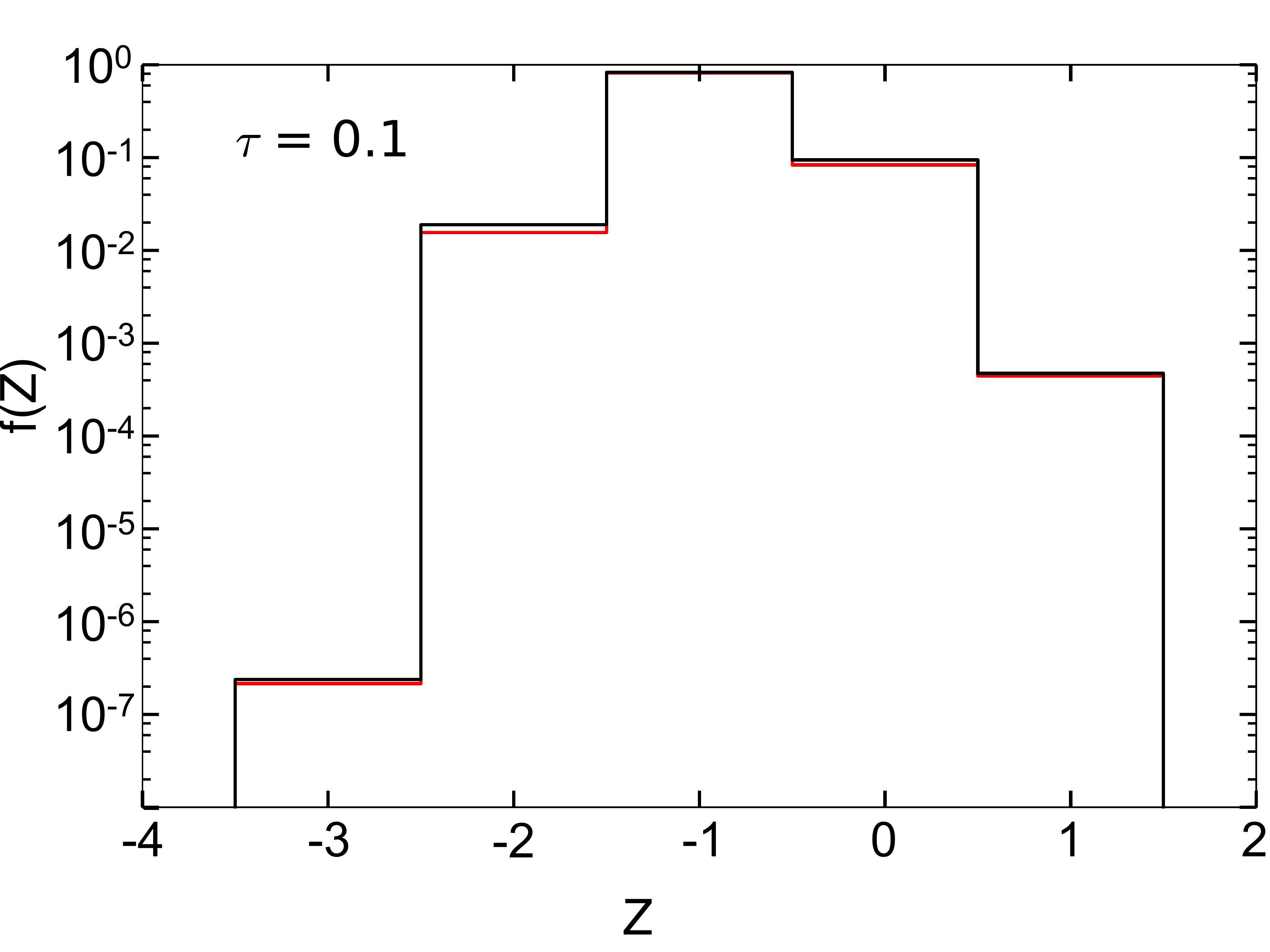} 
\includegraphics[width=0.32\textwidth]{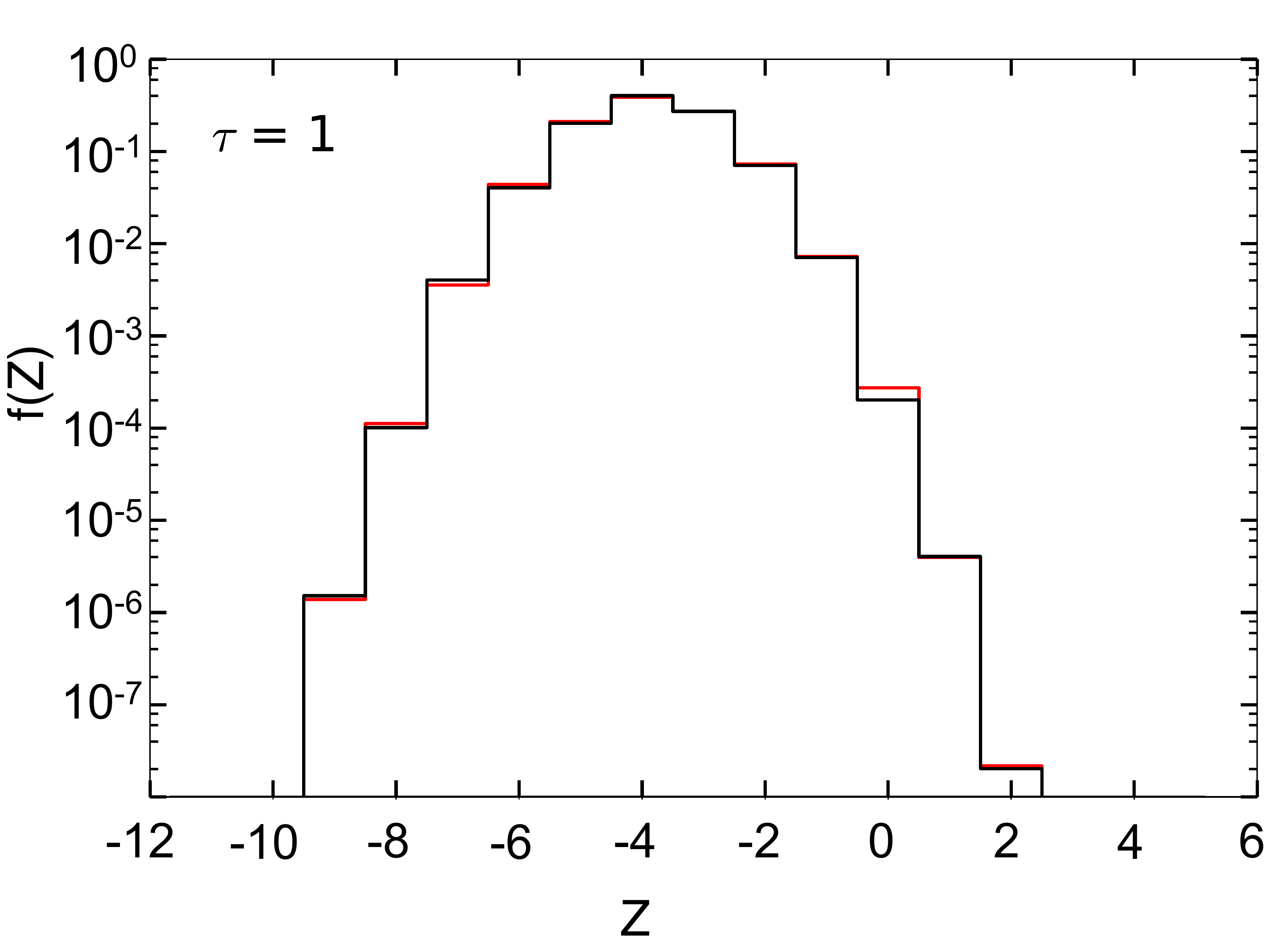} 
\includegraphics[width=0.32\textwidth]{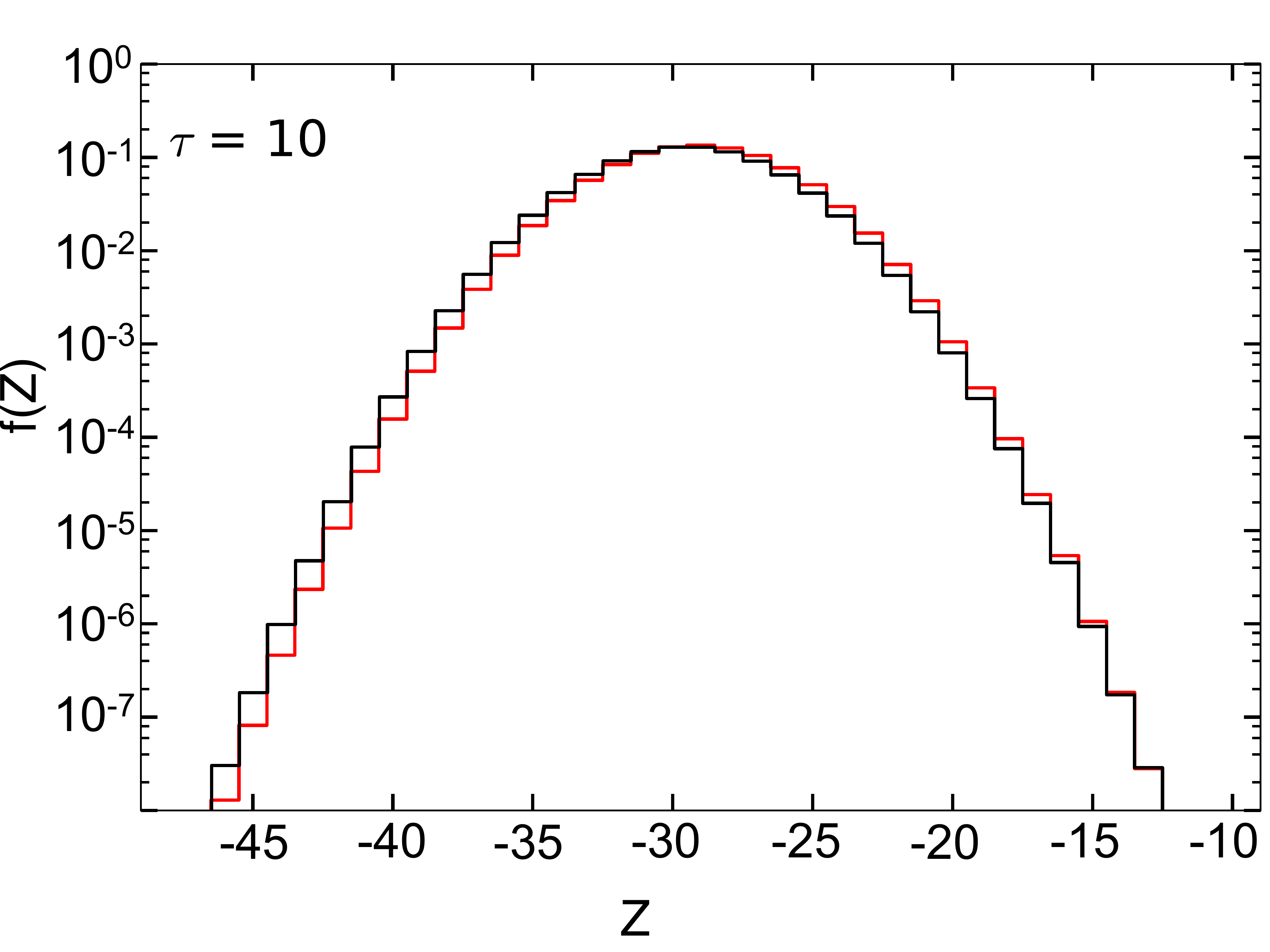} 
\caption{Charge distribution function for a dust grain accounting only for collisional charging for ({\emph{red}}) calculations using {\dustcode} and calculation by (\emph{black}) \citet{Draine1987CollisionalGrains}. Results are shown for an ``effective atomic weight'' $\mu = 1$, for three different values of the ``reduced temperature'' $\tau = akT/q^{2}$ of (\emph{left}) $\tau=0.1$, (\emph{middle}) $\tau=1.0$, and (\emph{right}) $\tau=10.$, where $a$ is the grain radius, $T$ the temperature of the plasma, and $q$ the charge of the colliding partner.
\label{fig:collisional_charging}}
\end{figure*}      

Figure \ref{fig:collisional_charging} show the charge distribution functions calculated with {\dustcode} and results by \citet{Draine1987CollisionalGrains}.
It is evident that both distributions are very similar, confirming that our calculations are performing as expected with respect to the collisional charging of dust grains. \\

\section{photoelectric charging by the interstellar radiation field}
\label{appendix_photoelectric_charging}

In this appendix we discuss the details about some of the components in the calculation of the photoelectric charging as presented in equation \ref{eq:Jpe}.

For the diffuse interstellar radiation field we adopt the one estimated by \citet{Mezger1982TheEmission} and \citet{Mathis1983InterstellarClouds} in the Solar neighborhood, also shown in equation 31 and table 1 in \citet{Weingartner2001PhotoelectricHeating}: 
\begin{strip}
\begin{equation}
\qquad\qquad\qquad\qquad\qquad \nu u_{\nu}^{\mathrm{ISRF}} = 
\begin{cases}
\begin{aligned}
&0, & h\nu > 13.6 \mathrm{~eV}\\
&3.328\times10^{-9} \mathrm{~ergs~cm}^{-3} \left(\frac{h\nu}{\mathrm{eV}}\right )^{-4.4172},\, & 11.2 \mathrm{~eV} < h\nu < 13.6 \mathrm{eV} \\
&8.463\times10^{-13} \mathrm{~ergs~cm}^{-3} \left(\frac{h\nu}{\mathrm{eV}} \right)^{-1},\, & 9.26\mathrm{ eV} < h\nu < 11.2\mathrm{~eV}\\
&2.055\times10^{-14} \mathrm{~ergs~cm}^{-3} \left( \frac{h\nu}{\mathrm{ev}} \right)^{0.6678},\, & 5.04\mathrm{~eV}< h\nu < 9.26\mathrm{eV}\\
&\frac{4\pi\nu}{c} \sum_{i=1}^{3} w_{i}B_{\nu}(T_{i}), \, & h\nu < 5.04 \mathrm{ eV}
\end{aligned}
\end{cases}
\label{eq:ISRF}
\end{equation}
\end{strip}
%
where blackbody temperatures, $B_{\nu}(T_{i})$, and dilution factors $w_{i}$ are given in table \ref{table:ISRF}.

In this work, we characterize the radiation field in units of the Habing field, $G = u^{\tiny{UV}}_{\mathrm{rad}} / u^{\tiny{UV}}_{\mathrm{\tiny{Hab}}}$, where $u^{\tiny{UV}}_{\mathrm{rad}}$ is the energy density of the radiation field integrated between 6 and 13.6 eV, and $u^{\tiny{UV}}_{\mathrm{\tiny{Hab}}} = 5.33 \times 10^{-14}$~ergs~cm$^{-3}$.
The integrated energy density of the radiation field by \citet{Mezger1982TheEmission} and \citet{Mathis1983InterstellarClouds} is equal to $G=1.13$, where the intensity of the radiation field used in the simulations by \citetalias{Joshi2018OnConditions} is equal to $G_{0}=1.7$.
For this reason we scale the radiation field by a factor of 1.5 to match that of the simulations.

Another important term in equation \ref{eq:Jpe} is the photoelectric yield, given by,
\begin{eqnarray}
Y(h\nu, Z, a) = y_{2}(h\nu, Z, a) \,\mathrm{min}[y_{0}(\theta) y_{1}(a, h\nu), 1].
\end{eqnarray}
This form of the photoelectric yield depends on three parameters: $y_{0}$, $y_{1}$ and $y_{2}$, given by equations 16, 13 and 11 respectively, in \citet{Weingartner2001PhotoelectricHeating}.
Here, $y_{0}$ corresponds to the yield of electrons that have enough energy to attempt to escape the grain, $y_{1}$ is responsible for the geometrical enhancement of the yield, and $y_{2}$ is the fraction of attempting electrons capable of escaping the grain to infinity. 

\begin{table}
  \centering
  \begin{tabular}{c*{3}{l}}
  	\multicolumn{3}{c}{Components of the ISRF.} \\
    \toprule
    i &	 $w_{i}$ &	 $T_{i}$~[K] \\
	\midrule
	1  & $1.00\times 10^{-14}$ & 7500  \\
	2  & $1.65\times 10^{-13}$ & 4000  \\
	3  & $4.00\times 10^{-13}$ & 3000  \\
   \bottomrule
  \end{tabular}
  \caption{Blackbody components for the ISRF in equation \ref{eq:ISRF} from \citet{Mathis1983InterstellarClouds}.}
  \label{table:ISRF}
  \vspace{-0.4cm}
\end{table}

The calculations of $y_{0}$ and $y_{2}$ are the same as the ones in \citet{Weingartner2001PhotoelectricHeating}; however, for the calculations of $y_{1}$ we use a fixed value for the electron escape length, $l_{e} = 10$~{\AA}, and photon attenuation length, $l_{a} = 100$~{\AA}, for both silicate and carbonaceous grains. 
This is equivalent to the calculation by \citet{Bakes1994TheHydrocarbons}, but differs from  the calculations by \citet{Weingartner2001PhotoelectricHeating}, where the authors include a variable photon attenuation length, equation (14) in \citet{Weingartner2001PhotoelectricHeating}, depending on the energy of the incident photon and the complex refractive index, Im(m), of the grain composition\footnote{Dielectric functions for astronomical silicate, graphite, and silicon carbide can be found at https://www.astro.princeton.edu/~draine/dust/dust.diel.html}. However, for the photon energies studied in this work, the full calculation of the photon attenuation length is of the order of 100~{\AA}.

\begin{figure*}
\centering 
\includegraphics[width=0.90\textwidth]{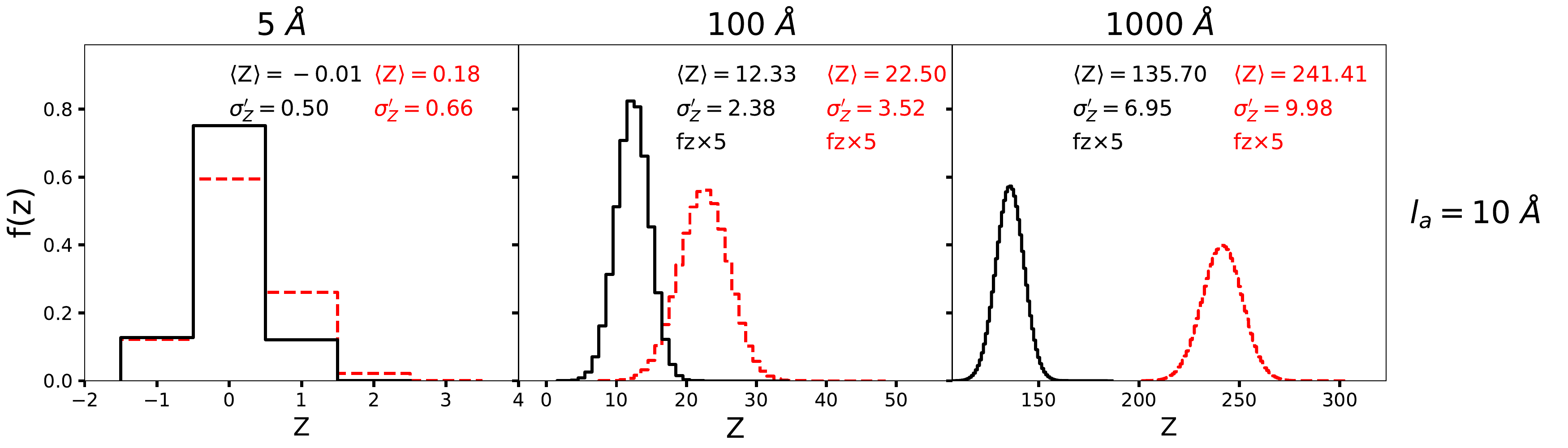}
\includegraphics[width=0.90\textwidth, trim={0 0 0 1cm },clip]{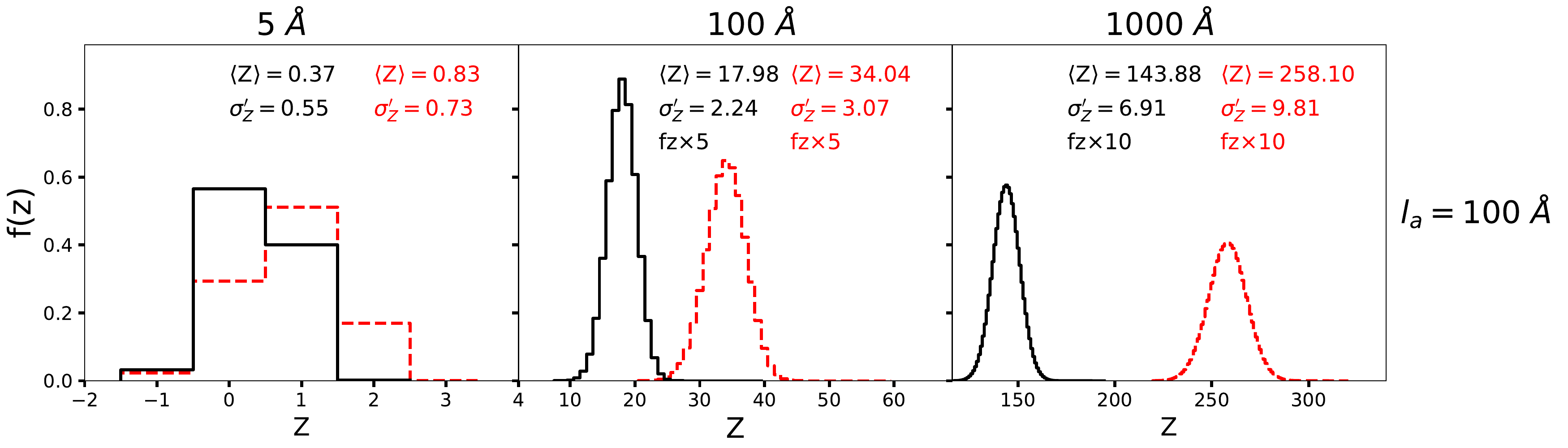}
\includegraphics[width=0.90\textwidth, trim={0 0 0 1cm },clip]{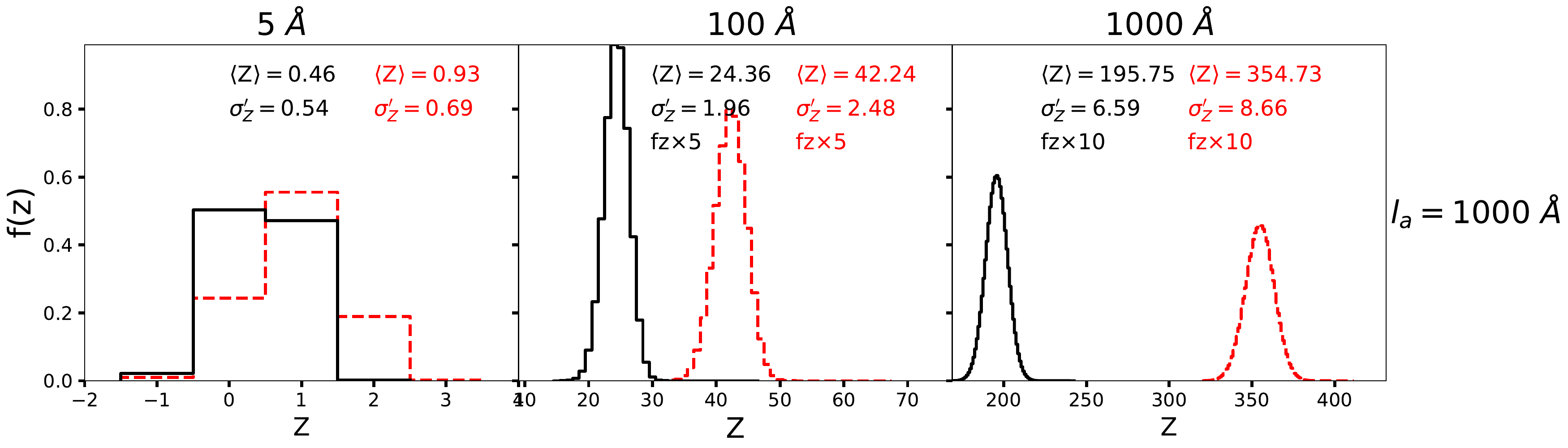}
\caption{Charge distribution of $5$~{\AA}, $100$~{\AA} and $0.1~\mu$m size, carbonaceous ({\emph{dashed red}}) and silicate ({\emph{solid black}}) grains, varying the attenuation length: (\emph{top}) $l_{a}=10$~{\AA}, (\emph{middle}) $l_{a}=100$~{\AA}, (\emph{bottom}) $l_{a}=1000$~{\AA}.
The local environment parameters for the calculations correspond to the WNM in Figure \ref{fig:fz_sample}: $n_{\mathrm{H}} = 0.9$~cm$^{-3}$, $T = 7000$~K, $\chi_{e}= 0.01$, $G=1.61$, $x_{\mathrm{H_{2}}} = 4.6\times10^{-5}$.
Each panel includes the centroid, $\langle Z \rangle$, and width, $\sigma_{Z}$, of the distributions. 
\label{fig:fz_attenuationlength}} 
\end{figure*}

\begin{figure*}
\centering 
\includegraphics[width=0.80\textwidth]{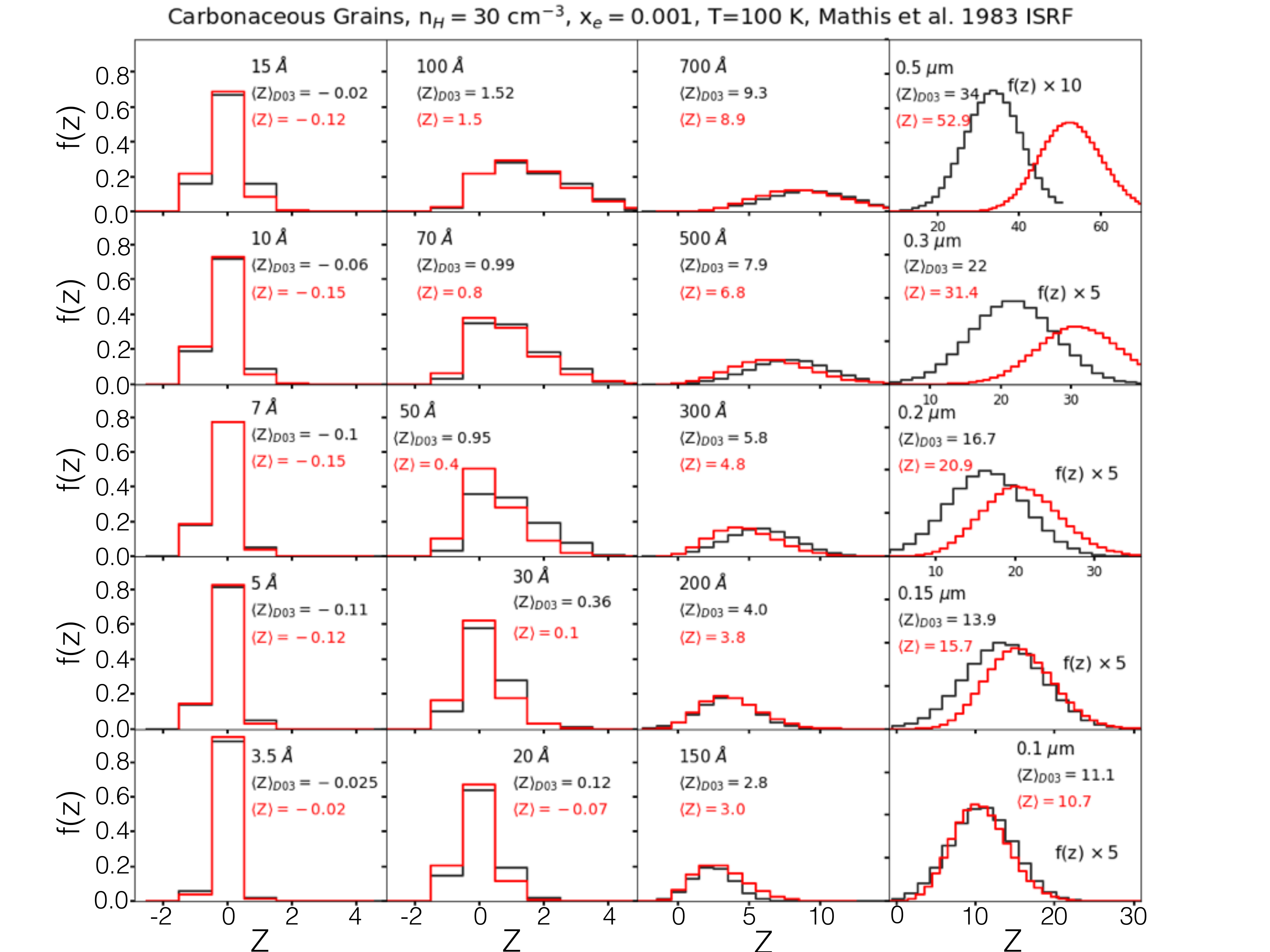}
\includegraphics[width=0.80\textwidth]{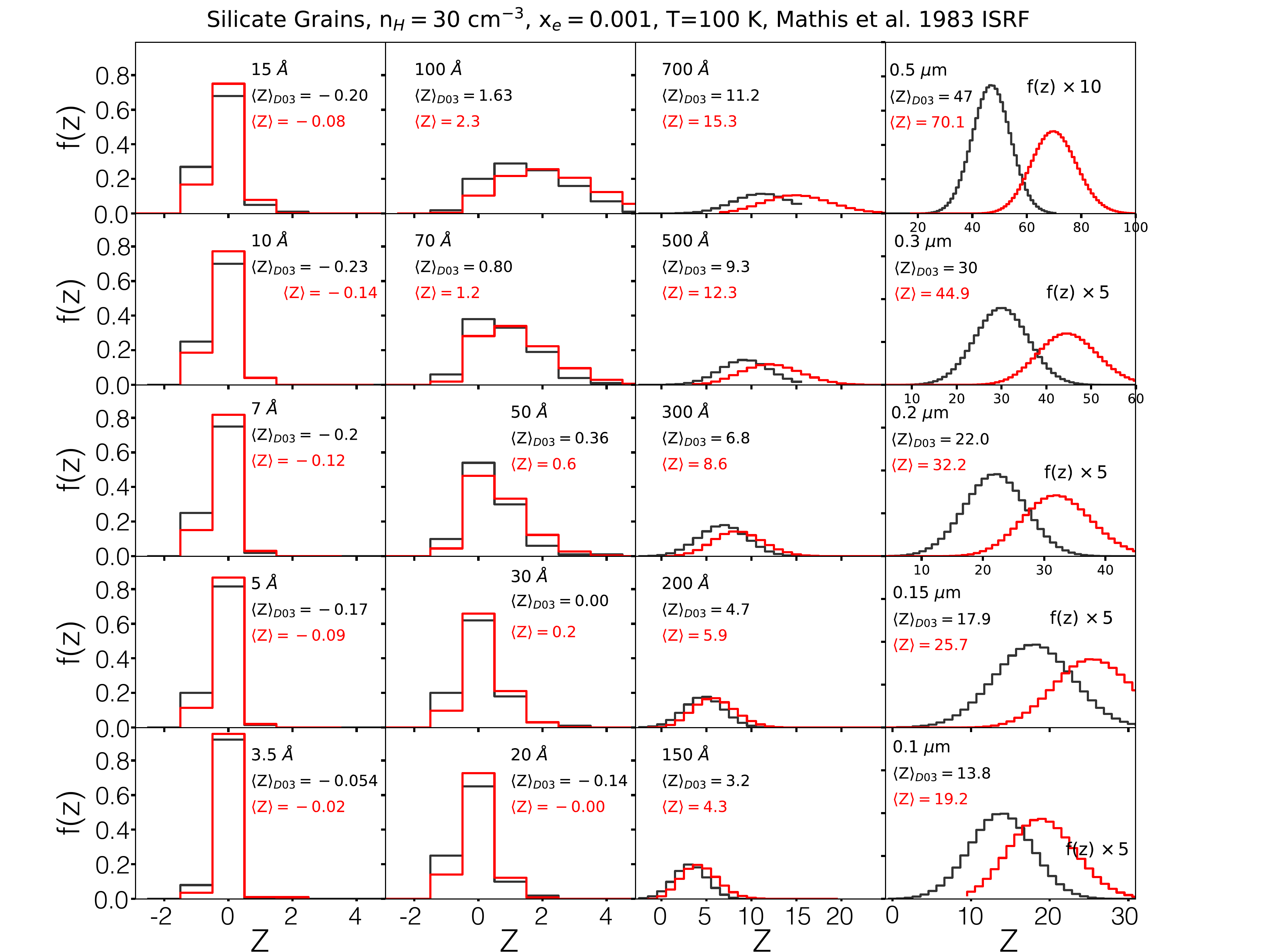} 
\caption{charge distribution function $f(Z)$ for carbonaceous grains of various sizes, between 3.5~{\AA} and 0.5~$\mu$m, with ambient conditions $n_{\mathrm{H}}=30$~cm$^{-3}$, $\chi_{\mathrm{e}}=10^{-3}$, $T=100$~K and $G=1.13$.
{\emph{Red lines}} correspond to the calculations performed in this work, and {\emph(black lines)} correspond to the calculations by \citep{Draine2011PhysicsMedium}. Within each panel we include the size of the grain, and the centroid, $\langle Z \rangle$ of the distributions, with the subscript $D03$ for the calculations by \citep{Draine2011PhysicsMedium}.
\label{fig:photoelectric_charging}}
\end{figure*}      

Figure \ref{fig:fz_attenuationlength} shows the influence of a variable photon attenuation length for a small, intermediate and large, silicate and carbonaceous grain in WNM conditions.
The attenuation length clearly modifies the photoelectric yield, thus influencing the photoelectric charging of grains.
Larger photon attenuation lengths produce higher photoelectric charging rates, which in turn results in higher positive charges.
This effect is clearly observed in the charge distribution of large grains, where the charge centroid for silicates (carbonaceous) grains is $\langle Z \rangle \approx 135$ ($\langle Z \rangle \approx 241$) for $l_{a}=10$~{\AA}, and increases to $\langle Z \rangle \approx 195$ ($\langle Z \rangle \approx 354$) for $l_{a}=1000$~{\AA}. 
For small grains, the photon attenuation length also influences the charge distribution, where the shape of the distribution function changes for the different values of $l_{a}$, and the values of the charge centroid vary from $\langle Z \rangle \approx 0.0$ ($\langle Z \rangle \approx 0.18$) for $l_{a}=10$~{\AA}, to $\langle Z \rangle \approx 0.46$ ($\langle Z \rangle \approx 0.93$) for $l_{a}=1000$~{\AA}.

Figure \ref{fig:photoelectric_charging} shows a comparison of the charge distribution functions calculated with {\dustcode} and \citet{Draine2011PhysicsMedium}.
It is observed that both distributions are very similar, with only minor variations in the charge centroid for grains between 3.5~{\AA} and $0.1~\mu$m. 
Above $0.1~\mu$m, the charge distributions calculated using {\dustcode} result in more positively charged grains than those by \citet{Weingartner2001PhotoelectricHeating}, with increasing difference with increasing size.
This can be attributed to the varying photon attenuation length previously discussed, where our fixed value of $l_{a} = 100$~{\AA}, may result in larger photoelectric yields compared to the calculations by \citet{Weingartner2001PhotoelectricHeating}.

\section{Effects of CR-induced charge}
\label{appendix_CR_charging}

\citet{Ivlev2015InterstellarEffects} discussed how CRs influence the charge distribution of dust grains within dense cold regions.
In this appendix, we show how including the effects of CR-induced charging influences the charge distribution of dust grains for different sizes within a dense molecular region. 

In order to compute the CR-induced H$_{2}$ fluorescence within molecular clouds, equation \ref{eq:CR_FUV}, we need to know the local CR ionization rate.
In this work, we implement the parameterised formula by \citet{Padovani2018Cosmic-rayDiscs}, which returns the total CR ionization rate as a function of the H$_{2}$ column density, and is given by:
\begin{eqnarray}
\mathrm{log}_{10}\frac{\zeta}{\mathrm{s}^{-1}} = \sum_{k\geq 0} c_{k} \mathrm{log}_{10}^{k}\frac{N_{\mathrm{H}_{2}}}{\mathrm{cm}^{-2}}
\label{eq:CR_ionization}
\end{eqnarray}
valid for column densities between $10^{19}$~cm$^{-2} \leq N_{\mathrm{H}_2} \leq 10^{27}$~cm$^{-2}$.
Table \ref{table_CRionization} gives the coefficients, $c_{k}$, for two CR-ionization rate models, ``high'' $\mathcal{H}$ and ``low'' $\mathcal{L}$.

\begin{table}
  \centering
  \begin{tabular}{c*{3}{c}}
    \toprule
    \multicolumn{1}{c}{} & \multicolumn{2}{c}{model} \\
    \cmidrule(lr){2-3}
    	$k$ &	$\mathcal{L}$  &	 $\mathcal{H}$  \\
	\midrule
	0 &	 -3.331056497233~$\times \; 10^{6}$     &	1.001098610761~$\times \; 10^{7}$  \\
	1 &	  1.207744586503~$\times \; 10^{6}$     &	-4.231294690194~$\times \; 10^{6}$  \\
	2 &	 -1.913914106234~$\times \; 10^{5}$     &	 7.921914432011~$\times \; 10^{5}$ \\
	3 &	 1.731822350618~$\times \; 10^{4}$      &	 -8.623677095423~$\times \; 10^{4}$ \\
	4 &	 -9.790557206178~$\times \; 10^{2}$     &	 6.015889127529~$\times \; 10^{3}$  \\
	5 &	 3.543830893824~$\times \; 10^{1}$      &	 -2.789238383353~$\times \; 10^{2}$  \\
	6 &   -8.034869454520~$\times \; 10^{-1}$   &	 8.595814402406~$\times \; 10^{0}$  \\
	7 &   1.048808593086~$\times \; 10^{-2}$    &	 -1.698029737474~$\times \; 10^{-1}$  \\
	8 &   -6.188760100997~$\times \; 10^{-5}$   &	 1.951179287567~$\times \; 10^{-3}$  \\
	9 &   3.122820990797~$\times \; 10^{-8}$    &	 -9.937499546711~$\times \; 10^{-6}$  \\
   \bottomrule
  \end{tabular}
  \caption{Coefficients $c_{k}$, for the polynomial equation \ref{eq:CR_ionization} for two models of the CR proton spectrum: a ``low'' $\mathcal{L}$ and ``high'' $\mathcal{H}$ spectrum.}
  \label{table_CRionization}
\end{table}

\begin{figure*}
\centering 
\includegraphics[width=0.85\textwidth]{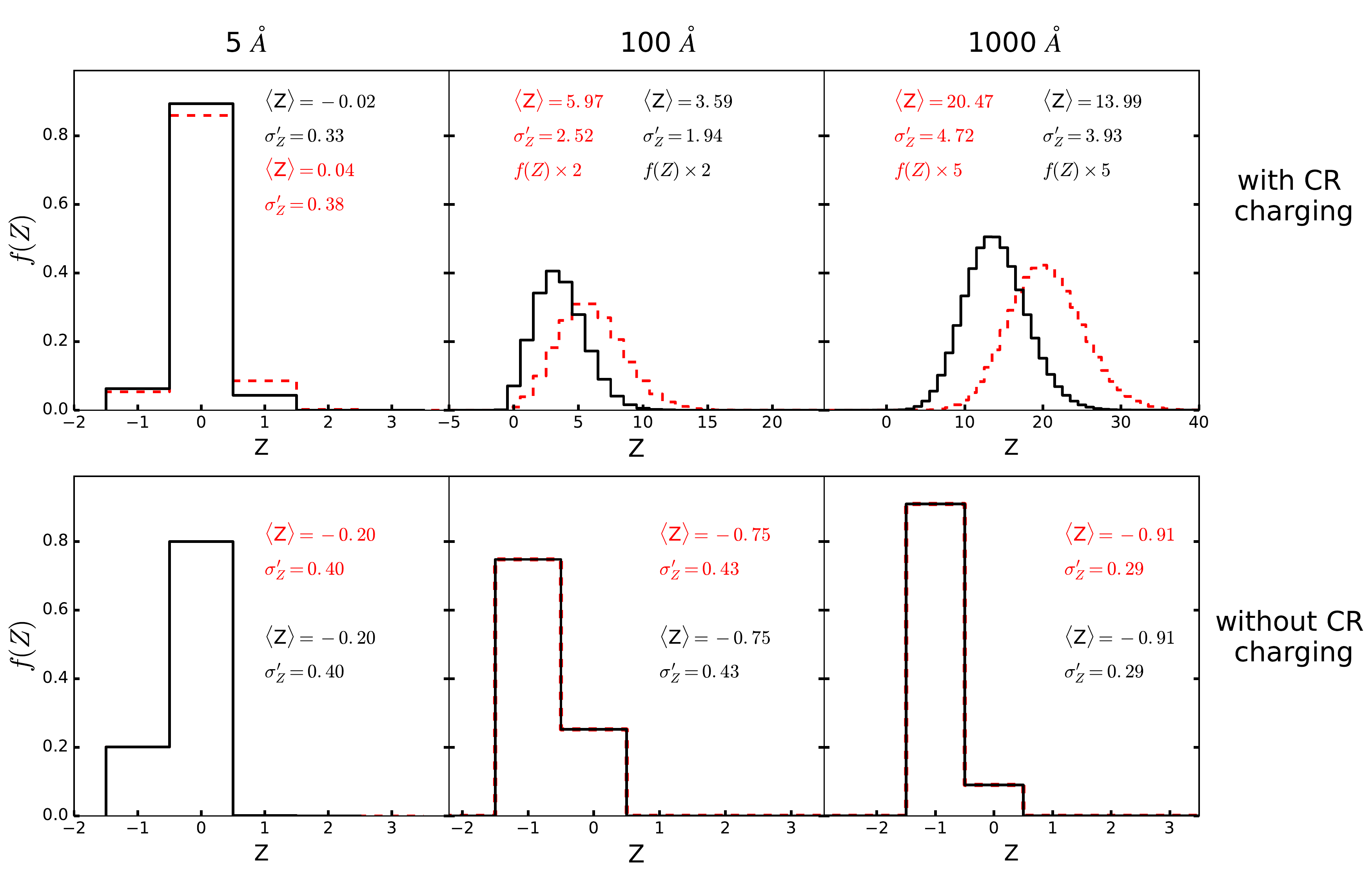} 
\vspace*{-0.4cm}
\caption{Charge distribution of $5$~{\AA}, $100$~{\AA} and $0.1~\mu$m size, carbonaceous ({\emph{dashed red}}) and silicate ({\emph{solid black}}) grains, with ({\emph{top row}}) and without ({\emph{bottom row}}) the effects of CR-induced charge rates.
The local environment parameters for the calculations correspond to the CMM in Figure \ref{fig:fz_sample}: $n_{\mathrm{H}} = 9 \times 10^{4}$~cm$^{-3}$, $T = 10$~K, $\chi_{e}=3.68\times10^{-8}$, $G=1.1\times10^{-4}$, $x_{\mathrm{H_{2}}} = 0.99$. 
Each panel includes the centroid, $\langle Z \rangle$, and width, $\sigma_{Z}$, of the distributions. 
\label{fig:CR_charging}}
\end{figure*}      

Figure \ref{fig:CR_charging}, shows the charge distribution of a small, intermediate and large, carbonaceous and silicate dust grain within a cold molecular region, with and without CR-induced charging effects. 
As discussed by \citet{Ivlev2015InterstellarEffects}, the local radiation field generated by H$_{2}$ fluorescence is the dominant charging mechanism, positively charging dust grains within cold-dense regions, thus significantly influencing the charge distribution of dust grains in these environments.
Although this might be only a minor effect for small dust grains, it has a strong influence on larger dust grains, where grain charges would be close to zero if CR-charging processes were not taken into account.

\section{Ionization fraction}
\label{appendix_ionization_fraction}

As discussed in section \ref{sec:colliding_flows}, the chemical network used in the simulations by \citetalias{Joshi2018OnConditions} produce very low ionization fractions within dense molecular regions.
This is because in the network, the electron recombination rate is set proportional to $G\sqrt{T}/n_{e}$, as suggested by \citet{Weingartner2001ElectronIonHydrocarbons}.
This assumes that the main recombination channel is a grain-assisted ion recombination.
However, as discussed by \citet{Ivlev2015InterstellarEffects}, the ionization equilibrium in dense molecular regions is determined by the balance between CR ionization and various recombination processes. 

In this work we use equation \ref{eq:caselli} to calculate the ionization rate within dense regions in a postprocessing step, where we request that for regions with $n_{\mathrm{H}} \geq 1000$~cm$^{-3}$, the new ionization fraction is the maximum between the one given by \citetalias{Joshi2018OnConditions}, or the one given by equation \ref{eq:caselli}.

\begin{figure*}
\centering 
\includegraphics[width=0.7\textwidth]{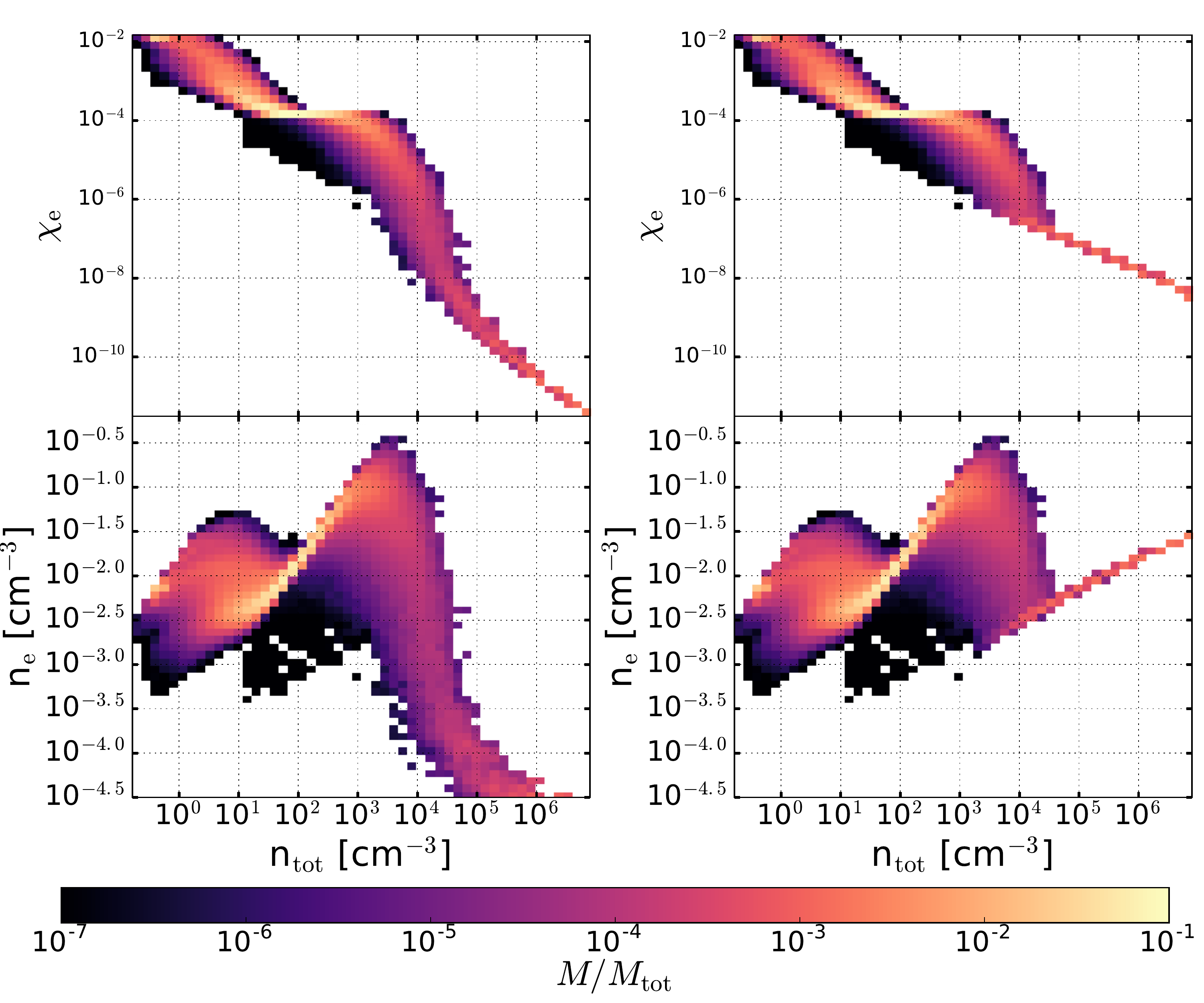} 
\vspace*{-0.2cm}
\caption{Probability distribution of the (\emph{top row}) ionization fraction as a function of total gas number density, and (\emph{bottom row}) electron density as a function of total gas number density, from the simulations by (\emph{left column}) \citetalias{Joshi2018OnConditions} and (\emph{right column}) after postprocessing using equation \ref{eq:caselli}. 
\label{fig:ionizationfraction_comp}}
\end{figure*}      

Figure \ref{fig:ionizationfraction_comp} shows a comparison between the ionization fraction and electron density in the simulations by \citetalias{Joshi2018OnConditions} (\emph{left column}), and the new ionization fraction and electron density after postprocess.
It is evident that the ionization fraction in the simulations by \citetalias{Joshi2018OnConditions} quickly drops at densities above $n_{\mathrm{H}}>1.0\times10^{3}$~cm$^{-3}$, due to the fast recombination rates used in the network.
However, after the postprocess, the ionization fraction drops proportional to $n_{\mathrm{H}}^{-0.56}$, such that the electron density increases towards dense regions, instead of decreasing.

\section{Completeness test}
\label{appendix_completeness_test}

In this work we stochastically sample 1\% of the total number of cells in the simulation to compute the charge distribution for silicate and carbonaceous grains between 3.5~{\AA} and 100~{\AA}, and 0.1\% of the cells for 500~{\AA} and 0.1~$\mu$m grains.
In this appendix we discuss the influence of these stochastic sampling and its influence on the calibration of the parametric equations.

We perform stochastic sampling of cells because of the extreme computational expense that is required to compute charge distributions.
The computation of $f(Z)$ requires to successively calculate and balance the charging rates between the minimum and maximum possible charges for a given dust grain.
While small grains allow only a narrow range of charges, e.g. 5~{\AA} silicate grain has $Z_{\mathrm{min}}=-1$ and $Z_{\mathrm{max}}=2$, large grains allow a very wide range of charges, e.g. 0.1~$\mu$m silicate grain has $Z_{\mathrm{min}}\approx-5000$ and $Z_{\mathrm{max}}\approx400$.
For this reason, the computation of the charge distribution becomes more and more expensive with increasing grain size.

\begin{figure*}
\centering 
\includegraphics[width=1.0\textwidth]{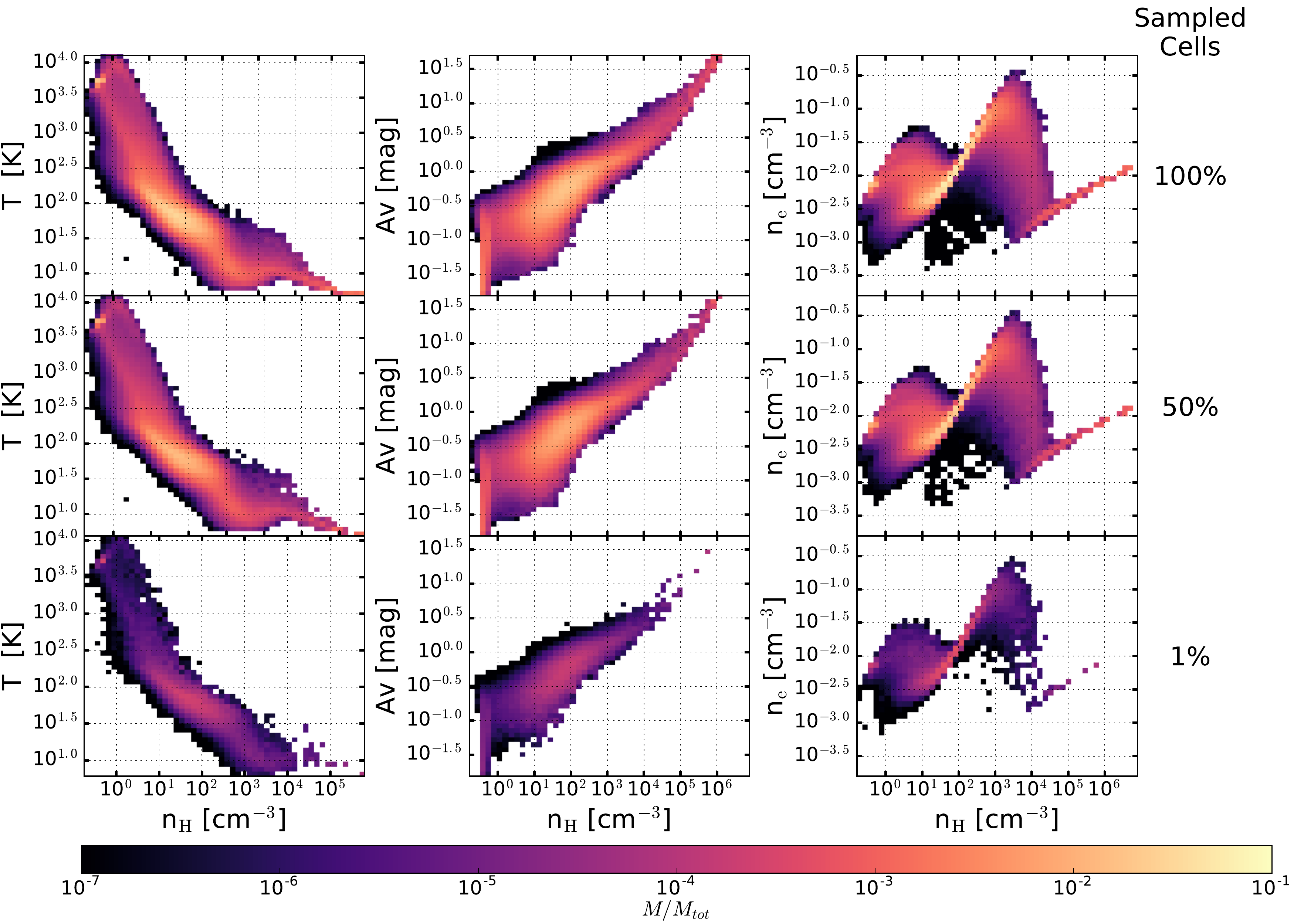} 
\caption{Mass weighted 2D PDF of the ({\emph{left}}) total gas number density vs temperature, ({\emph{middle}}) total gas number density vs visual extinction and ({\emph{right}}) total gas number density vs electron number density in the colliding flow simulation by \citetalias{Joshi2018OnConditions}. The columns present various percentages of stochastically sampled cells corresponding to {\emph{top}} 100\%, {\emph{middle}} 50\% and {\emph{bottom}} 1\%.
The pdf is normalized to the total mass in the 32~pc$^{3}$ box.
\label{fig:sampled_cells}} 
\end{figure*}

Figure \ref{fig:sampled_cells} shows the number density, temperature, local radiation field strength and electron number density distribution of all the cells, 50\% and 1\% stochastically sampled cells within a 32~pc$^{3}$ box centered in the plane of collision, see Figure \ref{fig:CF_projection}.
It is clear that, although there are significant differences in the coverage of these three cases, the 1\% sampled cells still recover the general trend of the fully sampled case.
However, in order to quantify the influence of the low coverage of the 1\% stochastically sampled cells in the fitted parameters in tables \ref{table:silicate}, and \ref{table:carbonaceous}, we compute the charge distribution of 50\% of the cells in the simulation for a 5~{\AA} silicate grain, and compare them to the predicted values using equation \ref{eq:charging_par} and the parameters in table \ref{table:silicate}.

\begin{figure}
\centering 
\includegraphics[width=0.5\textwidth]{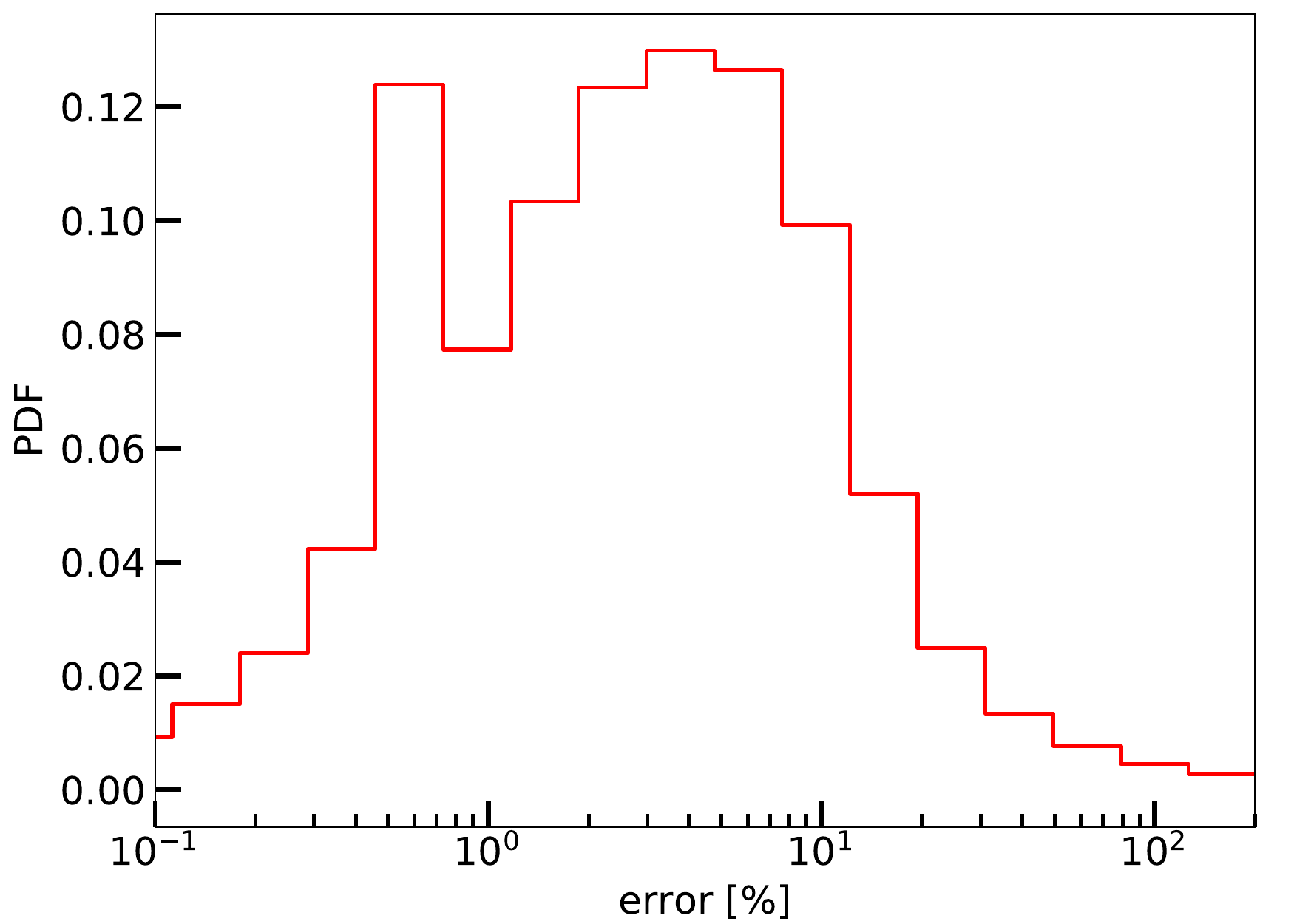} 
{\vspace{-0.4 cm}}
\caption{Distribution of errors between the predicted value of the charge centroid, using equation  \ref{eq:charging_par} and the parameters in table \ref{table:silicate}, and the expected value of the charge centroid, for the 50\% stochastically sampled cells.  
\label{fig:completeness_error}} 
\end{figure}

Figure \ref{fig:completeness_error} shows the distribution of errors between the predicted value of the charge centroid and the full calculation of the charge centroid, where the 25$^\mathrm{th}$, 50$^\mathrm{th}$ and 75$^\mathrm{th}$ percentile of the distribution correspond to an error of 1.3\%, 4.2\% and 10.3\%.
These result shows that, although we use only 1\% of the cells to calibrate the parameters, our parametric equations perform a good job at recovering the charge centroid of dust grains with only small errors in the calculation.

\section{Centroid and width distributions for Silicate grains}
\label{appendix_silicate_grains}

\begin{figure*}
\centering 
\includegraphics[width=0.90\textwidth]{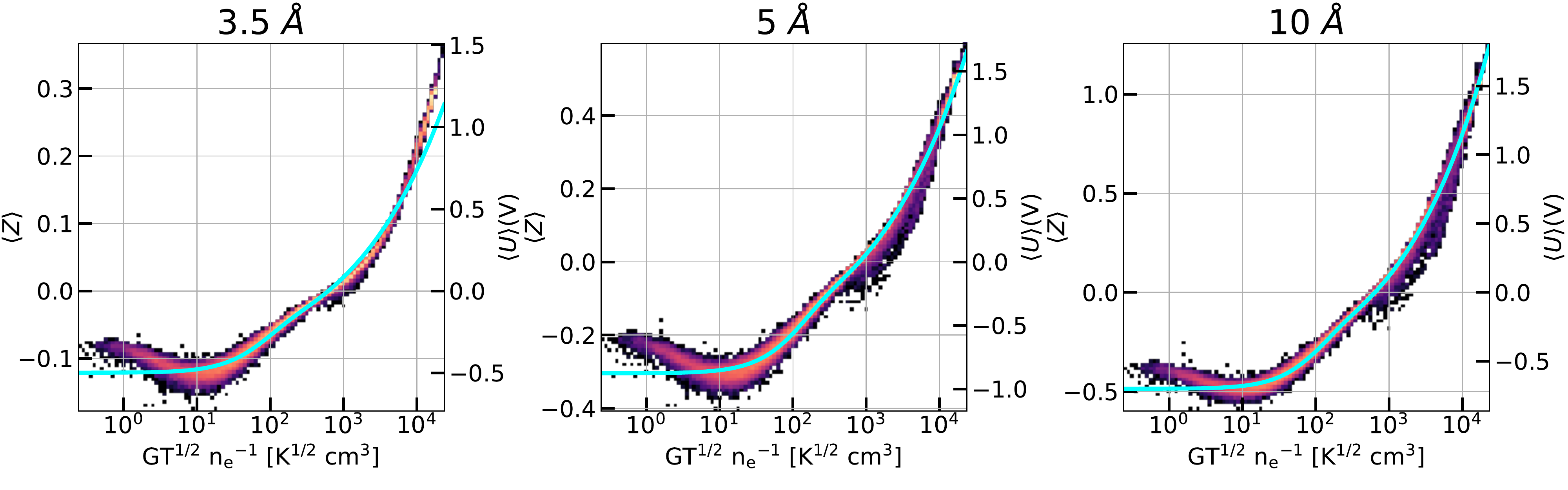} 
\includegraphics[width=0.90\textwidth]{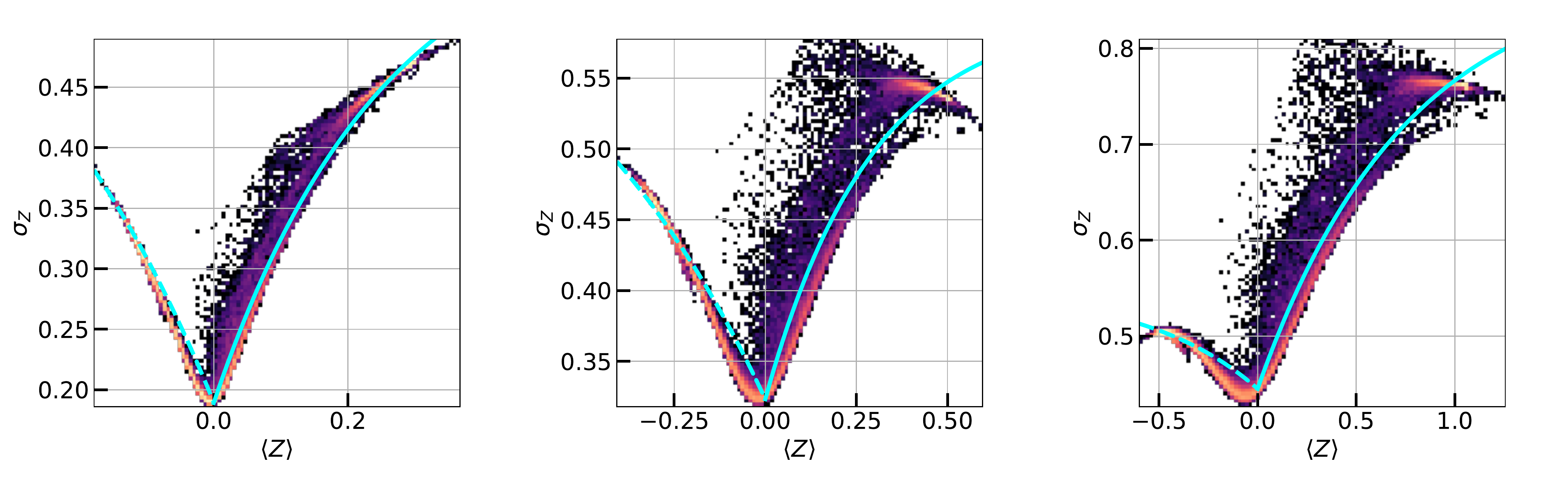} 
\includegraphics[width=0.90\textwidth]{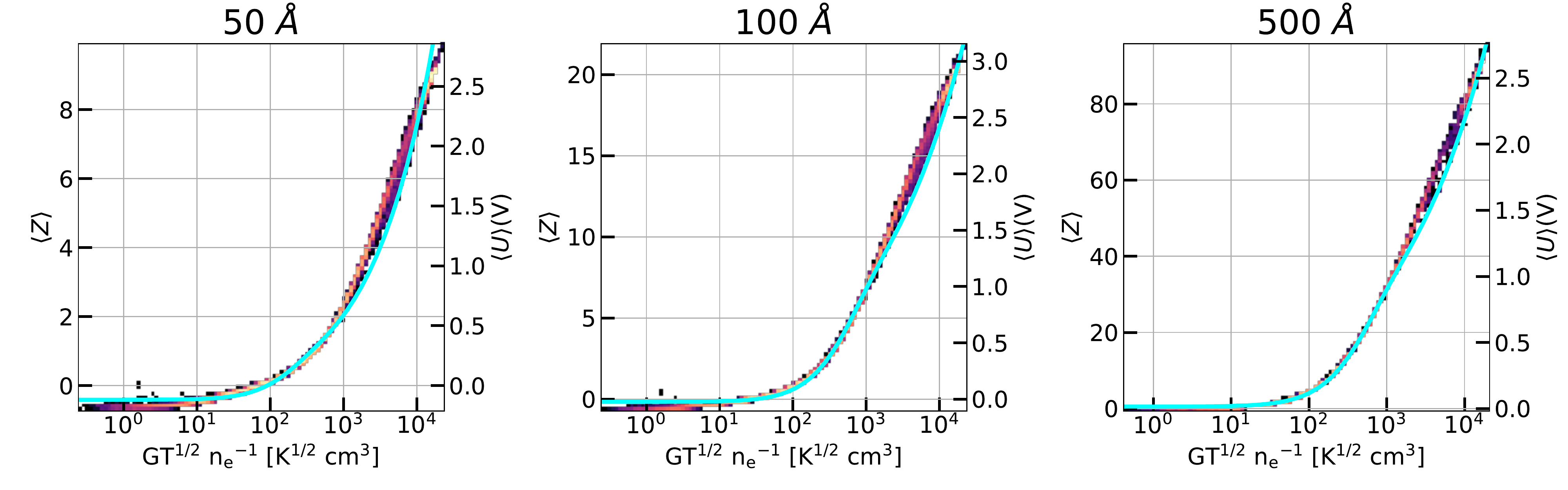} 
\includegraphics[width=0.90\textwidth]{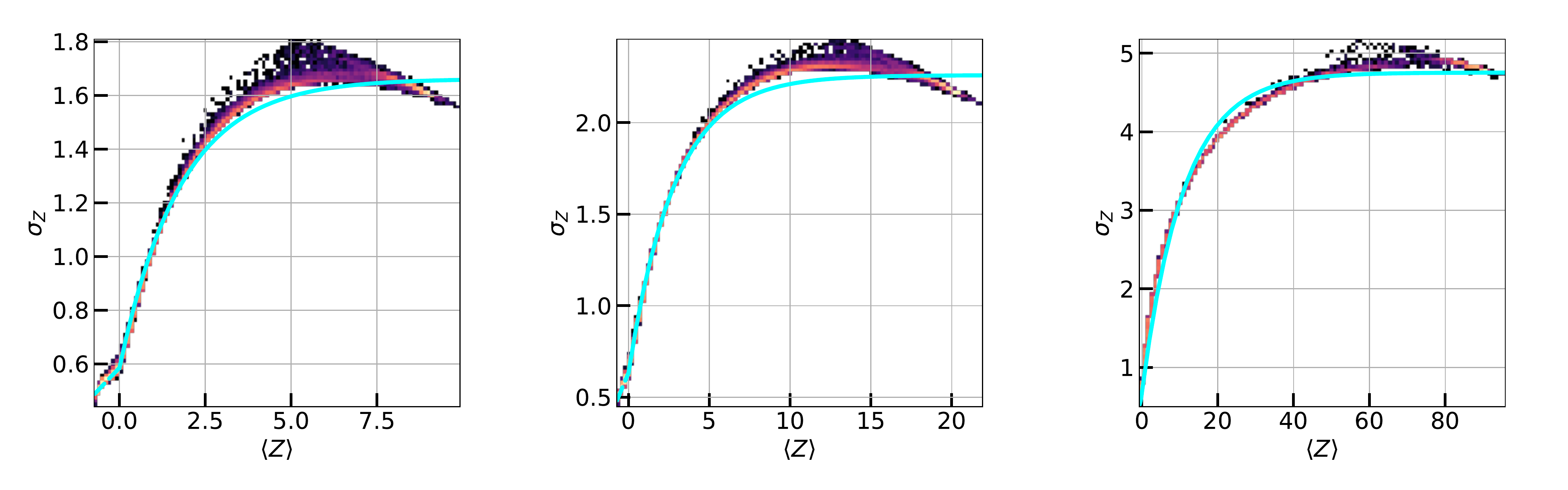} 
\caption{Distribution of charge centroids, $\langle Z \rangle$, and average electrostatic potential, $\langle U \rangle$(V) {\emph{upper rows}}, and widths, $\sigma_{Z}$, \emph{lower rows}, as a function of the charging parameter for silicate grains with sizes 3.5~\textrm{\AA}, 5~\textrm{\AA}, 10~\textrm{\AA},50~\textrm{\AA}, 100~\textrm{\AA}, 500~\textrm{\AA}.  
\label{fig:Zcent_all_silicate}} 
\end{figure*}

In this appendix, we show the distribution of charge centroids and widths for silicate grains with sizes 3.5, 5, 10, 50, 100, and 500 {\AA}. Figure \ref{fig:Zcent_all_silicate} shows these distributions including the parametric equation for the centroid and width, \ref{eq:charging_par}, \ref{eq:width_par_pos} and \ref{eq:width_par_neg} respectively, with the best fit parameters shown in Table \ref{table:silicate}.

\section{Charge distribution for Carbonaceous grains}
\label{appendix_carbonaceous_grains}

In this appendix, we show the figures of the distribution of charge centroids and widths, charging timescales and fits to the charge distributions for carbonaceous grains.

\begin{figure*}
\centering 
\includegraphics[width=0.95\textwidth]{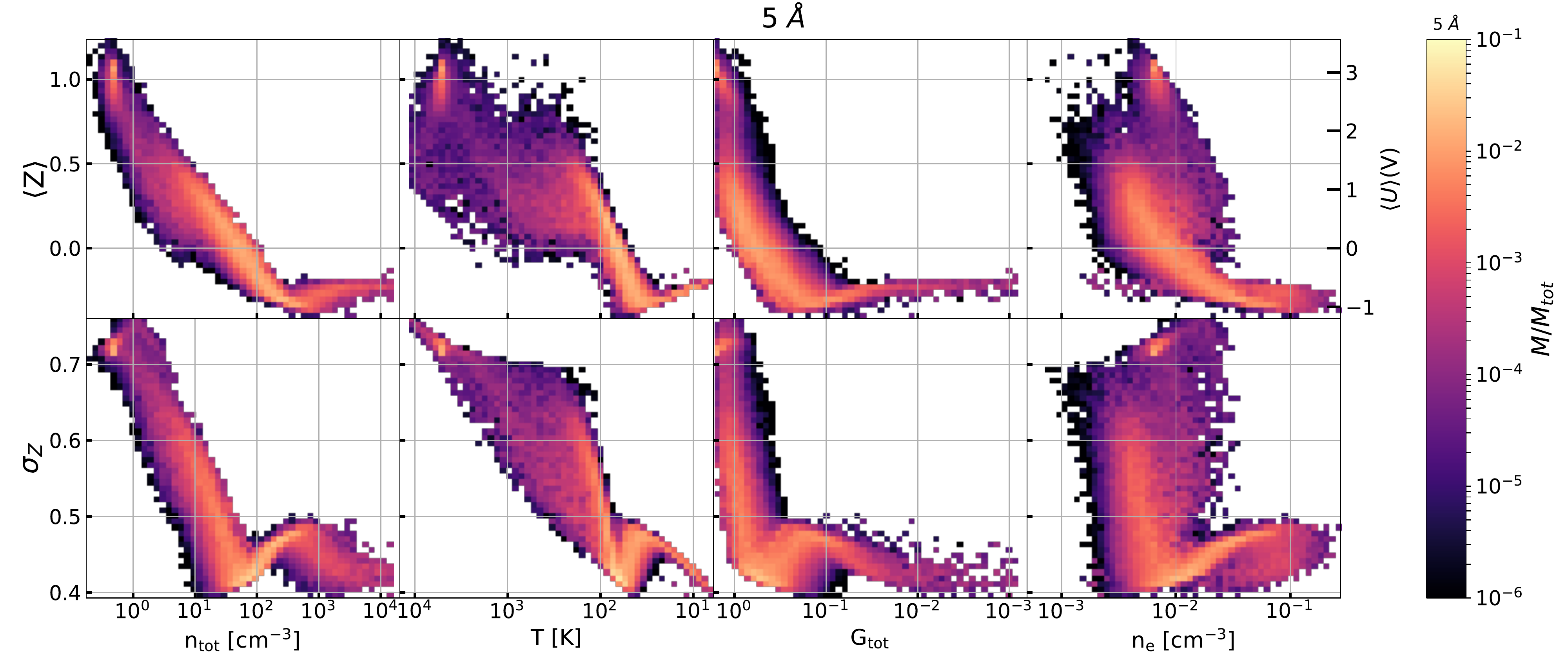} 
\includegraphics[width=0.95\textwidth]{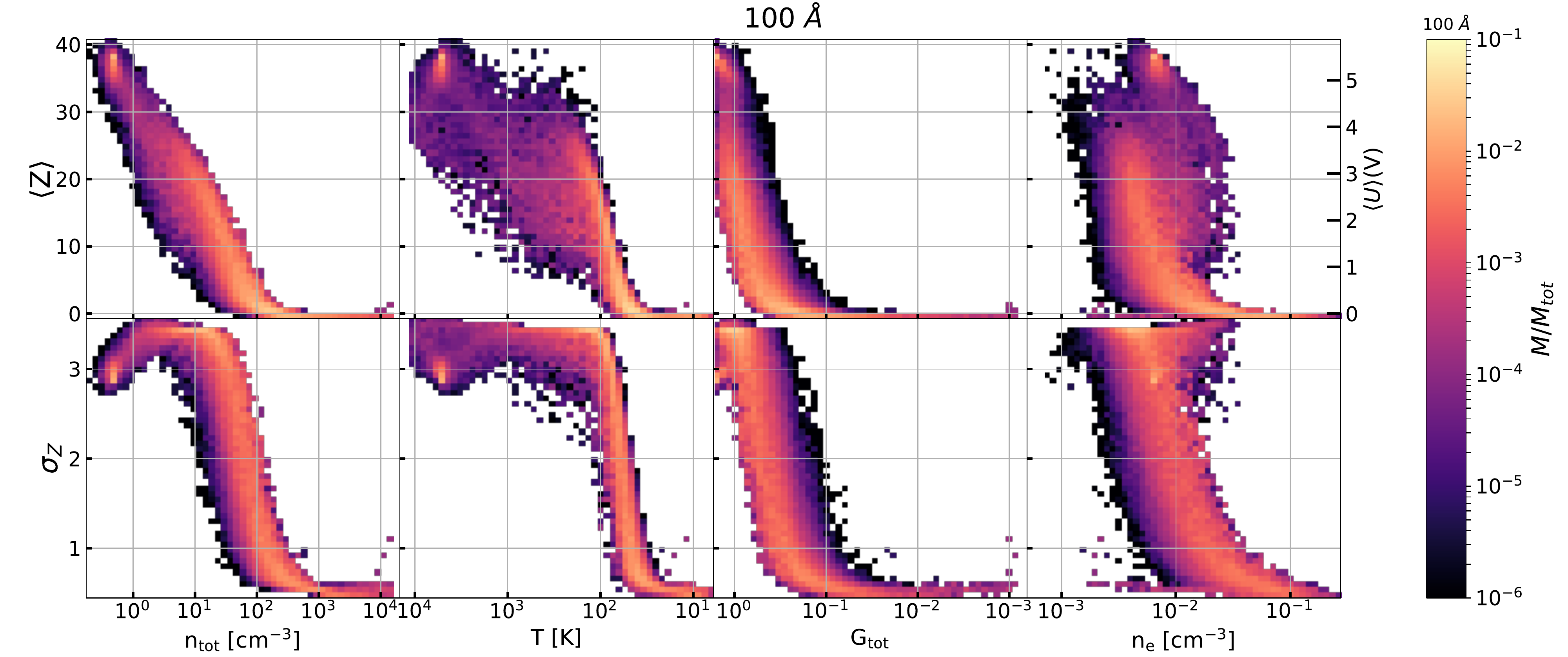} 
\includegraphics[width=0.95\textwidth]{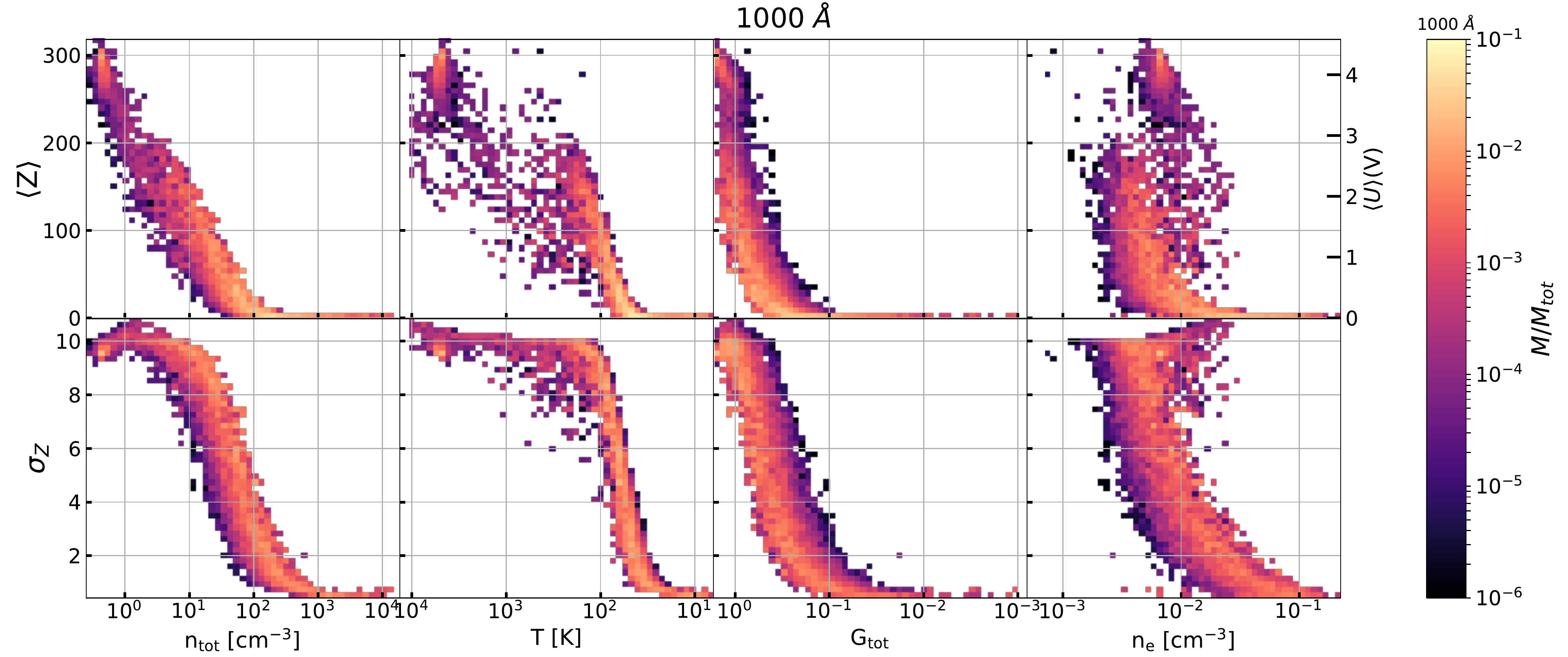} 
\caption{Distribution of charge centroids, $\langle Z \rangle$, and average electrostatic potential, {\emph{upper rows}}, and widths, $\sigma_{Z}$, \emph{lower rows}, as a function of total gas number density, temperature, strength of the interstellar radiation field, in units of Habing field $G$, and electron number density, from left to right, for the stochastically sampled cells within the MHD simulation, for carbonaceous grains of of sizes 5~\textrm{\AA} (\emph{top}), 100~\textrm{\AA} (\emph{middle}), and $1000~$\textrm{\AA} (\emph{bottom}).  
\label{fig:fz_carb}} 
\end{figure*}

Figure \ref{fig:fz_carb} shows the distribution of charge centroids and widths for a small, intermediate and large carbonaceous grains. 
Similar to the discussion in figure \ref{fig:fz_sil}, the shape of the distributions is similar for the different sizes, however the range of the minimum and maximum charge centroids and widths is very different for the different grain sizes.
Compared to the silicate grains, carbonaceous grains reach higher positive charges because they have a lower work function, such that the photoelectric charging is more effective at stripping electrons from the grain.

\begin{figure*}
\centering 
\includegraphics[width=0.90\textwidth]{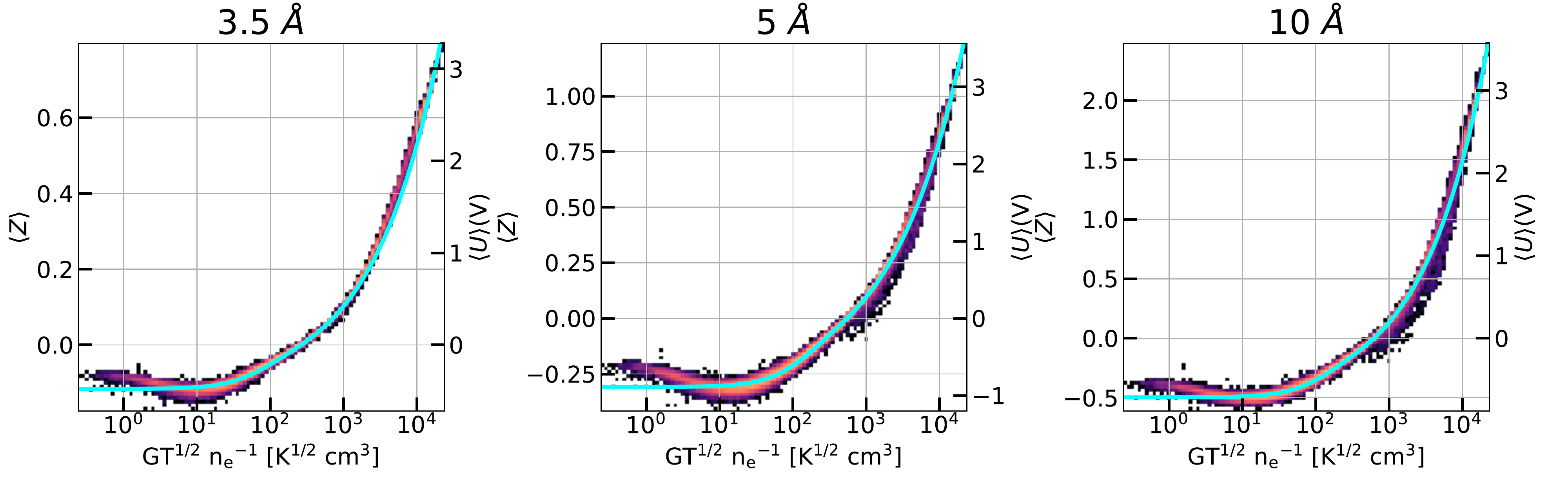} 
\includegraphics[width=0.90\textwidth]{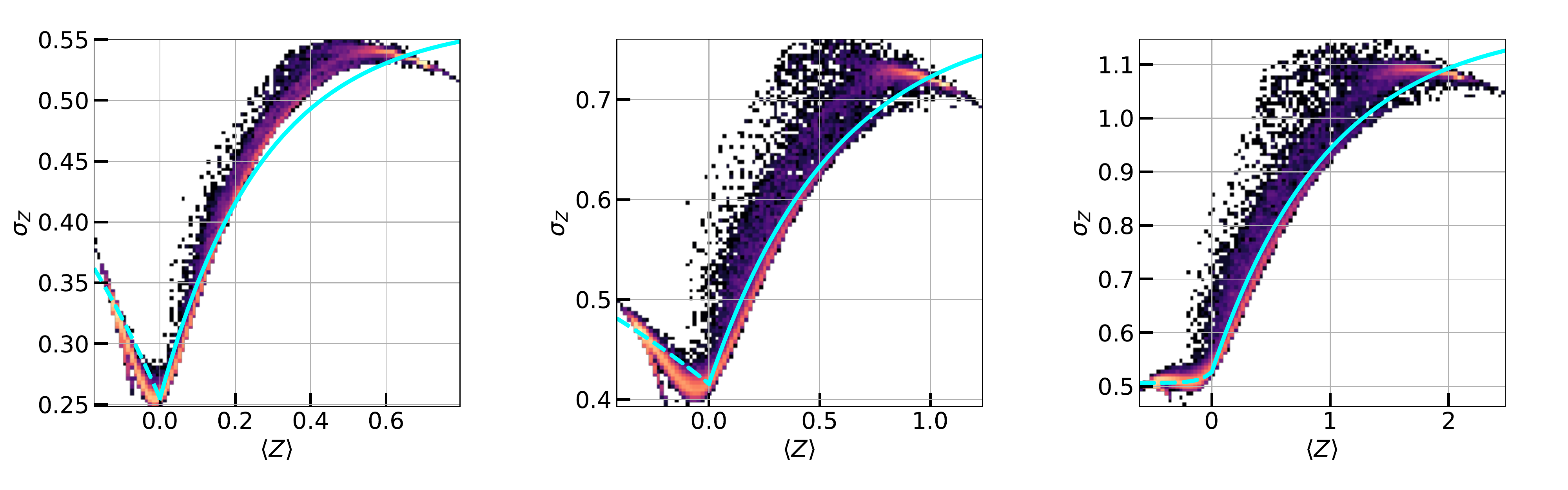} 
\includegraphics[width=0.90\textwidth]{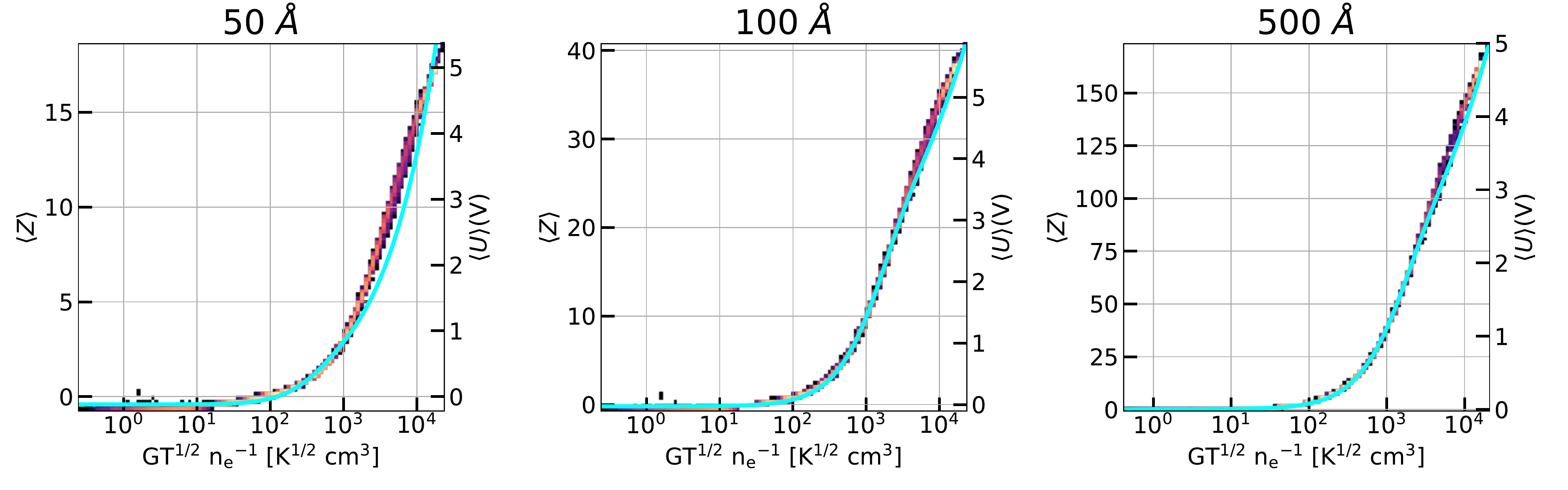} 
\includegraphics[width=0.90\textwidth]{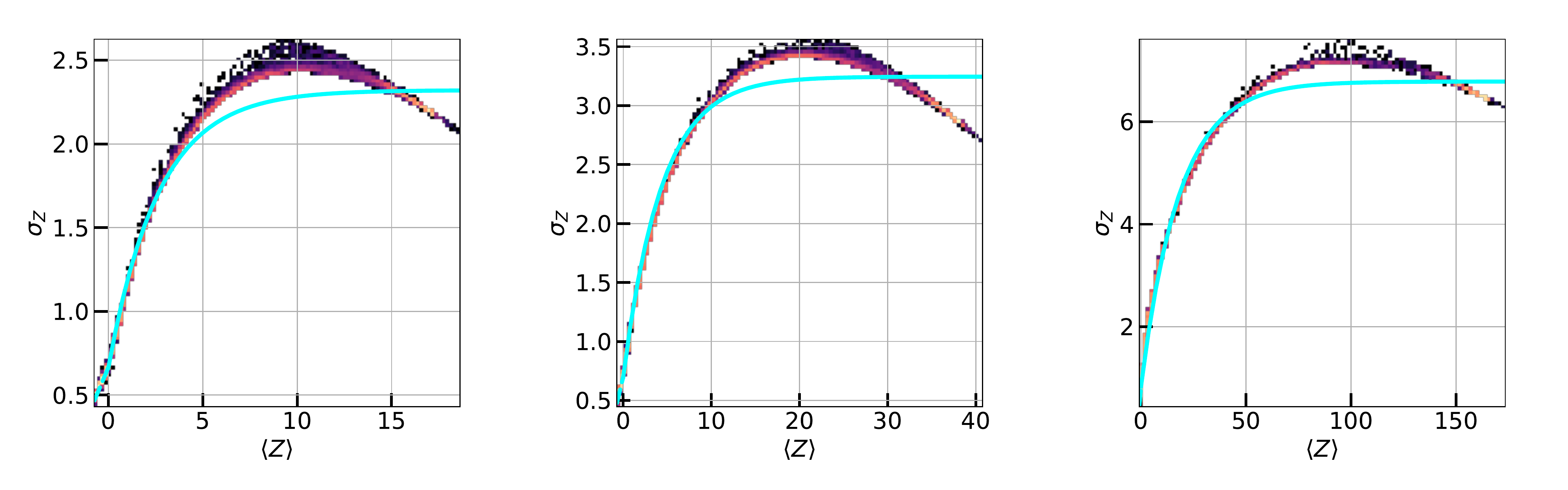} 
\caption{Distribution of charge centroids, $\langle Z \rangle$, and average electrostatic potential, {\emph{upper rows}}, and widths, $\sigma_{Z}$, \emph{lower rows}, as a function of the charging parameter including the new polynomial fit, equations \ref{eq:charging_par}, \ref{eq:width_par_pos} and \ref{eq:width_par_neg}, for carbonaceous grains with sizes 3.5~\textrm{\AA}, 5~\textrm{\AA}, 10~\textrm{\AA},50~\textrm{\AA}, 100~\textrm{\AA}, 500~\textrm{\AA}.  
\label{fig:Zcent_all_carbonaceous}} 
\end{figure*}

\begin{figure}
\centering 
\includegraphics[width=0.46\textwidth]{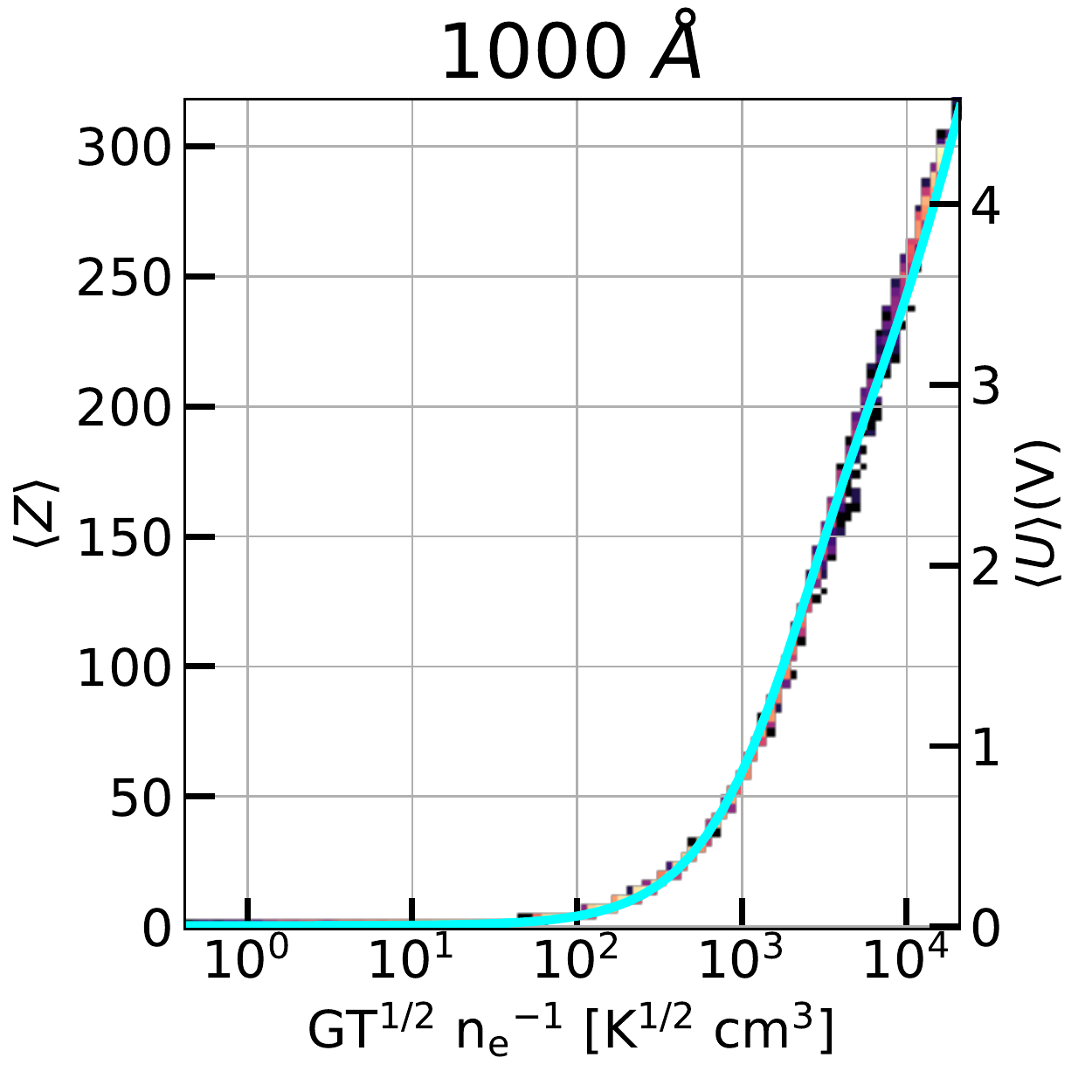} 
\includegraphics[width=0.45\textwidth]{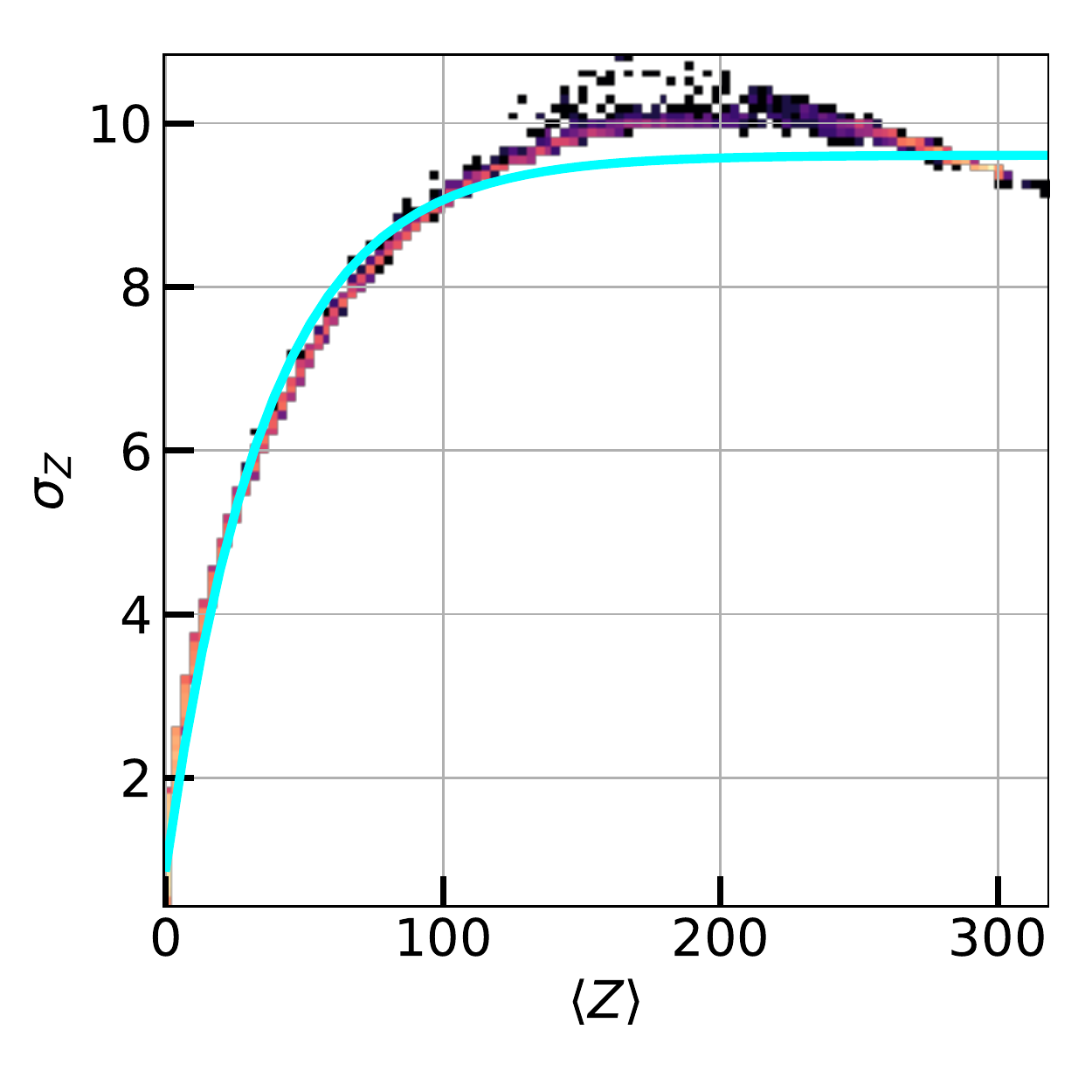} 
\caption{Distribution of charge centroid, $\langle Z \rangle$, and average electrostatic potential, {\emph{upper row}}, and width, $\sigma_{Z}$, \emph{lower row}, as a function of the charging parameter for 1000~{\AA} carbonaceous grain.  
\label{fig:Zcent_1000_silicate}} 
\end{figure}

Figures \ref{fig:Zcent_all_carbonaceous} and \ref{fig:Zcent_1000_silicate} show the distribution of charge centroids and widths for all the carbonaceous grains analyzed here, including the new parametric equations for the charge centroid, eq. \ref{eq:charging_par}, and the parametric equations for the width, eqs. \ref{eq:width_par_pos} and \ref{eq:width_par_neg}.

\begin{figure}
\centering 
\includegraphics[width=0.45\textwidth]{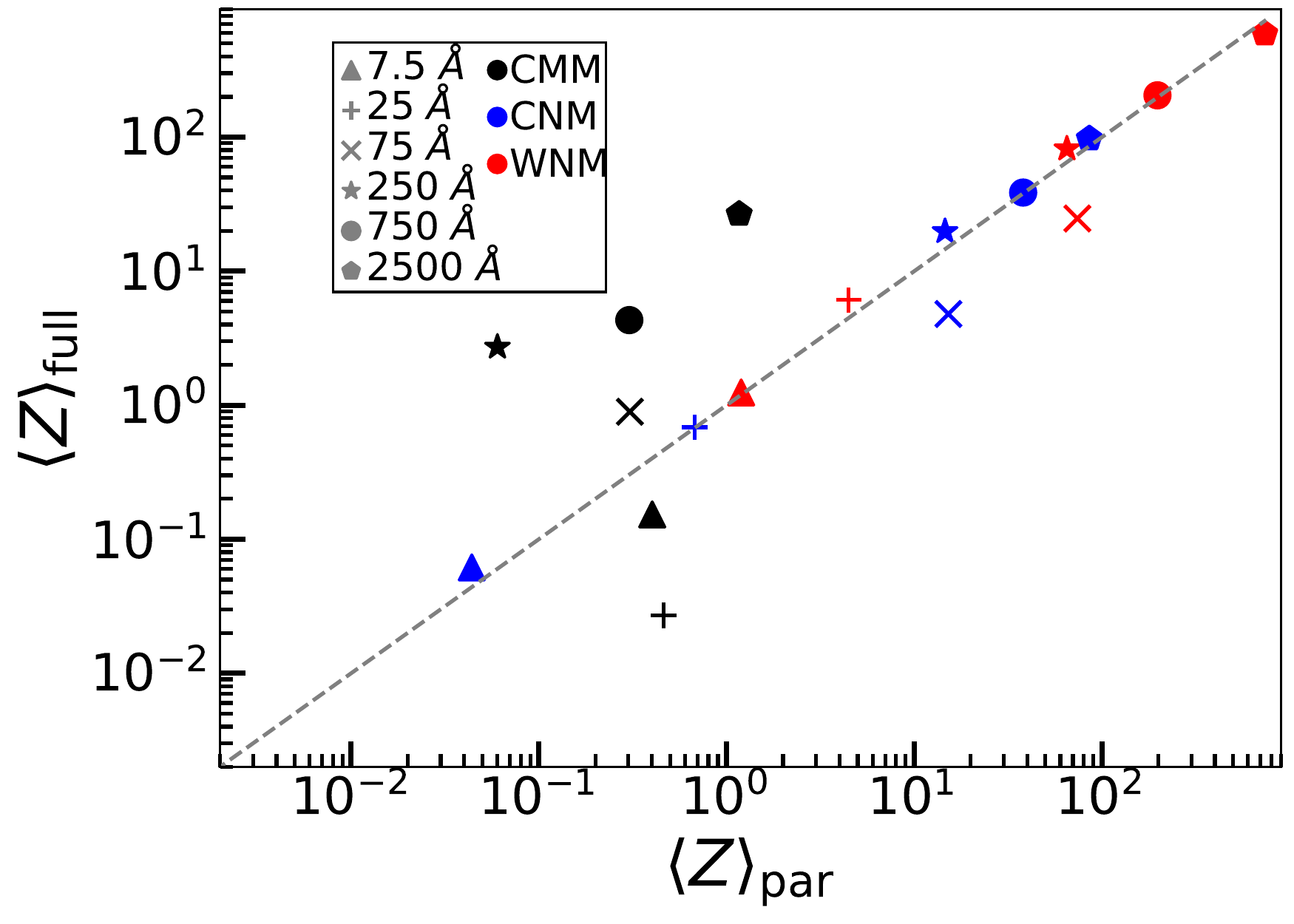}
\caption{Charge centroid calculated using the full calculation of the charge distribution, $\langle Z \rangle _{\mathrm{full}}$, versus the one calculated using the parametric equation, eq. \ref{eq:charging_par}, for carbonaceous grains of five different sizes in three different environments. Grain sizes are differentiated by markers, and ISM ambient parameters by colors. WNM, CNM, and CMM, conditions used for the calculation are the same as the ones used in figure \ref{fig:fz_sample}. 
\label{fig:test_carb}} 
\end{figure}

Figure \ref{fig:test_carb} shows the charge centroid computed using the full calculation of the charge distribution, $\langle Z \rangle _{\mathrm{full}}$, versus the centroid calculated using the parametric equation, eq. \ref{eq:charging_par}, $\langle Z \rangle _{\mathrm{par}}$.
The parameters for the grains between 3.5~{\AA} and 1000~{\AA} are calculated using linear interpolation of the values in table \ref{table:carbonaceous}, and the parameters for the grain with size 2500~{\AA} extrapolates these parameters.
We find good agreement between the calculations of the centroids in the cold and warm neutral medium.
Some deviations between the full calculation and the parametric equation values of the centroid are observed in the cold molecular medium, where the is the larger scatter of the distribution of centroids as a function of the charging parameter.
In particular for grains larger than 250~{\AA}, it appears that the parametric calculation systematically underestimates the centroid of the distribution.

\begin{figure}
\centering 
\includegraphics[width=0.48\textwidth]{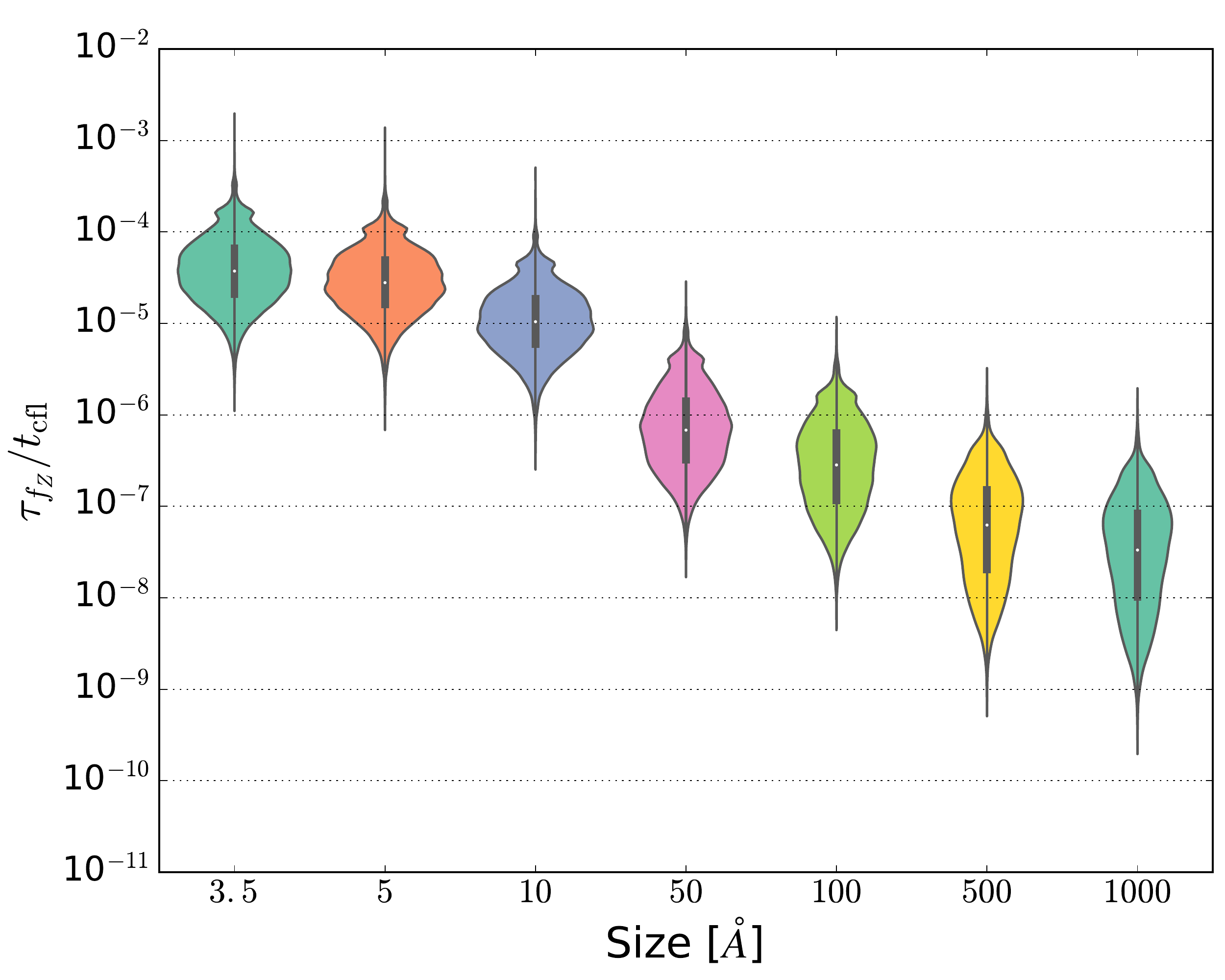} 
\vspace*{-2mm}
\caption{Ratio of charging timescale and local CFL timescale at each point where the dust charge distribution was evaluated for carbonaceous grains with sizes between 3.5~{\AA} and 1000~{\AA}. The shape of each object in the violin plot represents the distribution of relative timescales. The central white point corresponds to the median, the thick black line to the inter-quantile range and the thin black line to the 95\% confidence interval.}
\label{fig:charging_equilibrium_carb} 
\end{figure}

Figure \ref{fig:charging_equilibrium_carb} shows the charging timescale relative to the CFL timescale in the simulations for carbonaceous grains with different sizes.
Similar to the case for silicate grains, the timescales necessary for the charge distribution to be in equilibrium is several orders of magnitude smaller than the CFL timescale in our simulations, suggesting that this is a valid approximation in our model.

\label{lastpage}
\end{document}